\documentclass[a4paper,11pt,notitlepage]{article}

\usepackage{a4wide,epsfig,graphics,amsthm,amsfonts,amssymb,amsopn,amsmath,longtable,dcolumn,calc,verbatim,threeparttable,eurosym,lscape,rotating,multirow, color, ulem, threeparttablex, environ, booktabs, hyphenat}
\usepackage[paper=a4paper,left=25mm,right=25mm,top=28mm,bottom=28mm]{geometry}
\usepackage[utf8]{inputenc}
\usepackage{bm} 
\usepackage{lscape}
\usepackage{natbib}
\usepackage{caption}
\usepackage{pdfpages}
\usepackage{ragged2e,array,longtable}
\usepackage{arydshln}
\usepackage{booktabs}
\usepackage[onehalfspacing]{setspace}
\usepackage{bbm}
\newcolumntype{.}{D{.}{.}{-1}}
\newcolumntype{d}[1]{D{.}{.}{#1}}

\renewcommand{\baselinestretch}{1.5} 
\theoremstyle{definition}

\theoremstyle{plain}
\usepackage{diagbox}
\usepackage{graphicx}
\usepackage{subfig}
\usepackage[colorlinks=true,linkcolor=blue,citecolor=blue,backref=page]{hyperref}
\usepackage{paralist}
\usepackage[bottom]{footmisc}
\usepackage{titletoc}
\usepackage[en-US]{datetime2}

\newcommand{\bi}{\begin{itemize}}
\newcommand{\ei}{\end{itemize}}

\newcommand{\bc}{\begin{center}}
\newcommand{\ec}{\end{center}}

\newcommand{\bs}{\begin{scriptsize}}
\newcommand{\es}{\end{scriptsize}}
\newcommand{\beq}{\begin{equation}}
\newcommand{\eeq}{\end{equation}}
\newcommand{\ben}{\begin{enumerate}}
\newcommand{\een}{\end{enumerate}}
\newcommand{\bean}{\begin{eqnarray}}
\newcommand{\eean}{\end{eqnarray}}

\DeclareMathSymbol{\R}{\mathalpha}{AMSb}{"52}
\DeclareMathSymbol{\E}{\mathalpha}{AMSb}{"45}
\newcommand{\sym}[1]{$^{#1}$} 

\usepackage{amssymb}
\usepackage{pifont}
\usepackage{float}

\def\sym#1{\ifmmode^{#1}\else\(^{#1}\)\fi}

\usepackage[section]{placeins}
\newlength{\asdf}
\usepackage{xargs}   
\usepackage[pdftex,dvipsnames]{xcolor}  
\usepackage[colorinlistoftodos,prependcaption,textsize=tiny]{todonotes}
\newcommandx{\DCC}[2][1=]{\todo[inline, linecolor=red,backgroundcolor=red!25,bordercolor=red,#1]{#2}} 
\newcommandx{\MC}[2][1=]{\todo[inline, linecolor=blue,backgroundcolor=blue!25,bordercolor=blue,#1]{#2}} 

\begin{document}

\renewcommand{\baselinestretch}{1.12}
\title{\vspace{-10mm}\textbf{Do Unions Shape Political Ideologies at Work?} \thanks{We thank Marco Caliendo, Filipe Campante, David Card, Axel Dreher, Henry Farber, James Feigenbaum, Andreas Fuchs, Alexander Hertel-Fernandez, Katrin Huber, Simon Jäger, Ethan Kaplan, Krisztina Kis-Katos, Simon Luechinger, Dennis Quinn, Sulin Sardoschau, Thomas Siedler, and Samuel Young, as well as participants at the ifo Workshop on Political Economy (Dresden, 2022), BSoE Summer Workshop (Berlin, 2022), Research Seminar for Applied Microeconomics (FU Berlin, 2022), Beyond Basic Questions Workshop (Bern, 2022), GlaD-DENeB Workshop (Goettingen, 2022), Development Economics Seminar (Goettingen, 2022), Potsdam Research Seminar in Economics (Potsdam, 2023), IAAEU Colloquium on Economics (Trier, 2023), ESPE Annual Conference (Belgrade, 2023), EEA-ESEM Congress (Barcelona, 2023), EALE Conference (Prague, 2023), Society for Institutional \& Organizational Economics (Frankfurt, 2024), Harvard Political Economy Workshop (Cambridge, 2024), IZA Summer School (Bonn, 2024), and Verein für Socialpolitik (Berlin, 2024) for helpful comments. We are grateful to Maximilian Biller, Sunna Hügemann, Tilman Jacobs, Luise Koch, and Yasmin Zysk for excellent research assistance.}}

\author{
	\textbf{Johannes Matzat}  
	\thanks{University of Lucerne; e-mail: \texttt{johannes.matzat@unilu.ch}.} \hspace{2em}
	\textbf{Aiko Schmeißer} 
	\thanks{Columbia University: e-mail: \texttt{as8091@columbia.edu}.} \\
    \hspace{11.5em} (Job Market Paper)
	\bigskip
}

\date{
        \today
	\\ 
        \href{https://arxiv.org/pdf/2209.02637}{Latest Version Here}
	\vspace{-5mm}
}

\maketitle

\begin{abstract}
	
	\noindent

Labor unions influence economic outcomes not only through bargaining with employers over work contracts but also via political activities that can profoundly shape political systems. In unionized workplaces, they may mobilize and change the ideological positions of both unionizing workers and their non-unionizing management. In this paper, we analyze the workplace-level impact of unionization on workers' and managers' political campaign contributions. We link establishment-level union election data with transaction-level campaign contributions to federal and local candidates in the United States. Using a difference-in-differences design, validated through regression discontinuity tests and a novel instrumental variable approach, we find that unionization leads to a leftward shift of campaign contributions. Unionization increases support for Democrats relative to Republicans not only among workers but especially among managers, suggesting that managers converge toward workers' political preferences. The effects are stronger in settings with more cooperative union-employer interactions, such as when union elections are not contested by an unfair labor practice charge and result in a collective bargaining agreement.

	\vspace{0.5cm}
	
\end{abstract}
\thispagestyle{empty}
\clearpage
\setcounter{page}{1}

\section{Introduction}

Labor market institutions are typically evaluated by economists based on their ability to solve market failures and improve economic efficiency. Yet, the social welfare implications of labor market institutions also hinge critically on their potential ramifications for the political system \citep{Acemoglu2013}. Labor unions have historically functioned as institutions dedicated to redistributing both economic and political power within society. While a long tradition of research has focused on how unions affect work contracts of their members via collective bargaining \citep{Card1996, DiNardo2004, Farber2020, Frandsen2020, Knepper2020}, unions seek to shape economic outcomes more broadly by engaging in various political activities. They are often viewed as one of the few vehicles that give political voice to workers, counterbalancing the dominant influence of business elites in politics \citep{Acemoglu2023,Rosenfeld2014, Schlozman2015}.\footnote{Unions have been credited with shaping key welfare and labor market policies, such as the 8-hour workday, minimum wages, safety standards, and sickness and family leave \citep[e.g.,][]{Biden2021, King1965, Obama2010}. They invest significant resources into political activities. In 2010, labor unions in the United States employed over 3,000 full-time political workers and spent 700 million USD on political outreach, a figure that rose to 1.8 billion USD in 2020 \citep{NILRR2021, WSJ2012}.}

The greatest political leverage of U.S. labor unions likely stems from their connection with more than 14 million union members and their colleagues in unionized workplaces. After family and friends, the workplace is one of the most important arenas for social interaction and political discussion \citep{Hertel2020}, making it a particularly influential space for unions. By providing political information and training as well as facilitating communication networks between members, unions can mobilize workers and affect their ideological positions. Nonetheless, the overall political influence of unions in the workplace is far from clear. Even if unions successfully rally unionized workers around their political positions, it is unclear whether they can also persuade the firm's management. Heightened tensions between workers and managers, who act on behalf of ownership interests, may provoke adverse reactions to labor demands. Any backlash in the political behavior of this powerful out-group may prevent unions from shifting political power toward worker interests. 

In this paper, we examine the effect of labor unions on the political participation and political ideologies of workers and managers in the United States. We combine establishment-level data on 6,063 union elections with transaction-level data on 357,436 campaign contributions made to federal and local candidates from 1980 to 2016. In the campaign contribution data, we observe the employer, occupation, and address of individual donors, which allows us to match donors from different occupations with the union election results of their employing establishments. To estimate the causal effects of unionization, we employ a difference-in-differences (DiD) framework, comparing campaign contributions of employees in establishments that voted for unionization with those that voted against it (tests of the underlying parallel trends assumption and alternative sources of exogenous variation are described below). We assess the political effects of unions by examining two key outcomes: political mobilization, measured by changes in employees' total contribution amounts, and ideological shifts, captured by changes in the party composition of candidates they donate to. Linking these outcomes to union elections at the establishment level offers various new opportunities for studying the political influence of unions. 

To start with, our approach enables us to analyze the political effects of unions in the workplace, where both unionizing workers and their non-unionizing management can be affected. Exploiting the occupational information in the campaign contribution data, we can differentiate the political responses of workers and managers, allowing us to explore within-firm dynamics that have not been studied before. We first ask how workplace unionization alters the political behavior of workers. As \cite{Kerrissey2013} argue, unions provide their members with political capital -- they inform, engage, and mobilize members. Unions dedicate significant resources to political outreach and education, informing members on issues related to their working conditions and typically advocating Democratic Party positions \citep{Ahlquist2013, Iversen2015, Macdonald2021}.\footnote{Moreover, employee gatherings, voting for union officers, participation in hiring halls, and joint strike activities can improve communication networks between workers and create social experiences that turn them into more engaged citizens \citep{Lindvall2013, McAdam2001, Terriquez2011}.} Through these mechanisms, unionization can change the political participation and ideological positions of their members.

Our main results show that successful workplace unionization leads to notable increases in workers' political donations, particularly in favor of Democratic candidates. We find a significant rise in total campaign contributions of workers in the cycle of the union election, suggesting a short-term mobilization effect from the union campaign. Most importantly, when we examine the party composition of contributions, we find that unionization increases the percentage difference in donations from workers to Democrats versus Republicans by 12 percentage points in the six years following a union election.\footnote{The percentage interpretation of our results requires assumptions about the weighting of extensive- and intensive-margin changes in contribution amounts \citep{Chen2024}. For reasonable weights, we estimate an increase in the Democratic-Republican gap of workers' donations between 11 and 16 percentage points.} This result indicates a lasting shift in workers’ ideological positions towards the political left. 

Focusing solely on union members would overlook a critical out-group: the firm's management, whose responses can either amplify or counteract the political influence of unions. On one hand, unionization may foster greater understanding of worker issues. Unions give workers a collective voice, presenting their preferences on a more equal footing with management \citep{Freeman1970, Freeman1984}. The establishment of clear bargaining rules may improve both the quantity and quality of communication between managers and workers \citep{Verma2005}. According to contact theory, this increase in cooperative interactions can enhance perspective-taking and, ultimately, lead to an alignment of ideological positions \citep[e.g.,][]{Allport1954}. On the other hand, labor unions may provoke a backlash from management. Representing the interests of firm owners, managers are typically profoundly hostile to unionization.\footnote{In the run-up to union elections, employers frequently hire anti-union law firms and consultants, try to delay the election process, hold meetings in which employees are obligated to listen to the anti-unionization arguments, and -- although legally restricted -- threaten employees with dismissals and establishment closures \citep{Freeman1990a, Kleiner2001, Logan2002, Schmitt2009}.} The group threat hypothesis posits that the increased bargaining power of workers may be perceived by managers as undermining their own status, power, and economic interests \citep[e.g.,][]{Sherif1961}. When unionization heightens the salience of distributional conflicts and reinforces worker-manager identities, it may increase political polarization, as individuals increasingly adopt the stereotypes associated with these identities \citep{Bonomi2021}. Thus, it is unclear ex ante whether labor unions can persuade managers to align with workers' political positions or whether they reinforce the management's opposition. 

Our results notably reveal a leftward shift in campaign contributions not only for workers but also for managers. By exploiting occupational information in the campaign contribution data, we can directly identify the response of managers. We find that unionization increases the relative difference in managers’ donations to Democratic versus Republican candidates by 20 percentage points, without affecting their total spending.\footnote{The estimates vary between 19 and 26 percentage points for alternative extensive vs. intensive margin weights.} These results suggest that, instead of increasing tensions, unionization fosters a convergence in ideological positions between workers and their management.

Combining establishment-level political outcomes with variation in union elections also provides plausible identification strategies to identify the causal impact of unionization on employees' political behavior. Since we only consider establishments with union elections, i.e., where workers have shown an interest in unionization, our sample is likely more similar than a random sample of establishments. Within this sample, we compare campaign contributions between establishments where workers voted for and against unionization in a stacked DiD model.\footnote{The stacking approach addresses weighting issues that arise under staggered treatment timing and heterogeneous treatment effects \citep{Goodman-Bacon2021}. Our results are robust to employing alternative staggered DiD estimators \citep{Borusyak2021, Callaway2021}.}

We validate the underlying parallel trends assumption by complementing the DiD framework with tests originating from a regression discontinuity design (RDD) and with a novel instrumental variables (IV) approach. First, we test whether changes in outcomes are correlated with the pro-union vote share among establishments that lost the union election. Since the treatment status changes discontinuously at the 50\% threshold, there should be no differential trends across establishments with different vote shares below 50\%. Indeed, we do not find any evidence for differential changes across different vote shares, which helps us to rule out the possibility that any sizeable confounding factors correlated with the pro-union vote share and the timing of the election drive the results. Second, we restrict the sample to establishments with increasingly close elections, which are more likely to follow similar trends in contribution patterns. Our results are robust to a wide range of vote share bandwidths around the 50\% cutoff, even when focusing on elections decided by only a 5-10\% margin. Finally, we complement the DiD with arguably exogenous variation in union support, exploiting random shocks to the salience of workplace safety. Specifically, we use spikes in sector-level fatal work accidents shortly before the union election as an instrument for union support, and the results confirm our main findings. 

The establishment-level effects of unionization on employees’ campaign contributions could be explained by changes in the contributions of individual employees or by changes in the composition of the employed workforce. To differentiate between individual-level and compositional effects, we use donor identifiers and track each donor’s campaign donations over time. We study individual-level effects by focusing on incumbent employees who donated and were employed at the establishment before the union election. The results show a significant leftward shift in donations from individual workers and managers. Moreover, we examine compositional changes by estimating the effect on pre-election donation patterns of donors who were matched to an establishment after the election. We find no sizeable effect, indicating that the pre-existing political preferences of separated and newly hired employees do not differ across unionizing and non-unionizing establishments. In sum, these findings suggest that labor unions successfully persuade both workers and their management to support Democratic candidates. 

Our data also allows us to move beyond party preferences by considering candidates' ideological positions and endorsement by union organizations, as well as employees' support of interest groups. We document considerable within-party variation in the effects on contributions to different candidates. Liberal candidates gain and conservative candidates lose, while moderate candidates are not significantly impacted on average. This suggests that our findings are not only driven by an increased signal of Democratic over Republican party affiliation but reflect shifts between candidates with clearly distinguishable ideological positions. Moreover, we differentiate candidates by whether or not they are supported by unions' Political Action Committees (PACs). The results indicate that the partisan shift goes beyond union-endorsed candidates, reflecting a more general alignment with Democratic candidates. In addition, we show that our findings extend to contributions from employees to PACs: unions mobilize workers to increase donations to union and membership PACs, while managers reduce contributions to corporate PACs. The increased support for labor and civil society interest groups from workers and the reduced support for business interest groups from managers match with the observed ideological realignment in their contributions to candidates. In contrast, we do not find any effects on the donations that the PACs of unionized firms direct to candidates. These results are inconsistent with a potential strategic explanation of changes in donations under which individuals, in particular managers, merely signal goodwill toward the interests of unions.

Finally, we show that the effects of unionization on employees' campaign donations vary sharply with the broader labor relations environment that governs the unionization drive. We first examine the role of Right-to-Work (RTW) laws, which are typically advocated by pro-business activists to weaken unions. \cite{Feigenbaum2018} show that RTW laws diminish the political influence of unions, reducing voter turnout and Democratic vote shares at the local level. Consistent with this, we find that the establishment-level effects of unionization on support of Democratic over Republican candidates are smaller in states with RTW legislation. In addition, we reveal that in workplaces with strong managerial opposition to unionization -- as indicated by an unfair labor practice charge against the employer -- the political influence of unions is substantially smaller, in particular among managers. Similarly, we find that it is only when the union and the employer reach a collective bargaining agreement following the union election that unionization leads to increased support of Democratic over Republican candidates. These findings highlight the importance of cooperative labor-management interactions for unions’ ability to influence political behavior. Unions appear most effective in shaping political ideologies when they operate in environments conducive to cooperation rather than conflict.

Our results contribute to several strands of literature. First, we complement research on the economic impacts of unions by providing insights on the political channel. Numerous studies have examined the effects of unionization on wages and compensation at U.S. establishments, typically finding limited or no wage gains with some positive effects on fringe benefits \citep{DiNardo2004, Frandsen2020, Freeman1990, Knepper2020}. These modest establishment-level effects are hard to reconcile with evidence on the broader economic impact of unions. \cite{Stansbury2020} show that declines in worker power can explain the entire decrease in the labor share of income in the U.S. over the last decades. Similarly, \cite{Western2011} and \cite{Farber2020} find substantial negative effects of unions on income inequality, which they argue is hard to explain by income changes of union members alone, suggesting a potential link between unions and distributional legislation.\footnote{Several studies point toward an important role of unions in promoting greater political representation of the working class. \cite{Sojourner2013} shows that workers' likelihood of serving as state legislator increases with their occupation's unionization rate. Moreover, local union density is correlated with a more equal legislative responsiveness toward the poor versus the rich \citep{Flavin2018, Becher2021}. See also \cite{Ahlquist2017} for a review of how unions affect economic and political inequalities.} 

Second, we speak to the literature on the political influence of unions on their members (for a recent review on labor unions in political economy, see \cite{kaplan2024between}). By comparing union members to non-union members, several studies have documented a significant association with political outcomes, such as voting \citep{Freeman2003, Leighley2007}, preferences for redistribution \citep{Mosimann2017}, and trade liberalization support \citep{Ahlquist2014, Kim2017}.\footnote{\cite{yan2024} raises concerns about the causal interpretation and generalizability of these findings. Using several U.S. panel surveys, the paper finds that becoming a union member has no average effect on a range of political attitude and participation measures. In contrast to relying on self-reported changes in union membership status, we use administrative data on union elections and campaign contributions, thereby reducing measurement error issues. Moreover, our empirical design estimates the effect of first-time workplace unionization after winning a union election. This approach circumvents potential individual-level confounders correlated with becoming a union member (e.g., job transitions) and allows us to use lost union elections as a more comparable counterfactual for which we can test for parallel trends across the vote share distribution.} We add to these studies by assessing the causal impact of unions on workers' campaign contribution patterns. Campaign contributions are viewed as essential for candidates to win elections and have been shown to affect the selection and electoral performance of candidates \citep[e.g.,][]{Barber2016a, Battaglini2024}. Moreover, donors typically give to ideologically aligned candidates, such that campaign contribution patterns reveal the political ideology of donors \citep[e.g.,][]{Bonica2014, Bonica2018}. Analyzing campaign contribution patterns therefore provides insights into unions' influence on an important input into the political process and permits conclusions about shifts in political ideology.

Third, we shed new light on the spread of political preferences at work by combining establishment-level union election data with an individual-level political outcome. Existing research on the political impact of unions has focused either on individual union members and their households \citep[e.g.,][]{Freeman2003} or on aggregate outcomes comprising the whole county or state population \citep[e.g.,][]{Feigenbaum2018}. By focusing on the unionizing workplace, we are the first to consider within-firm dynamics, particularly the reaction of management –- an out-group that is likely indirectly affected by unionization and a key actor in political influence. Thereby, we expand on studies of politics in the workplace \citep{Colonnelli2022, Chinoy2024, Frake2024}, specifically those examining spillovers in political donations between managers and workers \citep{Babenko2020, Stuckatz2022} and the political effects of intergroup contact at work \citep{Andersson2021}.\footnote{Closely related to our findings, \cite{boudreau2024} and \cite{hertelfernandez2024} also show that union leaders can play a crucial role in coordinating workers by fostering consensus around unions' political positions and establishing norms of political participation.}

The paper is organized as follows. Section~\ref{sec:background} describes the institutional background, while Section~\ref{sec:data} introduces the data. The empirical approach is outlined in Section~\ref{sec:empirics}, after which Section~\ref{sec:results} presents the results. We study mechanisms in Section~\ref{sec:mechanisms} and conclude in Section~\ref{sec:conclusion}.

\section{Institutional Background}\label{sec:background}
\subsection{Unionizing through NLRB Elections}\label{sec:background_unionelections}

Since 1935, the National Labor Relations Act (NLRA) gives most private-sector workers in the U.S. the right to organize in unions and take collective action, such as bargaining and strikes. Collective bargaining between unions and employers takes place at the establishment level. Traditionally, workers unionize through a secret ballot election at their establishment that is administered by the National Labor Relations Board (NLRB).\footnote{While union elections are the primary means by which private-sector workers gain union representation, there are alternative procedures for unionization. First, employers may voluntarily recognize unions without an election through neutrality agreements and ``card checks". These cases are less common, however, since employers generally oppose union organization \citep{Schmitt2009}. Second, some workers' bargaining rights are not regulated by the NLRA. For example, the \textit{Railway Labor Act} determines bargaining rights of airline and railroad workers and several federal, state, and local laws regulate the organization of public-sector employees.} The unionization procedure involves three main steps: a petition drive, an election, and certification.

The organizing drive can be initiated either by the workers at an establishment or by a union organization. The initiator first needs to gather the signatures of at least 30\% of workers in the proposed bargaining unit who thereby express a desire for unionization. With these signatures, an election petition is filed to the NLRB. The NLRB decides whether to accept the petition by ascertaining whether workers in the proposed bargaining unit share common interests that can be adequately represented by the union. If the petition is accepted, the NLRB schedules a secret ballot election, which usually takes place in the workplace. The union wins the election if it obtains a strict majority of the votes cast. In case of union victory, the NLRB certifies the union as the sole authorized representative of employees in the bargaining unit. 

Union certification requires the employer to bargain ``in good faith" with the union. This bargaining generally aims at concluding a first contract between the union and the employer. However, as there is no legal obligation to reach such an agreement, only about 55\% of certifications yield a first contract within two years of the election \citep{Ferguson2008}. When both parties cannot reach a first agreement (or when subsequently they are disputing over the terms and conditions of the first contract), they can consult a neutral third party to resolve disputes via mediation or arbitration. After one year has passed since certification, employees can also decide to hold a decertification election to vote out the union.

The NLRA also lays out which employees may form a bargaining unit. While a bargaining unit can generally include all professional and nonprofessional employees at an establishment, managers and supervisors are always excluded.\footnote{The NLRA uses a rather broad definition for supervisors. It includes all individuals who have the authority to assign and direct the work of other employees, as long as this involves some independent judgment. There is no restriction as to the actual share of working time that involves supervisory duties. See Appendix \ref{sec:app_occclassify} for details.} These employees are considered to be part of a firm's management rather than its labor force and can therefore not join a union or be part of a bargaining unit. Representing the interests of capital owners, managers and supervisors typically oppose unionization and are thus treated as the ``out-group'' in our analysis. All other occupations form the ``in-group", as they are potentially in the bargaining unit and directly benefit from unionization.

\subsection{Campaign Contributions in U.S. Politics}\label{sec:background_contributions}

Money plays a dominant role in U.S. politics. Monetary resources are viewed as essential for political candidates in order to take part and be successful in the political process. There is indeed increasing evidence that campaign funding can influence who runs for and who wins elections \citep[e.g.,][]{Avis2022, Barber2016a, Battaglini2024, Broberg2023}. The large majority of campaign contributions in the U.S. originate from individual donors. The share of contributions to candidates that are donated by individuals has increased from 54\% in 2000 to 77\% in the 2020 congressional election cycle \citep{FEC2022a}. While political spending is certainly concentrated among the wealthy \citep{Bonica2018a, Hill2017}, it is a prevalent form of political participation for a substantial share of the U.S. electorate. \cite{Bouton2021} estimate that 12.7\% of the adult U.S. citizen population have made at least one campaign contribution between 2006 and 2020. 

Unlike corporations, which are prohibited from supporting candidates directly out of treasury funds, individual donors are allowed to make direct contributions to political candidates.\footnote{To make campaign donations, companies must set up a PAC, which may only solicit contributions from the firm's employees. The PAC can in turn donate directly to political candidates or other recipients.} There are, however, restrictions to the maximum amount that an individual can donate to a candidate. The limit varies by recipient type and election cycle. For the 2024 federal elections, for example, individuals were allowed to donate at most 3,300 USD to a single candidate and 5,000 USD to a PAC. Recipients are obligated to itemize all individual contributions greater than 200 USD and report the donor's identifying information along with the amount and date of the contribution. Donations smaller than 200 USD are not required to be itemized but are included in the total amount that the recipient reports to the Federal Election Commission.\footnote{Over the last years, increasing shares of donations are channeled through party's online fundraising platforms, such as Democrats' ActBlue (created in 2004) and Republicans' WinRed (created in 2019). Importantly, these platforms do not have a minimum reporting threshold and thus also have to itemize contributions below 200 USD. As a result, in recent years we can increasingly observe information on small donations \citep{Bouton2021}.} 

The literature differentiates two primary motivations for why individuals contribute to political candidates. First, contributions may serve as consumption goods, providing intrinsic value to donors who participate in politics and support candidates ideologically close to their own political position \citep{Ansolabehere2003}. Second, donors may view contributions as investment goods that can buy access to politicians and benefit their own material interests. There is extant evidence that individuals' donations are ideologically motivated. Donors report that candidate ideology has great importance when deciding to whom to give \citep{Barber2016}. Moreover, unlike access-seeking PACs, who prefer donating to moderate candidates, individuals tend to support more ideologically extreme candidates \citep{Barber2016a, Stone2010}. In merged survey-administrative data, contribution-based ideology measures also predict policy preferences of donors, even of donors from the same party \citep{Bonica2018}. That said, recent studies have also explored the role of influence-seeking motives in individual campaign contributions. \cite{Teso2023} shows that a business leader's likelihood of donating to a Member of Congress increases when the politician is assigned to a committee that is policy-relevant to the business leader's company. However, these effects are quantitatively limited: the estimates imply that less than 2\% of corporate leaders' donations to Congress members are driven by influence-seeking motives.\footnote{Further speaking against a strong role of influence-seeking motives, \cite{Bonica2016} finds that donations of corporate board members are ideologically quite diverse, both across and within companies. Compared to corporate PACs, business leaders also tend to support more non-incumbent candidates and less powerful legislators.} \cite{Stuckatz2022} analyzes how the workplace shapes employees' contribution behavior by examining alignment between donations of employees and those of their employer PACs. While the paper documents positive associations, the results indicate that only 13\% of donations from rank-and-file workers and 21\% of donations from executives go to candidates endorsed by their employer PACs, suggesting that a limited share of employees' donations can be influenced by corporate strategic objectives. Taken together, the evidence supports the interpretation that ideological preferences, rather than strategic influence-seeking, are the predominant motivation behind individuals' campaign contributions.

\section{Data}\label{sec:data}

Previous studies have been unable to assess the political impact of unions at the establishment level due to a lack of matched employer-employee data for political outcomes. Campaign contribution data are uniquely suited to overcome this constraint. To ensure transparency in politicians’ campaign funds, contributors are required to disclose their name, employer, address, and occupation. The employer and location information allows us to link donors to the union election results of their employers. Furthermore, we can use the occupation information to study the political effects of unionization not only on directly affected non-managerial workers but also on potentially indirectly affected managers and supervisors. In the following, we describe how we construct a new establishment-level dataset that links union elections to campaign contributions from employees.

\subsection{Union Elections}

We start with a comprehensive dataset on the universe of U.S. union representation elections between 1961 and 2018. Specifically, we combine data collected by \cite{Farber2016} with public data from NLRB election reports.\footnote{We obtain the dataset originally compiled by \cite{Farber2016} from the replication package of \cite{Knepper2020}, covering elections from 1961 to 2009, and add data from NLRB election reports for the years 2010 to 2018. See Appendix \ref{sec:app_dataunion} for details on the union election data, including the sample restrictions we impose.} Each data point represents a union election at a single establishment and contains vote counts for and against unionization, the dates of the petition filing and of the actual election, as well as the name of the union organization. Moreover, it includes the establishment's name and address, which we exploit to match campaign contributions.

\vspace{1em} \noindent \textbf{Sample restrictions.}\quad Before matching elections to campaign contributions, we impose several sample restrictions. First, we only consider elections held between 1985 and 2010. Given that our contribution data cover the years 1979-2016, this allows us to observe trends in contributions for three political election cycles (six years) before and after each union election. Second, we follow \cite{Frandsen2020} and restrict the sample to union elections where at least 20 votes were cast. This restriction ensures that winning establishments are affected by a non-trivial rise in union representation. Moreover, it helps to exclude small establishments, which are more likely to have come into existence recently and have a lower probability of survival over our period of analysis. Third, following \cite{Knepper2020} and \cite{Wang2021}, we only keep the first union election in each establishment.\footnote{In the election data, we identify an establishment as a unique address or a unique combination of the standardized firm name and commuting zone. For a firm that has multiple establishments within the same commuting zone, we thus only consider the first election among these establishments.} Excluding non-inaugural elections avoids having multiple observations for the same establishment with reversed treatment status over time, and helps alleviate election manipulation issues if managers or unions learn how to apply manipulation tactics in repeat elections. Our estimates should thus be interpreted as the effects of winning the first union election.\footnote{This does not perfectly correspond to the effect of union representation in all post-election periods for two reasons. First, establishments may lose representation after a decertification election. \cite{Wang2021} show that 5-10\% of establishments that win a first union election hold a decertification election within 5 years. Second, establishments, after losing the first election, can hold another successful election in subsequent years. According to \cite{DiNardo2004}, this is the case for around 10\% of lost first elections. By focusing on the effect of winning the first election, we thus accept an attenuation of our estimates relative to the true effects of union representation over all post-election periods.} These restrictions leave us with a sample of 28,823 union elections, which we seek to match to the campaign contribution data.   

\vspace{1em} \noindent \textbf{Summary statistics.}\quad Table \ref{tab:sumstat_election} shows summary statistics for characteristics of the matched union elections that are included in our baseline estimation sample (see details on the matching in the next subsection). 44\% of the elections were won by the union, with an average union vote share of 50\%. On average, 119 votes were cast in each election, which yields a total of 723,752 voters who participated in all elections of our sample.

\subsection{Campaign Contributions}

To measure the political mobilization and ideology of employees, we use the Database on Ideology, Money in Politics and Elections (DIME) compiled by \cite{Bonica2019}. DIME provides transaction-level data on campaign contributions registered with the Federal Election Commission and other state and local election commissions. We exploit all campaign contributions from individuals to candidates running for office at the federal and local level (specifically the President, House of Representatives, Senate, Governor, and upper and lower chambers of state legislature), as well as to all PACs (including party and interest group PACs). The dataset covers the 1979-2016 period and includes the amount and exact date of the donation, as well as identifying information on the donor and recipient.\footnote{Accurate reporting of the donor information (name, employer, address, occupation) is enforced by the Federal Election Commission through regular audits, as well as fines and further legal action in case of non-compliance. See \cite{FEC2022} for enforcement statistics.} 

\cite{Bonica2019} deploys identity resolution techniques to assign unique identifiers to each donor. The identifiers allow us to track donors' contributions over time, which we exploit to study whether establishment-level effects are driven by compositional changes from leaving and newly hired employees or by individual-level effects on incumbent employees. Further, the DIME includes measures for the political ideology of recipients and donors, so-called campaign finance (CF) scores, which are derived by \cite{Bonica2014} from solving a spatial model of contributions. The model formalizes the idea that donors contribute more to candidates with a similar ideological position and estimates ideal points of both recipients and donors along a typical liberal-conservative scale. Using the ideology scores, we can go beyond previous papers that only relate unions to Democratic versus Republican party affiliation and study how unionization affects ideological contribution patterns for candidates within the same party. 

\vspace{1em} \noindent \textbf{Matching algorithm.}\quad We link the campaign contributions to the employing establishments with union elections by combining a spatial match with a fuzzy match of firm names. We start by restricting potential matches to the same local labor market using 1990 commuting zones. 92\% of the population live and work in the same local labor market, making it very likely that a donor in our sample works at an establishment in the same local labor market \citep{Fowler2020}. The restriction substantially reduces the computational requirements for the fuzzy match and ensures that for multi-establishment firms we do not incorrectly match employees to establishments of the same firm in other locations.\footnote{We accept measurement error from assigning donors to the wrong establishment when a firm has several establishments within a commuting zone. Note that within-firm interactions may also generate spillover effects across establishments \citep{Knepper2020, Wang2021}.} To match the employer name in the contribution data to the establishment name in the union election data, we use the automated record-linkage program of \cite{Blasnik2010} and \cite{Wasi2015}. The linkage process first standardizes employer names and then calculates bigram scores for the similarity of each string pair. Lastly, we manually review all matches with a score above a minimum threshold.\footnote{See Appendix \ref{sec:app_matching} for details on the matching process. We have also experimented with matching employer names using a large language model which yields similar match accuracy in a test dataset.} 

To arrive at an establishment-level panel of employee contributions, we sum up all matched contributions within an establishment and two-year election cycle. Our period of analysis covers three cycles before to three cycles after each union election. In our baseline estimation sample, we only include establishments for which we observe at least one matched contribution over this period in order to reduce bias from false negative matches.\footnote{Note that we generally took a conservative approach in the manual review of matches and rejected potentially true matches if firm names in the union election data were misspelled or too generic to infer a unique firm per commuting zone (e.g. "community health center", "general construction", "support services corporation"). Including these establishments and assigning them zero amounts in the estimation sample, would lead to an attenuation bias in our estimates. Appendix Table \ref{tab:nonmatched} compares the characteristics of establishments with and without at least one matched contribution. Elections in our matched sample involve more voters, i.e., are likely to be larger, and tend to be held in more recent years as contribution numbers have sharply increased over time. At the same time, the matching does not strongly affect the selection of union elections in terms of voting outcome and industry composition.} In Appendix Table \ref{tab:any_contribution}, we verify that unionization does not affect the likelihood of observing any contribution after the union election, suggesting that the sample selection criterion is not related to our treatment of interest. Moreover, we will discuss the sensitivity of our main results to alternative sample restrictions. 

For our baseline sample, we are left with 6,063 matched establishments (and 42,441 establishment-cycle observations). As Table \ref{tab:sumstat_election} reports, the sample is built from 357,436 matched contributions that amount to 105.8 million USD spent by 46,719 different donors to 9,942 different recipients.\footnote{Extrapolating from our matched sample, we can estimate the total amount contributed by employees at all unionized establishments. We first calculate the cycle-specific average amount donated per eligible voter in newly unionizing establishments and then multiply it by the number of all workers represented by unions during that cycle. Summing up over all cycles from 1986 to 2010, we estimate that employees at unionized establishments have donated a total of 1.36 billion USD. Contrasting this number with the cost of running successfully for a seat in the House of Representatives -- which lies between \$777,000 and \$1.47 million \citep{persily2018campaign} -- highlights the sizable potential for political influence that unions could wield by impacting the donations of employees at unionized establishments.} 

\vspace{1em} \noindent \textbf{Classification of occupations.}\quad In order to differentiate between workers eligible for unionization and their managers and supervisors who are always excluded from the bargaining unit, we classify self-reported occupations of donors. We start by mapping the free-text occupation descriptions in the DIME to the 6-digit Standard Occupation Classification (SOC). For this, we combine an ensemble classifier called SOCcer \citep{Russ2016}, sub- and fuzzy string matching to an extensive crosswalk of laymen's occupation titles from O*NET, as well as manual reviews of the most common occupation titles.\footnote{Appendix Figure \ref{fig:donor_occup} shows the occupation distribution for the classified donations. While the largest share (44\%) is given by donors in management occupations, we also see substantial shares of contributions originating from lower-tier white-collar occupations such as healthcare, education, culture and sports, or financial operations workers. Blue-collar occupations, in contrast, account for small shares of the overall number of contributions, which is not surprising given that wealth is a strong predictor of political donating.} With the classified SOC codes at hand, we categorize donors into managers and supervisors versus non-managerial workers. We identify managers and supervisors by using all contributions from ``Management Occupations" (SOC group 11) and adding all occupations that involve a significant amount of supervising following the NLRA definition of supervisor tasks and leveraging occupational task descriptions from O*NET. Non-managerial workers are then defined as all remaining donors to whom we were able to assign a SOC code. 

The occupational composition in our final sample of candidate contributions looks as follows: 42\% of contributions originate from managers and supervisors (hereafter only termed ``managers"), 30\% from non-managerial workers (hereafter only termed ``workers"), and for 28\% we are unable to obtain an occupational classification. Due to the non-negligible share of unclassified occupations, we report results not only separately for managers and workers, but also for all employees together (including those without a classification). Moreover, as a robustness check, we classify workers and managers using information on the median household income in the census tract they live in.\footnote{See Appendix \ref{sec:app_occclassify} for details on the occupational classification. There, we also provide evidence that the likelihood of having a missing occupation classification is not affected by unionization and therefore unlikely to drive our results.}  

\vspace{1em} \noindent \textbf{Summary statistics.}\quad Table \ref{tab:sumstat_donorrecipient} reports mean values for the sum of all employees' contributions for a given establishment and election cycle. Managers donate on average 1,339 USD per cycle, while workers contribute 314 USD.\footnote{In Appendix Table \ref{tab:sumstat_worker_manager}, we show that the difference is driven by managers donating both higher numbers of contributions and higher average amounts per contribution than workers.} Both groups support different recipients. The majority of contributions by managers are donated to Republican candidates (54\%), whereas workers tend to favor Democratic candidates (65\% of the average amount is donated to Democrats). Moreover, managers give a larger share of donations to committees than to candidates. In contrast, workers more often contribute directly to candidates.

\section{Empirical Strategy}\label{sec:empirics}

\begin{figure}[htbp]
	\begin{center}
		\caption[gbadolite]{\label{fig:trend} Trends in Contributions for Won and Lost Union Elections}
		\centering
        \vspace{-1em}
		\mbox{\includegraphics[width=\linewidth]{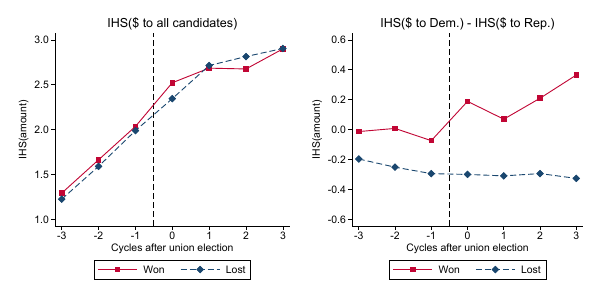}}
		\vspace{-0.5cm}
	\end{center}
	\vspace{-1em} 
	\begin{minipage}{\textwidth} {\scriptsize \smallskip
 \textbf{Notes:} The figure displays trends in average contribution amounts (in 2010 USD) from all employees in an establishment, separately by union election outcome and election cycle relative to the union election. The left graph shows the mean of IHS-transformed total contributions to all candidates. The right graph show the mean difference between IHS-transformed amounts contributed to Democratic and Republican candidates. $N=42,441$ establishment-cycle observations.\par}
    \end{minipage}
\end{figure}

Our objective is to estimate the causal effect of unionization on employees' campaign contributions. A simple comparison of employees in unionized and non-unionized workplaces will fail to account for differences between these groups along a number of dimensions. These arise because the decision to unionize is likely endogenous and correlated with many characteristics, among them potentially political behavior. Figure \ref{fig:trend} depicts average contribution amounts and their party composition across won and lost union elections before and after the election. Due to their shared interest in a union election at the same time, these establishments are expected to be more similar than a random sample of unionized and non-unionized establishments.\footnote{\cite{Dinlersoz2017} examine selection into union elections and find that elections are more likely to be held at younger, larger, more productive, and higher-paying establishments. Our strategy avoids such selection by comparing only establishments that hold union elections.} Pre-existing ideological differences are nevertheless visible: Workplaces that vote for unionization donate more to Democratic relative to Republican candidates even before the union election.

To account for pre-existing differences, we implement a difference-in-differences approach and compare campaign contribution patterns before and after the union election in establishments where the union won versus where it lost. We complement the DiD design with methods from the RDD literature to probe the validity of the underlying parallel trends assumption. In particular, we leverage the pro-union vote share, which discontinuously determines unionization at the 50\% threshold, to estimate placebo tests for differential trends by vote shares among losing union elections as well as to examine the robustness of our DiD estimates when restricting the sample to establishments with increasingly close election results.\footnote{Many papers on the effects of unionization follow an RDD by comparing establishments in which the union barely won versus where it barely lost \citep[e.g.,][]{Campello2018, DiNardo2004, Ghaly2021, Lee2012, Sojourner2015a, Sojourner2015}. This approach is complicated by the ability of both unions and employers to influence election outcomes even after the election, either by challenging the validity of individual ballots or by filing unfair labor practices charges. \cite{Frandsen2020} and \cite{Knepper2020} provide evidence for discontinuities at the 50\% threshold in the vote share distribution and in pre-election establishment characteristics. Appendix Figure \ref{fig:hist} verifies that, also in our matched sample of elections, there is a significant discontinuity in the vote share density at the 50\% cutoff, indicating a manipulation of closely contested elections. In addition, focusing on close elections may yield treatment effect estimates that are not representative for higher margin-of-victory elections. As \cite{Wang2021} show, barely won elections exhibit more decertification elections and delays in the election process, indicating a lower bargaining power of unions relative to elections with higher pro-union vote shares.} 

\vspace{1em} \noindent \textbf{Stacked DiD.}\quad As our main specification, we estimate the following stacked DiD model: 
\begin{equation}\label{model1}
y_{jk} = \alpha_j + \beta_{kc_j} + \delta_{\textrm{DiD}} \times \Big( \mathbbm{1}[k\geq 0] \times \mathbbm{1}[V_j > .5] \Big) + \epsilon_{jk},
\end{equation}
where $y_{jk}$ denotes a political outcome for employees in establishment $j$ and relative event time $k$. We observe each establishment from three cycles before to three cycles after the union election, i.e., $k=\{-3,-2,...,3\}$, where $k=0$ refers to the cycle in which the union election takes place. Our effect of interest is captured by $\delta_{\textrm{DiD}}$. It is the coefficient of an interaction term between a post-treatment dummy and a dummy indicating whether the election was won by the union, i.e., whether the pro-union vote share, $V_j$, is above 50\%. $\alpha_j$ denotes establishment fixed effects that capture all time-invariant differences between winning and losing establishments. Further, we introduce event-time $\times$ cohort fixed effects $\beta_{kc_j}$, where cohort $c_j$ refers to the political election cycle in which the union election was held, i.e., $c_j=\{1985/86,1987/88,...,2009/10\}$. Importantly, with these fixed effects our identifying variation only comes from comparing changes across won and lost elections within the same cohort. Thereby, our model is equivalent to the stacking approach used in staggered DiD settings \citep{Cengiz2019} which avoids ``forbidden comparisons" between late and early-treated establishments that may lead to negative weights when averaging potentially heterogeneous, cohort-specific treatment effects in these settings \citep{DeChaisemartin2020, Goodman-Bacon2021, Sun2021}.\footnote{Note that in our case the selection of appropriate control units for each cohort of treated units in the stacking approach is facilitated by the possibility that we can naturally compare treated establishments with a won union election to untreated establishments that have a lost union election in the same cycle.} Finally, we cluster standard errors at the establishment level, which is the unit of treatment assignment.  

Model (\ref{model1}) pools all periods after treatment, which maximizes power when estimating average treatment effects. To examine how treatment effects vary by event time, we also estimate the following stacked event-study model: 
\begin{equation}\label{model2}
y_{jk} = \alpha_j + \beta_{kc_j} + \sum_{s=-3, s\neq-1}^{s=3} \delta_s \times \Big( \mathbbm{1}[k=s] \times \mathbbm{1}[V_j > .5] \Big) + \epsilon_{jk},
\end{equation}
where the $\delta_s$ coefficients capture dynamic treatment effects relative to the cycle before the union election was held (the interaction with $k=-1$ is omitted). 

\vspace{1em} \noindent \textbf{Parallel trends assumption.}\quad The identifying assumption is that campaign contributions for winning establishments would have evolved in parallel to contributions in losing establishments had the union not won the election:
$$E[Y^0_{j,k\geq0} - Y^0_{j,k<0}|V_j>.5] = E[Y^0_{j,k\geq0} - Y^0_{j,k<0}|V_j\leq.5],$$
where $Y^0_{j}$ denotes the potential outcome of an establishment if the union loses the election. 

We run different tests to examine the validity of this assumption. First, we analyze whether outcomes developed in parallel before the election. Figure \ref{fig:trend} provides first visual evidence that pre-election changes in total contribution amounts and in their party composition are very similar across won and lost elections.\footnote{Note that the strong upward trends in total contribution amounts are explained by the increasing importance of campaign contributions in more recent election campaigns.} The pre-election $\delta_s$ coefficients estimated in the event study model will provide a formal test of pre-trends. 

Second, even in the absence of significant pre-trends, there may still be unobserved shocks that drive union voting results at the time of the election and that are related to changes in contribution patterns. To test whether such shocks likely violate our identifying assumption, we follow the approach of \cite{Wang2021} and analyze whether changes in outcomes are different among losing elections with different vote shares. If unobserved shocks were driving voting results that led to union victory or loss, we would also expect them to affect outcomes in losing elections with different union vote shares.\footnote{\cite{Wang2021} formulate the identifying assumption as parallel trends across all vote shares, i.e., $E[Y^0_{j,k\geq0} - Y^0_{j,k<0}|V_j] = E[Y^0_{j,k\geq0} - Y^0_{j,k<0}]$, which yields the testable implication that trends should be parallel between losing elections with different vote shares.} To implement this test, we modify the DiD model as follows:
\begin{equation}\label{model3}
y_{jk} = \alpha_j + \beta_{kg_j} + \sum_g  \delta_g \times \Big( \mathbbm{1}[k\geq 0] \times \mathbbm{1}[V_j \in \nu^g]\Big) + \epsilon_{jk},
\end{equation}
where $\nu^g$ denotes a complete set of vote share categories. In particular, we divide the vote share distribution into the following six groups: 0-20\%, 20-35\%, 35-50\%, 50-65\%, 65-80\%, 80-100\%. In the model we omit the 20-35\% vote share category, such that all estimated effects are interpreted relative to that group. Significant estimates for the 0-20\% or 35-50\% categories would then indicate the presence of unobserved shocks that drive both voting results in the union election and campaign contribution behavior.

Third, we restrict the sample to elections where the union won or lost by an increasingly close margin. Establishments with closer election results can be expected to be more similar not only in terms of baseline characteristics but also in terms of shocks that they are exposed to over time. Specifically, we examine the robustness of the DiD estimates when narrowing the sample to increasingly small vote share bandwidths around the 50\% cutoff. In the limit, when comparing establishments where the union barely lost versus where it barely won, we approach the discontinuity-in-differences model estimated by \cite{Frandsen2020} and \cite{Knepper2020}. For our baseline results from models (\ref{model1}) and (\ref{model2}), however, we follow \cite{Wang2021} and consider all elections with a pro-union vote share between 20\% and 80\%. This improves power and allows us to generalize effects for a broader sample of union elections. 

\vspace{1em} \noindent \textbf{Definition of outcome variables.}\quad Throughout the analysis, we consider two primary outcomes of employees' political behavior at the establishment level. The first is the total amount of campaign contributions to all political candidates which we interpret as a measure of employees' political participation and mobilization. Our second main outcome is the difference between the contribution amounts to Democratic and Republican candidates. Given the extant evidence on ideological motivations driving individuals' donation behavior, we interpret the party composition of donations as a measure of employees' ideological positions. 

In our baseline specification, we transform contribution amounts using the inverse hyperbolic sine (IHS) function to approximate log changes in amounts while retaining zero values. As demonstrated in recent econometric work \citep{Aihounton2021, Chen2024, Mullahy2024}, the IHS transformation yields marginal effects that can depend on the units in which the outcome is measured. To avoid arbitrary scaling of our outcomes, we always measure contribution amounts in 2010 USD before applying the IHS transformation. \cite{Chen2024} show that the scale dependence arises because the scale implicitly determines the value that IHS assigns to changes along the extensive versus intensive margin. If the treatment has larger extensive margin effects -- in our case, if unionization induces employees to donate positive amounts -- the estimates using IHS are more sensitive to scaling. In section \ref{sec:results_robust}, we thus analyze the extensive margin effects of unionization and check the robustness when explicitly assigning alternative weights to extensive versus intensive margin changes.\footnote{Following the recommendation of \cite{Chen2024}, explicitly weighting extensive and intensive margin changes allows us to value a percentage point change in campaign contributions equally for all individuals, regardless of their initial contribution levels. Intuitively, a 10 USD increase in contributions by a lower-income individual may reflect the same ideological shift as a 100 USD increase by a more affluent individual. Alternative estimation strategies that avoid scale dependence by estimating treatment effects in levels fail to capture decreasing marginal utilities and may understate ideological shifts among lower-income donors.}

\section{Results}\label{sec:results}
\subsection{Main Results}\label{sec:results_main}
\begin{figure}[htbp]
	\begin{center}
	\caption[gbadolite]{\label{fig:event_study} Effect of Unionization on Candidate Contributions}
	\centering
	\vspace{-0.5cm}
	\mbox{\includegraphics[width=\linewidth]{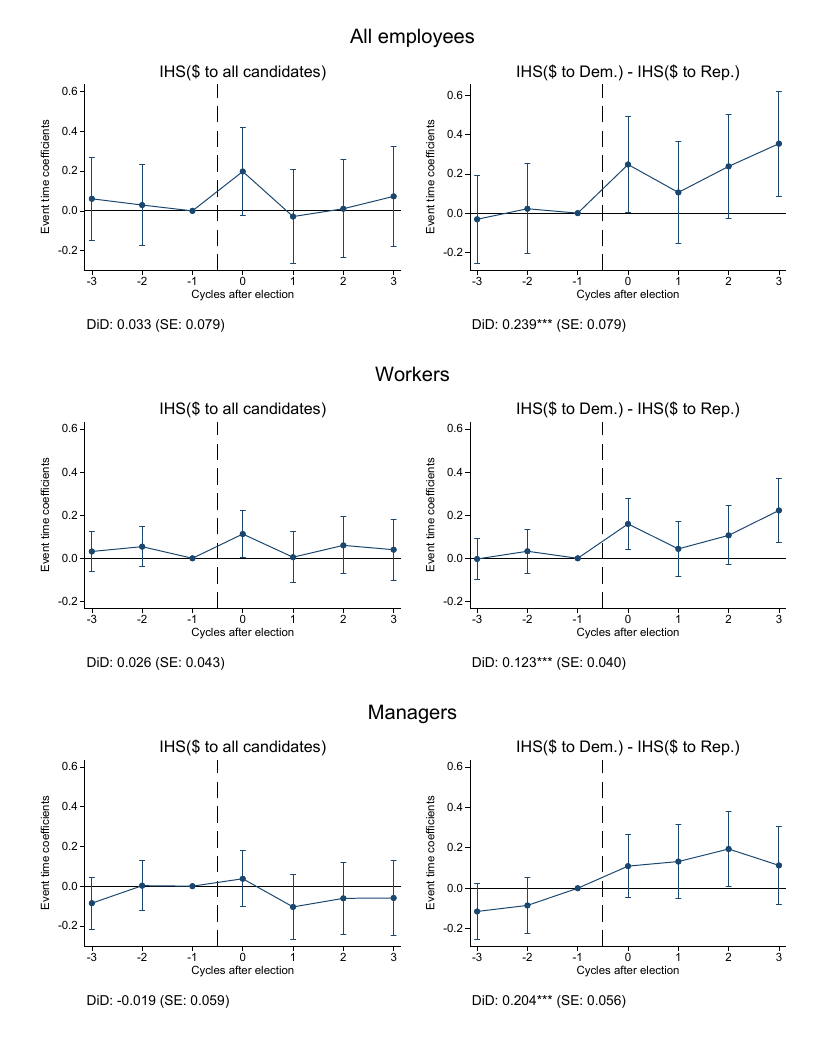}}
	\vspace{-1cm}
	\end{center}
	\begin{minipage}{\textwidth} {\scriptsize \smallskip 
 \textbf{Notes:} The figures report the event-study coefficients $\delta_s$ estimated in model (\ref{model2}). The sample includes all establishments with a pro-union vote share between 20\% and 80\% and covers three election cycles (six years) before and after the union election. $N=33,103$ establishment-cycle observations. Below each graph the DiD coefficient from model (\ref{model1}) is reported. In the graphs on the left side, the outcome is the IHS-transformed total amount contributed to all candidates. In the graphs on the right side, the outcome is the difference between the IHS-transformed amounts contributed to Democratic and Republican candidates. Results are reported for contributions from all employees (top part), from non-managerial workers (middle part), and from  managers and supervisors (lower part). 95\% confidence intervals are depicted for standard errors clustered at the establishment level.\par}
    \end{minipage}
\end{figure}

Figure \ref{fig:event_study} reports our main results for the effect of unionization on employees' campaign contributions, using the DiD and event study models (\ref{model1}) and (\ref{model2}). We start with the effects on the (IHS-transformed) total contribution amounts, which are depicted in the left-hand panels of the figure. The upper panel plots the results for all employees in an establishment. Note the absence of any significant differential trends between establishments winning and establishments losing the union election in the three cycles (six years) before the election. The effect of unionization on the amount of contributions is small and insignificant in all post-election periods, but we see a moderate spike in contributions in the cycle of the union election (which we are not able to estimate precisely, though). Differentiating between contributions made by workers and managers in the lower panels highlights that workers drive the increase in contributions. The event-study results indicate that unionization raises workers' contributions by 11\% in the cycle of the union election (significant at the 5\% level). This pattern is consistent with a short-term political mobilization of workers through a successful union campaign in the workplace. Overall, however, the DiD coefficients indicate that there is no significant average effect on the amount of contributions over the three cycles after a union election.

Next, we assess changes in the party composition of campaign contributions. If unions are able to change individuals' political views, campaign contributions will shift to different candidates. The right-hand panels of Figure \ref{fig:event_study} plot estimates for the effect of unionization on the difference in (IHS-transformed) amounts spent to Democratic versus Republican candidates. Starting with all employees, we again see no differential trends in contribution composition before the election. After the election, however, there is a significant increase in contributions donated to Democratic relative to Republican candidates. The partisan shift appears to be strongest in the long term, i.e., six years after the union election. The DiD estimate indicates that, over all post-election periods, unionization increases the Democratic-Republican contribution gap by 24 percentage points (significant at the 1\% level). Differentiating again between workers and workers in the lower two panels reveals that the effect occurs among both groups. Not only workers but also managers significantly shift contributions from Republican to Democrat candidates in response to successful unionization. Quantitatively, the DiD estimates show that winning the union election increases donations to Democrats relative to Republicans by 12 percentage points for workers and by 20 percentage points for managers (both significant at the 1\% level).\footnote{Appendix Table \ref{tab:did_main_results} also reports results separately for amounts contributed to Democratic and Republican candidates. For both workers and managers, the coefficients for contributions to Democrats are positive and those for Republicans are negative. However, among workers only the increase in contributions to Democrats is statistically significant, while among managers only the decrease in contributions to Republicans is significant. 
That said, our main results show that these party differences are not large enough to generate significant effects on total contribution amounts.} The relatively stronger response among managers may seem surprising, as they are not the beneficiaries of unionization and are typically not targeted by unions' political activities. One possible explanation is that managers begin from more conservative ideological positions -- see, again, the average donation patterns in Table \ref{tab:sumstat_donorrecipient} -- which increases the scope for changes after unionization. Workers, by contrast, already show strong Democratic support prior to unionization, leaving less room for change. Overall, our results point a political realignment within the workplace, where unionization narrows ideological gaps between managers and their unionizing workers.

\subsection{Addressing Identification Challenges}\label{sec:results_ident}

\begin{figure}[htbp]%
	\caption{\label{fig:rdd_tests}\centering Vote Share Tests}%
	\subfloat[\centering Vote Share Heterogeneity]{{\includegraphics[width=.475\linewidth]{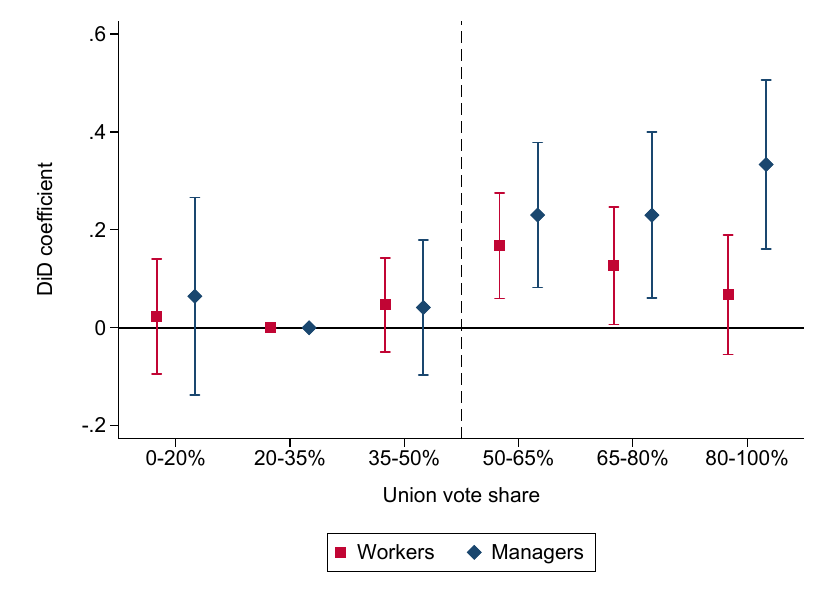} }}%
	\qquad
	\subfloat[\centering Vote Share Bandwidth Sample Restrictions]{{\includegraphics[width=.475\linewidth]{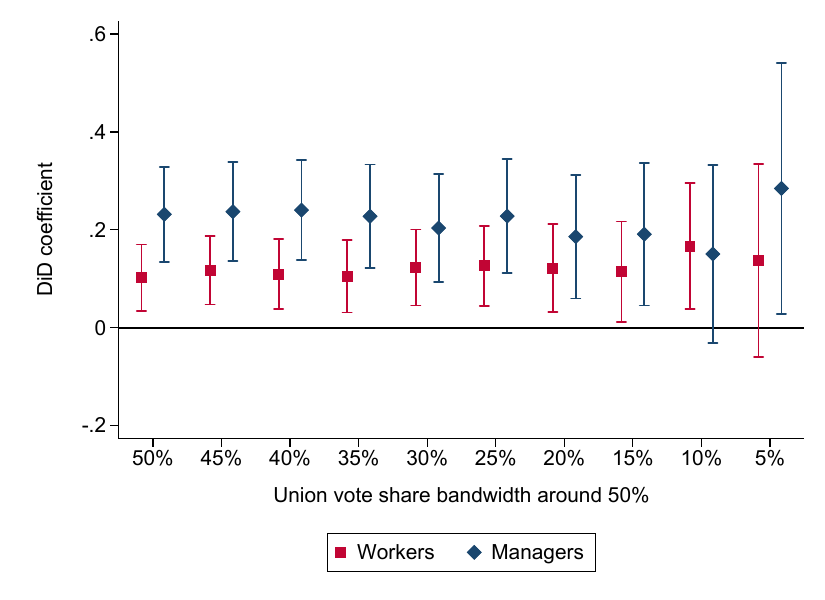} }}%
	\vspace{0.1em}
	\begin{minipage}{\textwidth} {\scriptsize \smallskip 
 \textbf{Notes:} The graphs show RDD-motivated placebo and robustness tests for the effect of unionization on the difference between the IHS-transformed amounts contributed to Democratic and Republican candidates. Panel (a) reports the $\delta_g$ coefficients estimated in model (\ref{model3}). The vote share distribution is partitioned into six bins, indicated on the x-axis. The omitted reference group is 20-35\%. Panel (b) reports DiD coefficients estimated in model (\ref{model1}). Each dot refers to a single DiD coefficient that is estimated among elections with a union vote share in a given bandwidth around the 50\% cutoff. Estimates from smaller bandwidths compare changes between increasingly close elections. Results are always shown separately for contributions from non-managerial workers (``workers") and from managers and supervisors (``managers"). 95\% confidence intervals are depicted for standard errors clustered at the establishment level.\par}
    \end{minipage}
\end{figure}

\noindent \textbf{Vote share tests.}\quad We continue presenting results for our RDD-motivated tests to probe the validity of the underlying parallel trends assumption of the DiD model.\footnote{One particular concern for the parallel trends assumption would arise if union elections were endogenously timed around federal election dates. Appendix Figure \ref{fig:cyclicality} investigates whether union elections follow political cycles. Across years with and without federal elections, there are no strong differences in the number of union elections held and the probability of winning a union election, in particular not around the week of federal elections. Thus, we do not see evidence that employers or unions successfully manipulate union election dates to change union support around federal election cycles.} Figure \ref{fig:rdd_tests} focuses on the partisan composition of contributions, while effects on the total amount of contributions are presented in Appendix Figure \ref{fig:rdd_tests_total}. Results are always reported separately for workers and managers. We first analyze the heterogeneous effects of unionization across the vote share distribution. Panel (a) of Figure \ref{fig:rdd_tests} displays the $\delta_g$ coefficients from model (\ref{model3}) for the interaction between the post-election dummy and different vote share categories. The results show that there are no significantly different trends among losing elections with a vote share of 0-20\% or 35-50\% relative to those with 20-35\%, for contributions from both workers and managers. The partisan composition of contributions thus appears to evolve similarly across losing establishments with different vote shares, suggesting that unobserved shocks correlated with voting results are unlikely to drive our results.\footnote{In Appendix Figure \ref{fig:rdd_tests_prepost}, we also investigate whether pre-trends in the contribution composition are similar across the vote share distribution. For this, we estimate the following modified version of model (\ref{model3}): 
\begin{equation}\label{model3b}
y_{jk} = \alpha_j + \beta_{kg_j} + \sum_g  \delta^{PRE}_g \times \Big( \mathbbm{1}[k<-1] \times \mathbbm{1}[V_j \in \nu^g]\Big) + \sum_g  \delta^{POST}_g \times \Big( \mathbbm{1}[k\geq 0] \times \mathbbm{1}[V_j \in \nu^g]\Big) + \epsilon_{jk}
\end{equation}
The results show that none of the $\delta^{PRE}_g$ coefficients are significantly different from zero, indicating that, also before the union election, contribution patterns evolved similarly across establishments with different voting results.} Moreover, the results indicate whether treatment effects are heterogeneous across vote shares among won union elections. For the composition of contributions from managers, the estimate is significant across all vote share categories above 50\%. Thus, the political response of managers does not appear to depend on whether workers won the union election with large or small margins of victory. For workers, the effect on partisan support is significant only for vote shares between 50 and 80\% and appears smaller for elections won by a large margin.   

We further examine the robustness of our DiD estimates when focusing on close union elections. Establishments with narrowly decided votes are likely to be more similar in characteristics and to be exposed to similar shocks, making the parallel trends assumption more credible. Panel (b) of Figure \ref{fig:rdd_tests} presents coefficients from  model (\ref{model1}), estimated on subsamples restricted to increasingly narrow union vote share bandwidths (in 5\% steps) around the 50\% cutoff. Our baseline results from Figure \ref{fig:event_study} include only elections with a pro-union vote share between 20 and 80\%, i.e., a bandwidth of 30\%. Figure \ref{fig:rdd_tests} shows that treatment effects are very similar when instead using all elections. Importantly, the results are also very stable when focusing on closer elections. Even when restricting the sample to establishments that won with a maximum vote margin of 5\%, we see a positive and significant effect on the composition of campaign contributions for managers. Similarly, for workers a maximum vote margin of 10\% already yields a positive and significant effect.

\vspace{1em} \noindent \textbf{Alternative source of variation.}\quad To further check the causal interpretation of our results, we also complement the DiD strategy with a novel IV approach. For this, we exploit variation in unionization resulting from the timing of exogenous shocks to the salience of safety at work that are triggered by unexpected fatal workplace accidents shortly before the union election. After the NLRB accepts a petition to hold a union election, it sets the timeline of the unionization process and fixes an election date. Any random unexpected shocks between petition and election that shift union support are potential candidates for an instrument. We focus on sector-level fatal work accidents in the 30 days before a union election. Unions often campaign on safety issues and are found to improve safety conditions in the workplace \citep[e.g.,][]{AFLCIO2022, Hagedorn2016, Li2022}. An increase in the salience of serious work-safety issues thus plausibly strengthens union support in the union election. 

The exclusion restriction of the instrument relies on the notion that a shock in fatal work accidents in the same sector affects political behavior in subsequent years only through its impact on the likelihood that an establishment will unionize. Two points are worth highlighting in that regard. First, while all individuals in our sample are potentially exposed to the information on fatal work accidents, only some vote on unionization in the following 30 days. We control for accidents in the same sector and year, such that we only exploit variation in the \textit{timing} of the information shocks relative to the union election. Second, we focus on the medium-term impact of spikes in fatal accidents, excluding observations in the cycle of the union election. The result that common shocks in fatal work accidents influence political behavior \textit{years afterward} in some but not other establishments would be difficult to explain other than through the path dependency triggered by the increase in the likelihood of unionization shortly after the accidents.

Appendix \ref{sec:app_iv} describes the implementation of the IV approach and presents the results. While the IV estimates confirm our main finding -- that unionization leads to a leftward shift in campaign contributions -- they are less precise than the DiD estimates. We therefore consider the IV approach as a supplementary strategy that lends additional support to our main results.

\subsection{Robustness}\label{sec:results_robust}

We now discuss further robustness checks for our main DiD estimates. Results are presented in Appendix Tables \ref{tab:extensive} to \ref{tab:federal_local} and \ref{tab:losing}.

\vspace{1em} \noindent \textbf{Extensive versus intensive margin weights.}\quad The percentage effect interpretation of the IHS results implicitly assumes a weight of extensive and intensive margin changes in contribution amounts \citep{Chen2024}. To understand the role of both margins, we perform two analyses. First, in Appendix Table \ref{tab:extensive}, we directly study extensive margin effects using as outcome variable a dummy for any positive donation by an employee at the establishment. The DiD estimates show no significant effect of unionization on the likelihood that at least one employee in the establishment donates a positive amount to any candidate. However, for the partisan composition, we find that unionization results in an increased likelihood of employees donating to Democratic relative to Republican candidates of 2.8 percentage points. This effect is observed for both workers' and managers' donations and suggests that extensive margin changes may explain at least some of the partisan shift in campaign contributions. 

Second, we analyze the robustness of our main results when placing explicit weights on extensive and intensive margin changes. For that, we follow the suggestion in \cite{Chen2024} and use the following outcome transformation: 
\begin{equation}\label{value_extensive_intensive}
	m(\$) = 
	\begin{cases} 
	\ln(\$) & \text{for } \$ > 0 \\
	-x & \text{for } \$ = 0. 
	\end{cases}
\end{equation}
This transformation assigns equal weight to the extensive margin effect of moving from 0 to 1 and the intensive margin effect of increasing donations by 100$x$ log points (approximately 100$x$\% percent) for establishments with positive donations. Appendix Table \ref{tab:extensive_intensive} reports results for alternative weights $x \in \{0, 0.1, 0.7, 1, 3\}$. Setting $x = 0.7$, we find the same DiD estimates as for the IHS-transformed outcomes, which indicates that our main results implicitly weight an extensive margin change from 0 to 1 as equivalent to an intensive margin change in donations by 70 log points.\footnote{This equivalence is expected, given that 
\begin{equation*}
\lim_{y \to \infty} \left( \text{arsinh}(y) - \ln(y) \right) 
= \lim_{y \to \infty} \left( \ln \left(y+\sqrt{y^2+1}\right) - ln(y) \right) 
= \ln(2) \approx 0.7.
\end{equation*}} Importantly, the results remain similar, growing only slightly in magnitude, for alternative weights of 10, 100, or 300 log points. For $x = 0$, which shuts off extensive margin changes by treating zero amounts as equal to one USD, we also find very similar results. Across all weights, we estimate increases in the Democratic-Republican gap between 11 and 16 (19 and 26) percentage points among workers (managers).

Taken together, the results show that unionization has a significant extensive margin effect, increasing the likelihood of at least one employee donating to a Democratic rather than a Republican candidate. However, our results do not vary strongly with alternative extensive-intensive margin weights, indicating that the extensive margin plays only a minor role in explaining the partisan shift in overall contribution amounts resulting from unionization. 

\vspace{1em} \noindent \textbf{Outcome transformations.}\quad \cite{Roth2021} point out that different transformations of the outcome variable may imply different parallel trends assumptions in a DiD design. We therefore test the sensitivity of our results to alternative outcome transformations. In addition to transformations using IHS and equation (\ref{value_extensive_intensive}), we consider the log function after adding one to the amounts, take the quartic root of amounts\footnote{As shown by \cite{Thakral2023}, power transformations -- unlike quasi-logarithmic transformations (such as the IHS or log(y+1) function) -- maintain scale invariance of the implied semi-elasticities.}, as well as leave amounts untransformed (in 2010 USD). Results, shown in Panels B, C, and D of Appendix Table \ref{tab:robust}, yield the same conclusions as the results for the IHS-transformed outcomes.

\vspace{1em} \noindent \textbf{Staggered DiD estimators.}\quad The recent econometrics literature has proposed different methods to circumvent issues of treatment effect heterogeneity in staggered DiD designs. All the proposed estimation strategies have in common that they restrict the set of effective comparison units by ruling out the use of early-treated units in the estimation of treatment effects for currently-treated units. They differ, however, in terms of how exactly comparison units are identified and used in the estimation, as well as in terms of how cohort- or individual-specific treatment effect estimates are aggregated.\footnote{In our stacking approach of model (\ref{model1}), we effectively only compare won elections to lost elections that were held in the same period, i.e., we only use never-treated units in the comparison group. The strategies by \cite{Borusyak2021} and \cite{Callaway2021}, in contrast, also allow including not-yet-treated units in the comparison group. Both approaches differ in that \cite{Borusyak2021} use the average pre-treatment outcome over all pre-treatment periods, whereas \cite{Callaway2021} only use the outcome one period before treatment start. In terms of aggregation, \cite{Gardner2021} shows that the stacking approach identifies a convexly weighted average of cohort-specific treatment effects where the weights are given by the number of treated units and the variance of treatment within each cohort. In comparison, \cite{Borusyak2021} and \cite{Callaway2021} first estimate unit- or cohort-specific effects and then aggregate through a simple average across treated units. \cite{Callaway2021} also allow other weights, but we use the default option where cohort-specific estimates are weighted by the number of treated units in each cohort.} In Panels E and F of Appendix Table \ref{tab:robust}, we present results from the imputation approach of \cite{Borusyak2021} and the estimator developed by \cite{Callaway2021}. The estimates are very similar to our stacked DiD results.

\vspace{1em} \noindent \textbf{Sample restrictions.}\quad For our baseline results, we have restricted the sample to establishments with at least one matched contribution over the observation period. If employees have not contributed before the union election and start contributing in response to unionization, concerns of sample selection bias may arise. In Appendix Table \ref{tab:any_contribution}, we show that there are no effects of unionization on the likelihood of observing any contribution after the union election. Nevertheless, Panel G of Appendix Table \ref{tab:robust} also shows results when restricting the analysis to establishments that have at least one matched contribution in any cycle before the union election. This reduces our sample size but leaves results largely unchanged. Another concern may arise from selected establishment survival \citep{Frandsen2020, Wang2021}. In particular, we may wrongly assign zero contribution amounts to establishments that do not yet exist or that have already been closed. In order to restrict the analysis to a window of known survival, we exclude for each establishment the cycles before the first matched contribution and the cycles after the last matched contribution. As shown in Panel H, this restriction increases the size of our estimates suggesting that our baseline results can be interpreted as lower bounds.

\vspace{1em} \noindent \textbf{Manager-worker classifications.}\quad In Appendix Table \ref{tab:robust_classification}, we check whether our results are sensitive to the exact definition of managers and supervisors versus non-managerial workers. To see whether the political response is different for lower- and upper-tier managers, we use more stringent definitions of managers/supervisors. First, we vary the cutoff for the importance of supervisor tasks (Panels B and C). Second, we only consider ``Management Occupations" (SOC group 11) and treat all other occupations (including those with high importance of supervisor tasks) as workers (Panel D). The results do not change much with these alternative classifications. Even for more upper-tier managers we find an increase in the support for Democrats over Republicans. As the contribution data contains a non-negligible share of occupations that we were unable to classify, we also use donors' exact residence location as an alternative predictor of occupational status that is available for the complete data. Specifically, we identify managers and workers by whether they live in a census tract with a median household income higher or lower than the 80\textsuperscript{th} or 90\textsuperscript{th} percentile of the state-specific distribution of census-tract incomes (Panels E and F). We consistently find positive effects of unionization on Democratic vs. Republican support for donors who live in high- and low-income neighborhoods.

\vspace{1em} \noindent \textbf{Federal versus state candidates.}\quad Next, we examine whether our effects are limited to contributions to candidates for either federal or state offices. Focusing on federal races offers the advantage of analyzing a set of comparable candidates who are potentially relevant to citizens nationwide. At the same time, U.S. legislation on labor issues, which unions may focus on when endorsing candidates, is enacted not only at the federal but also at the state level (e.g., state-specific minimum wages, right-to-work laws). Appendix Table \ref{tab:federal_local} shows that our estimates are driven by contributions to federal and state candidates. The effect sizes are somewhat larger for contributions to candidates running for federal office, but at both levels, we observe a significant shift in donations from Republicans to Democrats in response to unionization.

\vspace{1em} \noindent \textbf{Effects of losing a union election.}\quad Our DiD results measure the differential change in contributions between establishments with won and lost union elections. The observed relative shift in donations may reflect not only unionization effects after winning the election but also effects from holding and losing an election. Interactions with union organizers and heightened awareness of worker issues and distributional conflicts could affect employees' political behavior, in particular in the short term, even if the union election is lost. To test this, we estimate the effects of losing an election compared to holding no election, leveraging variation in the timing of elections. Specifically, we implement a stacked DiD model, using as the control group establishments that hold and lose an election in the future (see Appendix \ref{sec:app_losing} for details of the stacking implementation). Results, presented in Appendix Table \ref{tab:losing}, show small and insignificant estimates for our two main outcomes and for both workers and managers, with a precision similar to our baseline results. This suggests that losing an election serves as a valid counterfactual, confirming that our results capture the effects of unionization after winning a union election.

\section{Mechanisms and Interpretation}\label{sec:mechanisms}

\subsection{Composition versus Individual-Level Effects}
We now examine whether the establishment-level effects of unionization on employees' campaign contributions indeed reflect changes in contributions of individual employees. Alternatively, the total contribution amount of an establishment may be affected by a change in the number of employees and the partisan leaning of employees' contributions may be driven by compositional changes regarding what type of employees separate from and are newly hired into unionized establishments. \cite{Frandsen2020} shows that unionization leads older and higher-paid workers to leave and younger workers to join union jobs. Separations and hirings may also be selective in terms of political ideology. For example, conservative union-avoiding managers may want to leave unionized workplaces and be replaced by more liberal ones. To disentangle compositional and individual-level changes, we exploit donor identifiers in the DIME, which allows us to track donors' contributions over time. 

\vspace{1em} \noindent \textbf{Composition effects.}\quad We first study pure composition effects. In other words, we take out any direct effect on individuals in unionized workplaces. For this, we modify the construction of our establishment-level aggregates of employee donations in the following way. For each post-election event time $k \geq 0$, we still consider the set of donors that have at least one contribution matched to the respective establishment in that period. Then, instead of using these donors' contributions in that period, we trace their contributions before the election (in the three pre-election cycles) and use them in the establishment-level aggregation. As a result, the post-election aggregates only reflect pre-existing contribution patterns. We use them along with the actual pre-election aggregates, which are constructed as before from the actual matched contributions in those periods. Panel A of Table \ref{tab:composition} presents results from the DiD model, while event study results are shown in Appendix Figure \ref{fig:event_study_composition}. In both models, we find very small and almost always insignificant coefficients, indicating that the set of post-election employees does not differentially change in unionized versus non-unionized establishments in terms of pre-existing contribution amounts. Only for workers do we see a marginally significant estimate in line with more Democratic workers entering union jobs (or fewer Democratic workers leaving union jobs). The effect size, however, is much smaller than in our main estimates, which suggests that composition effects are unlikely to fully explain the results.\footnote{Note that the compositional analysis is complicated by the fact that we only observe employees if they contribute. In principle, our compositional test may thus also pick up changes in the extensive margin in terms of which employees stop donating after the union election. As regards candidates' party affiliation, we would expect that unionization decreases [increases] the likelihood that employees stop donating to Democrats [Republicans]. Then, the extensive margin channel would yield a positive effect on contributions to Democrats relative to Republicans that post-election employees donated before the election, in line with what we expect for the actual compositional effect. Our results show that the sum of both effects is small, suggesting that both effects play a minor role.}

\vspace{1em} \noindent \textbf{Individual-level effects.}\quad Next, we study individual-level effects of unionization, i.e., we consider the direct effect of unionization on employees. For this, we focus on a sample of incumbent employees who were employed at the establishment prior to the union election. These employees are identified based on having made at least one contribution in the year before the union election that is matched to the establishment.\footnote{We focus on donors with at least one matched contribution to a PAC in the year prior to the union election to avoid the sample selection criterion directly affecting our outcome of interest -- contributions to candidates. In Appendix Table \ref{tab:robust_individual}, we show results for alternative sample definitions: extending the pre-election period to two years and including donors with at least one matched contribution to either a PAC or a candidate.} We pull all the contributions of these donors, regardless of whether or not they are matched to the establishment, and aggregate them per cycle. With that, we estimate individual-level versions of the DiD and event study models (\ref{model1}) and (\ref{model2}).\footnote{Specifically, we estimate
\begin{equation}\label{did_individual}
y_{ijk} = \alpha_i + \beta_{kc_j} + \delta_{\textrm{DiD}} \times \Big( \mathbbm{1}[k\geq 0] \times \mathbbm{1}[V_j > .5] \Big) + \epsilon_{ijk},
\end{equation}
and 
\begin{equation}\label{event_study_individual}
y_{ijk} = \alpha_i + \beta_{kc_j} + \sum_{s=-3, s\neq-1}^{s=3} \delta_s \times \Big( \mathbbm{1}[k=s] \times \mathbbm{1}[V_j > .5] \Big) + \epsilon_{ijk},
\end{equation}
where $y_{ijk}$ denotes the outcome for individual $i$ in establishment $j$ at event time $k$ relative to the cycle of the union election. The models include individual fixed effects $\alpha_i$ that capture unobserved heterogeneity across incumbent employees.
} 
To ensure that the estimates can be readily compared to the establishment-level results, each incumbent employee is weighted by the inverse of the total number of donating incumbent employees in the establishment, and standard errors are again clustered at the establishment level. Panel B of Table \ref{tab:composition} presents results from the DiD model, while event study results are shown in Appendix Figure \ref{fig:event_study_individual}. For all employees jointly, we find no significant effect on the total contribution amounts but a significant increase in the amount donated to Democratic relative to Republican candidates. When restricting the sample to workers, we see a significant rise in total donations, which is entirely driven by an increase in support for Democrats. For managers, the results indicate a significant shift from Republicans to Democrats without a change in total amounts. Note that the estimated effects are substantially larger in magnitude than our baseline establishment-level effects. This may be explained by our focus on incumbent employees who have made at least one donation before the union election. These individuals are more politically engaged, making it likely that they respond more strongly to unions' political activities and to changes in workplace relations resulting from unionization. 
 
Taken together, the results indicate that our establishment-level effects are driven by individual-level changes in donation patterns rather than by compositional effects.

\subsection{Ideological versus Strategic Motives}\label{sec:ideo_vs_strategy}
As discussed in Section \ref{sec:background_contributions}, campaign contributions can be driven by either expressive (ideological) or investment-oriented (strategic) motives. While a large strand of work on money in politics finds that individual donors allocate their contributions primarily based on ideological preferences \citep{Stone2010, Bonica2014, Bonica2018, Barber2016, Barber2016a}, more recent research has found that parts of employees' donations are also related to their employers' strategic interests \citep{Stuckatz2022, Teso2023}. Along that line, the observed shift in managers' donations from Republican to Democratic candidates could reflect, for example, a strategic signaling of goodwill and understanding of unions' and workers' interests instead of a true shift in managers' preferences. Moreover, unionization could change workers' efforts to support candidates in exchange for future legislative votes on worker-friendly policies. While it is difficult to fully disentangle donors' motives, in the following we present several pieces of evidence that all point toward changes in ideological preferences being the main driver of our results. 

\vspace{1em} \noindent \textbf{Differentiating candidates by within-party ideology.}\quad To examine the ideological patterns in campaign contributions in more detail, we study the changing support for ideologically different candidates within the same party. \cite{Kuziemko2023} document the growing influence of conservative factions within the Democratic Party that place less emphasis on predistributive economic policies, such as higher minimum wages, protectionism, and strong labor unions. They trace the origins of this shift to the 1970s, coinciding with the period in which unionization rates began to decline more sharply. To analyze how unionization affects the support of different party factions, we make use of \citeauthor{Bonica2014}'s (2014) CF scores that assign each recipient an ideal point along a liberal-conservative scale. Democratic candidates are defined ``liberal'' (as compared to ``moderate'') if their CF score lies below the median CF of all Democrats observed in our sample of matched contributions. Similarly, we distinguish between ``conservative'' and ``moderate'' Republicans using the median Republican CF score. Table \ref{tab:partyideo} shows the DiD results for the effect on the amount contributed to each of the candidate types. Considering first all employees jointly, we see strong differences in the effects of unionization by the within-party ideological positions of candidates. Unionization significantly increases employees' support for the most liberal Democrats and decreases support for the most conservative Republicans. In contrast, contributions to moderate Democrats or Republicans are not significantly affected. These results are similar when we focus on donations from managers only, and also for workers the increased support for Democrats is more pronounced for more liberal Democrats. Overall, our results do not seem to be driven by a mere signal of affiliating with the Democratic Party but instead reflect a shift in contributions between clearly distinguishable conservative and liberal candidates.

\vspace{1em} \noindent \textbf{Differentiating candidates by union support.}\quad We continue by examining whether the partisan shift in employee contributions can be solely explained by a shift from candidates not supported to candidates supported by the union organization. To identify union-supported candidates, we match donations from PACs associated with the union organization that was on the ballot in a given union election (including PACs of local union branches). Appendix Table \ref{tab:union_ideo} reports for each union organization in our sample the share of matched contributions to Democratic (as opposed to Republican) candidates. On average, union PACs give 94\% of their donations to Democrats, which demonstrates the strong alliance between labor unions and the Democratic Party \citep{Dark2001}. In Table \ref{tab:union_support}, we use this information to estimate the effects of unionization on employees' contribution amounts to candidates supported and not supported by the respective union in a given cycle. Unions may encourage their members to contribute to the same candidates as their PAC, either directly through providing information about these candidates or indirectly through employees contributing to the union PAC and observing which candidates are supported by the union PAC. While the estimated coefficients for the effect of unionization on employees' contribution amounts to union-supported candidates are positive, they are small and insignificant.\footnote{When taking the difference in the amounts going to union-supported and non-union-supported candidates, the effect of unionization is significant only for contributions from managers and only at the 10\% level (not shown).} Moreover, we find a shift in contributions from Republican to Democratic candidates that remains significant even after conditioning on whether or not candidates are supported by the union. In other words, among both union-supported and non-union-supported candidates we observe significant effects on contribution amounts to Democratic relative to Republican candidates. These results suggest a limited role for unions' direct influence in channeling employees' contributions toward candidates in which the union has strategic interests. Instead, unionization appears to cause a more general leftward shift in employees' donation behavior.

\vspace{1em} \noindent \textbf{Contributions to PACs.}\quad In Table \ref{tab:sumstat_donorrecipient}, we have shown that contributions to PACs account for a large share of employees' political contributions. Besides changing support for specific candidates, unionization may also affect employees' donations to intermediary committees with varying ideological positions. Most evidently, unions may solicit contributions from their members to their own PACs. Table \ref{tab:committees} reports DiD estimates for the effect of unionization on employees' donations to party PACs and interest group PACs, where the latter are further disaggregated into labor union, membership organization, corporate, and trade association PACs. Besides considering the total amount given to these committees, we also measure partisan support by the difference in contribution amounts to Democratic versus Republican PACs. For interest group PACs, party affiliation is determined from the recipients of the PAC's own campaign contributions.\footnote{To track contributions that PACs donate themselves, we exploit that \cite{Bonica2019} has matched recipient identifiers to contributor identifiers for recipients' own contributions. Based on the matched outgoing contributions from PACs, we define an interest group PAC as ``Democratic" (``Republican") if more (less) than 50\% of its campaign contributions goes to Democratic candidates in a given election cycle.} Considering first the contributions from all employees of an establishment to party PACs, the results mimic those for candidate contributions. While there is no effect on total amounts, unionization leads to a significant shift from Republican to Democratic committees. Among interest group PACs, there is a significantly positive effect on donations to union PACs and a significantly negative effect on donations to corporate PACs. When distinguishing between donations from workers and managers, results differ somewhat. For workers, we see a significant increase in the total amounts donated to both party and interest group PACs, which implies that unions are successful in mobilizing PAC contributions from workers. The increase in donations appears to be driven by unions and membership organizations, pointing toward increased support for labor and civil society interest groups. In contrast to our results on candidates, however, we do not see a significant shift across party affiliations. For managers, the results are similar to those on candidate contributions. While there is no effect on overall party PAC spending, managers increasingly donate to Democratic rather than Republican PACs. Moreover, donations to corporate PACs drop, which highlights that unionization can decrease managers' support for business interest groups.\footnote{Note that the absence of significant effects on managers’ donations to union PACs, in contrast to the significant increase observed among workers, provides supporting evidence for our occupational classification. If individuals classified as managers were in fact misclassified union-eligible workers, we would expect to observe increased donations to union PACs among them as well. The lack of such an effect suggests that our manager group largely consists of individuals who are correctly identified as excluded from unionization.} Overall, these results match with the observed pro-liberal shift in workers and managers' contributions to political candidates.

\vspace{1em} \noindent \textbf{Contributions from firm PACs.}\quad Finally, we directly examine whether unionization influences strategic donations of unionized firms. A growing body of evidence suggests that corporations allocate their PAC contribution in ways that are consistent with an attempt to buy access to relevant politicians \citep{Kalla2016, Powell2016, Fouirnaies2018}. For each union election in our estimation sample, we link firm PAC contributions given to candidates and aggregate them at the firm level.\footnote{We link firm PAC contributions using a fuzzy match of the firm name. Given that the location of the firm PAC may differ from that of the establishment, we do not match on the commuting zone of the establishment. For 1,902 out of the 6,603 establishments in our estimation sample, we observe at least one firm PAC contribution over our observation window. In Appendix Table \ref{tab:firm_pac}, we report results for the full sample of establishments when considering unmatched observations as zeros, as well as for the reduced sample of establishments with at least one matched firm PAC contribution.} The results of our DiD model, reported in Appendix Table \ref{tab:firm_pac}, show that unionization does not significantly affect firm PACs' total contribution amounts or their partisan composition. We conclude that the leftward shift in employees' donations does not coincide with a corresponding shift in employers' strategic donations.

\subsection{Labor Relations Environment}
Our findings indicate that unionization increases the alignment of workers’ and managers’ political preferences, shifting support toward Democratic candidates. This influence can operate through an information channel, where unions provide political messaging that persuades both workers and managers, but it may also stem from shifts in workplace dynamics. If unionization facilitates cooperative contact between workers and managers, it may promote perspective-taking and foster managers’ understanding of workers’ political positions \citep{Allport1954}. The contact hypothesis literature suggests that interactions between different groups are most likely to reduce cleavages when they occur on an equal footing, involve cooperation toward common goals, and are governed by mutually accepted norms and regulations \citep{pettigrew2006}. 

Case studies from health care and manufacturing underscore how these dynamics can unfold in practice. At Kaiser Permanente and in joint ventures like NUMMI (Toyota-GM) and Saturn (GM-UAW), labor was embedded in systems of shared governance, including joint committees, co-developed performance metrics, and continuous problem-solving mechanisms \citep[e.g.,][]{BrownReich1989, Rubinstein2000, Kochan2008}. These arrangements closely reflect the conditions described by contact theory, fostering ongoing interaction under cooperative norms. Consistent with this, political behavior at Kaiser Permanente shifted soon after the establishment of its labor–management partnership in 1997. Individual political contributions from employees were nearly evenly split in the late 1990s but became overwhelmingly Democratic—reaching 83\% by 2004 and nearly 90\% thereafter \citep{OpenSecrets2025}. At the same time, institutional design alone does not seem to guarantee such outcomes. GM’s Van Nuys plant, which sought to replicate NUMMI-style collaboration, reveals the fragility of such efforts in the absence of trust and mutual commitment. Despite adopting similar team-based structures, the initiative faltered amid adversarial labor relations and management’s reluctance to cede meaningful authority \citep{BrownReich1989}. In that setting, formal cooperation did little to foster genuine perspective-taking, suggesting that the political consequences of unionization depend not only on structural features, but on the quality and depth of the interaction itself.

To investigate whether workplace contact can help explain the shifts in campaign donations in response to unionization, we examine how effects vary across different labor relations environments.

\vspace{1em} \noindent \textbf{Right-to-Work laws.}\quad
As a first measure of labor relations climate, we study the role of right-to-work (RTW) legislation. RTW laws allow workers in unionized establishments to forgo union membership and dues payments while still benefiting from union representation, thereby encouraging free-riding of workers and reducing unions' financial resources. RTW measures are typically advocated for by conservative activists as a means of reducing union strength and have been most successfully adopted in Southern U.S. states with more pro-business policies \citep{holmes1998, rao2011} and conflicts between unions and employers \citep{dixon2008, haskett2024}. Enacting RTW laws is indeed consequential: at the local level, it is found to reduce union membership rates \citep{ellwood1987, eren2016} and depress voter turnout and vote shares for Democratic candidates \citep{Feigenbaum2018}. We complement this evidence by studying how RTW laws moderate the effect of unionization on employees' campaign contributions at the establishment level.

Our results confirm that the political effects of unionization are significantly stronger in non-RTW states. In Table \ref{tab:hetero}, Panels A.1 and A.2, we split our estimation sample based on whether or not the union election takes place in a state that has an RTW law in force at the time of the election. In states without RTW laws, we see significantly positive effects of unionization on support for Democratic over Republican candidates, while for RTW states the coefficients are smaller and not significant. This is true for all employees, as well as for workers and managers separately. Thus, unions’ ideological influence on employees seems to be substantially lower in states with reduced union power and more adversarial labor relations.

\vspace{1em} \noindent \textbf{Unfair labor practice charges.}\quad
A second factor shaping union-employer relations is the extent of employer opposition to unionization. A long-standing literature documents the widespread use of anti-union campaigns by employers in the United States, often involving aggressive tactics to deter unionization \citep{Freeman1990a, Kleiner2001, Logan2002, Schmitt2009}. One measure of employer opposition is whether a union election is accompanied by a charge for unfair labor practices (ULPs) against the employer. The NLRA allows workers to file ULP complaints when employers interfere with their rights to organize and collectively bargain \citep{Bronfenbrenner2009, McCammon2001, Stansbury2021}.\footnote{These interferences can include restraining employees in their rights to organize in a union or collectively bargain (NLRA § 8(a)(1)), dominating or controlling a union (§ 8(a)(2)), discharging or otherwise discriminating workers involved in organizing (§§ 8(a)(3) and 8(a)(4)), and failing to bargain in good faith with the union (§ 8(a)(5)). To focus on employer misconduct in the process of the organizing drive, we restrict our analysis to charges for violations of §§ 8(a)(1), 8(a)(3), and 8(a)(4).} We gather and merge data on ULP charges against employers that were filed to the NLRB between 1984 and 2020.\footnote{The data stems from the CHIPS archive for 1984-2000, from the CATS archive for 1999-2011, and from the NLRB website for 2007-2020. Since charges must be filed within six months of an alleged violation, we consider all ULP charges that were filed between six months before to six months after a union election. See Appendix \ref{sec:app_otherdata} for more details on the data preparation.} In our estimation sample, 44\% of all union elections involve a ULP charge.

Our results in Panels B.1 and B.2 of Table \ref{tab:hetero} reveal that the political effects of unionization are substantially weaker in establishments where ULP charges are filed. In union elections without ULP charges, we observe a large and statistically significant increase in contributions to Democratic over Republican candidates among both workers and managers. However, when elections involve ULP charges, the estimated effects are smaller and remain marginally significant only for workers. These findings suggest that when unions must combat aggressive employer opposition, they may struggle to exert political influence on employees. When managers strongly resist unionization, the potential for cooperative contact and perspective-taking may diminish, reducing the likelihood that managers shift their political preferences in response to unionization.

\vspace{1em} \noindent \textbf{Collective bargaining contracts.}\quad Finally, we study heterogeneity by whether the union and employer reach an agreement after a union election is won, in particular whether they successfully negotiate a collective bargaining contract. Collective bargaining is a defining feature of unionized workplaces, yet not all successful unionization drives are characterized by cooperative labor-management interactions that lead to an agreement. Compiling and merging data on contract negotiation notices from the Federal Mediation and Conciliation Service (FMCS), we only observe a bargaining contract for 48\% of the won union elections.\footnote{We combine contract data compiled by \cite{Holmes2006} for 1985-2003 and by \cite{Gregg2024} for 2004-2020. The data includes notices of contract expirations that are provided to the FMCS for it to be able to prepare mediation services for the renegotiation of a contract. We consider contract expiration notices in the five years following the union election. See Appendix \ref{sec:app_otherdata} for more details on the data preparation.} 

We find strong political effects of unionization only in establishments where a collective bargaining agreement is reached. In Panels C.1 and C.2 of Table \ref{tab:hetero}, we divide our treatment group into won union elections with a reported bargaining contract and those without.\footnote{In both cases, we compare the won union elections to the control group of all lost union elections.} For union certifications that result in a collective bargaining agreement, both workers and managers exhibit a significant shift toward supporting Democratic over Republican candidates, while we find no significant effects in establishments without a bargaining agreement. For workers, the results may suggest that organizing alone is not enough -- unions must also deliver on the bread and butter issues to be able to politically influence their members. For managers, the results are particularly noteworthy. Although they do not directly benefit from a collective bargaining contract, they nonetheless exhibit a larger ideological shift in their donations in establishments where an agreement is reached. This result is again consistent with the idea that sustained, cooperative contact -- during successful contract negotiations -- can help explain why managers align their political positions with workers and the union.

\section{Conclusion}\label{sec:conclusion}

Labor unions employ enormous resources to shape labor policies and welfare regulations through political activities such as lobbying legislators or supporting candidates financially. However, lasting change requires changes in the political preferences of the electorate. To understand the political power of labor unions, it is crucial to examine whether they can mobilize and influence the ideologies of millions of individuals at the unionized workplace. This paper analyzes the political effects of unionization, building on an establishment-level dataset that combines union elections with campaign contributions from employees spanning the 1980-2016 period in the United States. Comparing establishments that won and lost the union election in a stacked DiD model, we find that unionization increases contributions to Democratic candidates relative to Republican candidates by 12 percentage points for workers and 20 percentage points for managers, while we do not find a lasting impact on the overall amount of contributions. Overall, we show that labor unions influence the political preferences not only of union members but also of their firms' management.
 
We also document that unionization leads to larger ideological shifts in settings where labor-management relations are more cooperative, such as in union elections without unfair labor practice charges and where the employer and union reach a collective bargaining agreement. These patterns suggest an important role of workplace contact, in particular for explaining why managers align their political views with those of workers and the union. The contact hypothesis posits that cooperative interactions can enhance perspective-taking and reduce biases in beliefs about opposing groups that may prevail among managers. Biased beliefs may also help explain why many employers in the U.S. remain strongly opposed to unionization. The literature finds little evidence that unionization strongly increases wages \citep{DiNardo2004, Frandsen2020, Freeman1990} or reduces productivity \citep{Dube2016, Sojourner2015a}, which could harm firms profitability; yet strong employer resistance persists. We welcome future research that explicitly studies how managers form beliefs about the impact of unionization and what factors sustain anti-union sentiment among employers.

Our findings may have implications for broader developments in U.S. politics. The longstanding decline in private-sector union density implies that millions of individuals have forfeited the engagement with unions, which has led to lasting shifts in political preferences. The erosion of unionization can be an important contribution to the increased alignment of workers with the political right that has been observed over the last decades \citep{Gethin2022}. More recently, prominent examples of strikes and union petition drives in Starbucks shops, Amazon warehouses, and healthcare facilities suggest a moment of resurgence for labor organization. Whether this trend persists may be consequential for the balance of political power and support for pro-labor politics in the United States.

\clearpage
\newpage\clearpage
\renewcommand{\baselinestretch}{1.05}
\setlength{\baselineskip}{12pt}
\addcontentsline{toc}{section}{References}
\bibliography{unions}

\clearpage 
\renewcommand{\baselinestretch}{1.5}
\setlength{\baselineskip}{20pt}

\section*{Figures and Tables}

\clearpage

\clearpage

\begin{table}[htbp]\centering
	\caption{Election and Contribution Descriptive Statistics}
	\label{tab:sumstat_election}
	\begin{tabular}{lccc}
		\toprule[1.5pt]
		& All & Union Loss & Union Win \\
		\midrule
		& \multicolumn{3}{c}{[A] Election characteristics} \\
		Number of elections & 6,063 & 3,397 & 2,666 \\
		Union vote share (average) & .4950 & .3204 & .7175 \\
		Number of votes (average) & 119.37 & 135.31 & 99.06 \\
		Number of votes (total) & 723,752 & 459,661 & 264,091 \\
		[0.5em]
		& \multicolumn{3}{c}{[B] Contribution characteristics} \\
		Amount (total, in million 2010 USD) & 105.82 & 65.38 & 40.43 \\
		Number of contributions (total) & 357,436 & 204,797 & 152,639 \\
		Number of donors (total) & 46,719 & 26,661 & 20,243 \\
		Number of recipients (total) & 9,942 & 7,208 & 5,681 \\
		[0.5em]
		\hline\hline
		\multicolumn{4}{l}{
		\begin{minipage}{0.8\linewidth} \smallskip \scriptsize
		\textbf{Notes:} Data from NLRB union certification elections, which have at least one employee contribution matched in any of seven election cycles around the union election (three before, cycle of union election, three after). Contribution characteristics refer to the total numbers over all these seven election cycles.   
		\end{minipage}} \\
	\end{tabular}
\end{table}

\vspace{5em}

\begin{table}[htbp]\centering
	\caption{Contributions by Donor and Recipient}
	\label{tab:sumstat_donorrecipient}
		\begin{tabular}{lccc}
			\toprule[1.5pt]
			\diagbox{Recipient:}{Donor:} 
				& All employees & Workers & Managers \\
			\midrule
			All & 2,493.24 & 313.80 & 1,339.38 \\
			[0.5em]
			Candidates & 1,181.96 & 173.42 & 594.44 \\
			\hspace{0.5em} Democratic candidates & 575.85 & 112.79 & 261.76  \\
			\hspace{0.5em} Republican candidates & 586.98 & 56.61 & 320.66 \\
			[0.5em]						
			Political action committees & 1,311.28 & 140.38 & 744.94 \\
			\hspace{0.5em} Party PACs &  364.92 &  52.52 & 192.77 \\
			\hspace{0.5em} Interest group PACs & 937.22 &  86.37 & 549.31 \\			
			\hline\hline
			\multicolumn{4}{l}{
				\begin{minipage}{0.7\linewidth} \smallskip \scriptsize
					\textbf{Notes:} The table reports mean values for the amount contributed in each of the 42,441 establishment-cycle combinations in the estimation sample. All amounts are in 2010 USD. Values are reported separately for contributions from all employees, from only non-managerial workers (``workers"), and from only managers and supervisors (``managers"). The difference in the amounts from all employees and the total from workers and managers is driven by contributions for which we were unable to classify the occupation.
			\end{minipage}} \\
		\end{tabular}
\end{table}

\clearpage

\begin{table}[htbp]\centering
	\caption{Composition versus Individual-Level Effects}
	\label{tab:composition}
		\begin{tabular}{l*{6}{c}}
			\toprule[1.5pt]
			&\multicolumn{3}{c}{IHS(\$ to all candidates)}&\multicolumn{3}{c}{IHS(\$ to Dem.) $-$ IHS(\$ to Rep.)} \\ \cmidrule(lr){2-4} \cmidrule(lr){5-7}
			&All&Workers&Managers&All&Workers&Managers \\	
			&(1)&(2)&(3)&(4)&(5)&(6) \\	
			\midrule
			\multicolumn{7}{l}{[A] Composition effects} \\  
			$\delta_{\textrm{DiD}}$    
			  &       -0.027          &       0.046    &   -0.067     &      0.071  &       0.053\sym{*} &       0.037  \\
                &     (0.070)           &    (0.036)     &    (0.051)   &     (0.064) &    (0.029)       &     (0.045)          \\
			[0.5em]
                N    & 33,103 	  & 33,103		& 33,103 	  & 33,103		& 33,103 		  & 33,103		    \\
                [0.5em]
			\midrule 
			\multicolumn{7}{l}{[B] Individual-level effects} \\  
			$\delta_{\textrm{DiD}}$    
			  &        0.047      &       0.507\sym{*}   &   0.183      &  0.633\sym{***}  & 0.727\sym{**} &  0.594\sym{**}   \\
                &      (0.162)      &    (0.261)      &    (0.260)    &   (0.160) &     (0.275)         &    (0.274)         \\
			[0.5em]
                N    & 22,799	  &  5,243 		& 13,104	  &  22,799	& 5,243		  &  13,104 	    \\
                [0.5em]
			\hline\hline
			\multicolumn{7}{l}
			{\footnotesize 
				\begin{minipage}{0.77\linewidth} \scriptsize \smallskip 
					\textbf{Notes:} The table reports DiD coefficients for the composition and individual-level effects of unionization on the IHS-transformed total amount contributed (columns (1) - (3)) and on the difference between the IHS-transformed amounts contributed to Democratic and Republican candidates (columns (4) - (6)). Panel A shows results from the establishment-level model (\ref{model1}). The establishment outcomes for the post-election periods are computed from the pre-election contributions of donors matched to an establishment in the respective post-election period. Outcomes for the pre-election periods are constructed as before from the actual contributions in those periods. Panel B shows results from the individual-level model (\ref{did_individual}). The sample includes incumbent employees who made at least one contribution to a PAC in the year before the union election that is matched to the establishment. The individual $\times$ cycle panel aggregates all the contributions of these donors three cycles before to three cycles after the union election, irrespective of whether or not they are matched to the initial establishment. In the regressions, each incumbent employee is weighted by the inverse of the total number of donating incumbent employees in the establishment. All samples include establishments with a pro-union vote share between 20 and 80\%. Standard errors clustered at the establishment level are in parentheses. \sym{*} \(p<0.10\), \sym{**} \(p<0.05\), \sym{***} \(p<0.01\)
				\end{minipage}
			}\\
		\end{tabular}
\end{table}	

\begin{table}[htbp]\centering
	\def\sym#1{\ifmmode^{#1}\else\(^{#1}\)\fi}
	\caption{Differentiating Candidates by Within-Party Ideology}
	\label{tab:partyideo}
	\begin{tabular}{l*{4}{c}}
			\toprule[1.5pt]
			&\multicolumn{2}{c}{IHS(\$ to Democrats)} &\multicolumn{2}{c}{IHS(\$ to Republicans)} \\ \cmidrule(lr){2-3}\cmidrule(lr){4-5}
			&  Liberal & Moderate & Moderate & Conservative     \\
			&\multicolumn{1}{c}{(1)} &\multicolumn{1}{c}{(2)}&\multicolumn{1}{c}{(3)}&\multicolumn{1}{c}{(4)}     \\
			\midrule
			\multicolumn{5}{l}{[A] All employees} \\
			$\delta_{\textrm{DiD}}$     
			&       0.119\sym{***}&       -0.017         &      -0.0690    &      -0.153\sym{***}\\
			&    (0.046)         &    (0.054)         &    (0.055)         &    (0.049)         \\
			[0.5em]
			\midrule
			\multicolumn{5}{l}{[B] Workers} \\
			$\delta_{\textrm{DiD}}$    
			&      0.053\sym{*}  &      0.031         &     -0.015         &     -0.031         \\
			&    (0.030)         &    (0.024)         &    (0.022)         &    (0.026)         \\
			[0.5em]
			\midrule
			\multicolumn{5}{l}{[C] Managers} \\
			$\delta_{\textrm{DiD}}$     
            &       0.088\sym{**} &       0.014         &      -0.056      &      -0.123\sym{***}\\
            &     (0.035)         &     (0.039)         &     (0.040)      &     (0.037)         \\
			[0.5em]
			\hline\hline
			\multicolumn{5}{l}
			{
				\begin{minipage}{0.58\linewidth} \scriptsize \smallskip 
					\textbf{Notes:} The table reports DiD coefficients, estimated in model (\ref{model1}), for the effect of unionization on IHS-transformed amounts contributed to different candidate groups. Liberal (moderate) Democrats refer to Democratic candidates with a CF score below (above) the median CF score of all Democratic candidates observed in our sample of matched contributions. Conservative and moderate Republicans are differentiated accordingly using the median Republican CF score. The sample includes establishments with a pro-union vote share between 20 and 80\%. $N=33,103$ establishment-cycle observations. Standard errors clustered at the establishment level are in parentheses. \sym{*} \(p<0.10\), \sym{**} \(p<0.05\), \sym{***} \(p<0.01\)
				\end{minipage}
			}\\
	\end{tabular}
\end{table}

\begin{table}[htbp]\centering
	\def\sym#1{\ifmmode^{#1}\else\(^{#1}\)\fi}
	\caption{Differentiating Candidates by Union Support}
	\label{tab:union_support}
		\begin{tabular}{l*{6}{c}}
			\toprule[1.5pt]
			&\multicolumn{3}{c}{IHS(\$ to all candidates)}&\multicolumn{3}{c}{IHS(\$ to Dem.) $-$ IHS(\$ to Rep.)} \\ \cmidrule(lr){2-4} \cmidrule(lr){5-7}
			&All&Workers&Managers&All&Workers&Managers \\	
			&(1)&(2)&(3)&(4)&(5)&(6) \\	
			\midrule  			
			\multicolumn{7}{l}{[A] Candidates supported by union PAC} \\
			$\delta_{\textrm{DiD}}$    
            &  0.065         &       0.030         &       0.047         &       0.100\sym{*}  &       0.047\sym{*}  &       0.102\sym{**} \\
            &  (0.057)         &     (0.030)         &     (0.043)         &     (0.057)         &     (0.028)         &     (0.042)         \\
			[0.5em]
			\midrule 
			\multicolumn{7}{l}{[B] Candidates not supported by union PAC} \\
			$\delta_{\textrm{DiD}}$    
            &   -0.052         &       0.008         &      -0.046         &       0.225\sym{***}&       0.075\sym{**} &       0.134\sym{***}\\
            &   (0.070)        &     (0.036)         &     (0.053)         &     (0.072)         &     (0.033)         &     (0.052)         \\
			[0.5em]
			\hline\hline
			\multicolumn{7}{l}
			{
				\begin{minipage}{0.78\linewidth} \scriptsize \smallskip 
					\textbf{Notes:} The table presents DiD coefficients, estimated in model (\ref{model1}), for the effect of unionization on the IHS-transformed total amount contributed (columns (1) - (3)) and on the difference between the IHS-transformed amounts contributed to Democratic and Republican candidates (columns (4) - (6)). Panels A and B distinguish contribution amounts to candidates supported versus not supported by PACs associated with each election's union organization. See Appendix Table \ref{tab:union_ideo} for a list of all union organizations and the party composition of their matched donations. The sample includes establishments with a pro-union vote share between 20 and 80\%. $N=29,757$ establishment-cycle observations. Standard errors clustered at the establishment level are in parentheses. \sym{*} \(p<0.10\), \sym{**} \(p<0.05\), \sym{***} \(p<0.01\)
				\end{minipage}
			}\\
		\end{tabular}
\end{table}

\clearpage

\begin{table}[htbp]\centering
	\caption{Contributions to Political Action Committees}
	\label{tab:committees}
	\resizebox{\textwidth}{!}{
		\begin{tabular}{l*{8}{c}}
			\toprule[1.5pt]
			&\multicolumn{2}{c}{Party PACs}&\multicolumn{6}{c}{Interest group PACs} \\ \cmidrule(lr){2-3} \cmidrule(lr){4-9}
			&All	&Dem $-$ Rep	&  All& Union &Member  & Corporation  & Trade  &Dem $-$ Rep \\
			&		&			&  		& 			  &orga. &    & assoc. & \\
			& (1) & (2) & (3) & (4) & (5) & (6) & (7) & (8) \\
			\midrule
			\multicolumn{9}{l}{[A] All employees} \\
			$\delta_{\textrm{DiD}}$     
            &      -0.025         &       0.097\sym{**} &      -0.082         &       0.022\sym{*}  &      -0.011         &      -0.093\sym{**} &      -0.026         &       0.060         \\
            &     (0.052)         &     (0.048)         &     (0.064)         &     (0.011)         &     (0.031)         &     (0.041)         &     (0.044)         &     (0.041)         \\
			[0.5em]
			\midrule
			\multicolumn{9}{l}{[B] Workers} \\
			$\delta_{\textrm{DiD}}$    
            &       0.062\sym{*}  &       0.010         &       0.088\sym{**} &       0.024\sym{***}&       0.044\sym{**} &      -0.020         &       0.021         &       0.024         \\
            &     (0.032)         &     (0.028)         &     (0.035)         &     (0.007)         &     (0.019)         &     (0.021)         &     (0.016)         &     (0.027)         \\
			[0.5em]
			\midrule
			\multicolumn{9}{l}{[C] Managers} \\
			$\delta_{\textrm{DiD}}$     
            &      -0.001         &       0.102\sym{***}&      -0.093\sym{*}  &       0.006         &      -0.002         &      -0.082\sym{**} &      -0.026         &       0.081\sym{**} \\
            &     (0.034)         &     (0.031)         &     (0.049)         &     (0.007)         &     (0.018)         &     (0.034)         &     (0.033)         &     (0.032)         \\
			[0.5em]
			\hline\hline
			\multicolumn{9}{l}{
				\begin{minipage}{1.03\linewidth} \scriptsize \smallskip 
					\textbf{Notes:} The table reports DiD coefficients, estimated in model (\ref{model1}), for the effect of unionization on IHS-transformed amounts contributed to different committee groups. In columns (2) and (8) the dependent variable is the difference between the IHS-transformed amounts contributed to Democratic and Republican committees. Interest group PACs are categorized as ``Democratic" (``Republican") if more (less) than 50\% of their own campaign contributions goes to Democratic candidates. The sample includes establishments with a pro-union vote share between 20 and 80\%. $N=33,103$ establishment-cycle observations. Standard errors clustered at the establishment level are in parentheses. \sym{*} \(p<0.10\), \sym{**} \(p<0.05\), \sym{***} \(p<0.01\)
				\end{minipage}
			}\\				
		\end{tabular}
	}
\end{table}

\clearpage

\begin{table}[htbp]\centering
	\def\sym#1{\ifmmode^{#1}\else\(^{#1}\)\fi}
	\caption{Heterogeneous Effects by Labor Relations Environment}
	\label{tab:hetero}
		\begin{tabular}{l*{6}{c}}
			\toprule[1.5pt]
			&\multicolumn{3}{c}{IHS(\$ to all candidates)}&\multicolumn{3}{c}{IHS(\$ to Dem.) $-$ IHS(\$ to Rep.)} \\ \cmidrule(lr){2-4} \cmidrule(lr){5-7}
			&All&Workers&Managers&All&Workers&Managers \\	
			&(1)&(2)&(3)&(4)&(5)&(6) \\	
			\midrule  			
                \multicolumn{7}{l}{[A.1] States without Right-to-Work law} \\
		      $\delta_{\textrm{DiD}}$ 
                &       0.045         &      0.066         &     -0.039         &       0.284\sym{***}&       0.131\sym{***}&       0.218\sym{***}\\
			&    (0.090)         &    (0.050)         &    (0.067)         &    (0.088)         &    (0.046)         &    (0.064)         \\
			[0.5em]
			N      &       26,208         &       26,208         &       26,208         &       26,208         &       26,208         &       26,208         \\
			\midrule 
			\multicolumn{7}{l}{[A.2] States with Right-to-Work law} \\
			$\delta_{\textrm{DiD}}$    
			&     -0.055         &      -0.119         &     0.008         &      0.016         &      0.070         &       0.142         \\
			&     (0.170)         &    (0.082)         &     (0.125)         &     (0.177)         &    (0.077)         &     (0.117)         \\
			[0.5em]
			N      &        6,895         &        6,895         &        6,895         &        6,895         &        6,895         &        6,895         \\
			\midrule  			
                \multicolumn{7}{l}{[B.1] Elections without unfair labor practice charge} \\
		      $\delta_{\textrm{DiD}}$ 
                &       0.056         &       0.093         &      -0.052         &       0.347\sym{***}&       0.153\sym{***}&       0.290\sym{***}\\
                &      (0.111)         &     (0.062)         &     (0.084)         &     (0.110)         &     (0.059)         &     (0.079)         \\
			[0.5em]
			N    &      17,402         &      17,402         &      17,402         &      17,402         &      17,402         &      17,402         \\
			\midrule 
			\multicolumn{7}{l}{[B.2] Elections with unfair labor practice charge} \\
			$\delta_{\textrm{DiD}}$    
		      &       0.054         &      -0.044         &       0.001         &       0.167         &       0.102\sym{*}  &       0.132         \\
                &     (0.120)         &     (0.065)         &     (0.091)         &     (0.121)         &     (0.057)         &     (0.085)         \\
			[0.5em]
			N  &      14,658         &      14,658         &      14,658         &      14,658         &      14,658         &      14,658         \\
			\midrule  			
                \multicolumn{7}{l}{[C.1] Won elections with collective bargaining contract} \\
		      $\delta_{\textrm{DiD}}$ 
			&      0.070         &       0.112\sym{*}  &       0.080         &       0.468\sym{***}&       0.207\sym{***}&       0.356\sym{***}\\
			&     (0.104)         &     (0.058)         &     (0.077)         &     (0.105)         &     (0.053)         &     (0.076)         \\
			[0.5em]
			N     &      26,089         &      26,089         &      26,089         &      26,089         &      26,089         &      26,089         \\
			\midrule 
			\multicolumn{7}{l}{[C.2] Won elections without collective bargaining contract} \\
			$\delta_{\textrm{DiD}}$    
			&       0.002         &      -0.047         &      -0.109         &       0.037         &       0.054         &       0.076         \\
			&     (0.098)         &     (0.053)         &     (0.074)         &     (0.095)         &     (0.049)         &     (0.067)         \\
			[0.5em]
			N   &      26,859         &      26,859         &      26,859         &      26,859         &      26,859         &      26,859         \\
			\hline\hline
			\multicolumn{7}{l}
			{
				\begin{minipage}{0.75\linewidth} \scriptsize \smallskip 
					\textbf{Notes:} The table presents DiD coefficients, estimated in model (\ref{model1}), for the effect of unionization on the IHS-transformed total amount contributed (columns (1) - (3)) and on the difference between the IHS-transformed amounts contributed to Democratic and Republican candidates (columns (4) - (6)). Panels A.1 and A.2 distinguish between establishments in states with versus without Right-to-Work laws in the union election year. Panels B.1 and B.2 differentiate union elections by whether the union filed an unfair labor practice charge (for violation of sections 8(a)(1), 8(a)(3), or 8(a)(4) of the NLRA) in the six months before or after the election. In Panels C.1 and C.2, we differentiate won union elections that have a contract expiration notice reported to the FCMS in the 5 years following the election and those that do not. In both C.1 and C.2, the control group consists of all lost elections. The sample includes establishments with a pro-union vote share between 20 and 80\%. Standard errors clustered at the establishment level are in parentheses. \sym{*} \(p<0.10\), \sym{**} \(p<0.05\), \sym{***} \(p<0.01\)
				\end{minipage}
			}\\
		\end{tabular}
\end{table}

\newpage\clearpage

\begin{appendix}
	
	\bigskip
	\begin{center}
		\section*{{\LARGE Supplemental Appendix: \\ Do Unions Shape Political Ideologies at Work?}}
	\end{center}

\startcontents[sections]
\printcontents[sections]{l}{1}{\setcounter{tocdepth}{2}}

\newpage

\renewcommand{\thesection}{A}
\renewcommand{\thetable}{A.\arabic{table}}
\setcounter{table}{0}
\renewcommand{\thefigure}{A.\arabic{figure}}
\setcounter{figure}{0}


\section{Additional Figures and Tables}

\begin{figure}[htbp]
	\begin{center}	
		\caption[gbadolite]{\label{fig:donor_occup} Donor Occupations}
		\vspace{-1em}
		\centering
		\mbox{\includegraphics[width=0.8\linewidth]{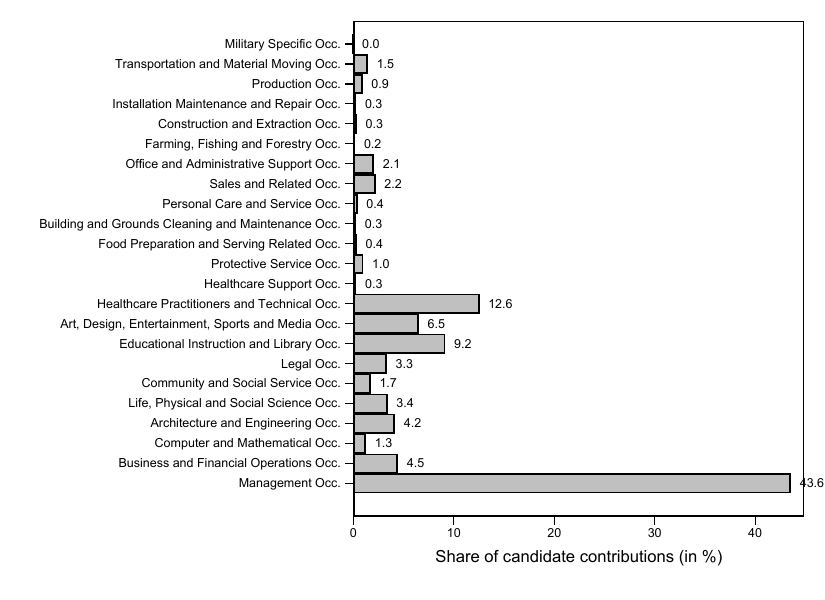}}
	\end{center}
	\vspace{-2em} 
	\begin{minipage}{\textwidth} {\scriptsize \smallskip 
 \textbf{Notes:} The figure shows the distribution of occupations for all candidate contributions that are included in our matched estimation sample and have a classified occupation. For 28.1\% of the contributions we were not able to assign an occupation code. Occupation groups are 2-digit codes of the 2018 Standard Occupational Classification (SOC). See Appendix \ref{sec:app_occclassify} for details on the occupation classification procedure.\par}
    \end{minipage} 
\end{figure}

\begin{figure}[htbp]
	\begin{center}	
\caption[gbadolite]{\label{fig:hist} Vote Share Distribution}
		\vspace{-1em}
\centering
\mbox{\includegraphics[width=0.8\linewidth]{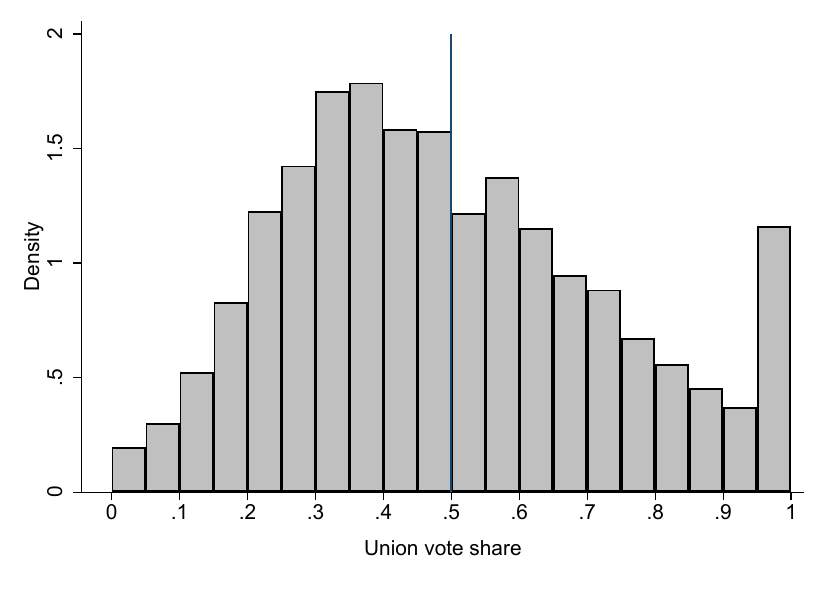}}
	\end{center}
\vspace{-2em} 
\begin{minipage}{\textwidth} {\scriptsize \smallskip 
\textbf{Notes:} The figure plots the density of union vote shares for all 6,063 union elections included in our matched estimation sample. The \cite{Frandsen2017} test strongly rejects continuity in the union vote share density at the 50\% cutoff (p-value $= .002$ for $k=0$ and p-value $= .003$ for $k=.02$).\par}
    \end{minipage}
\end{figure}

\begin{figure}[h!]
	\begin{center}
		\caption{\label{fig:cyclicality} Cyclicality of Union Elections}%
		\subfloat[\centering Number of Union Elections per Week of the Year]{{\includegraphics[width=.7\linewidth]{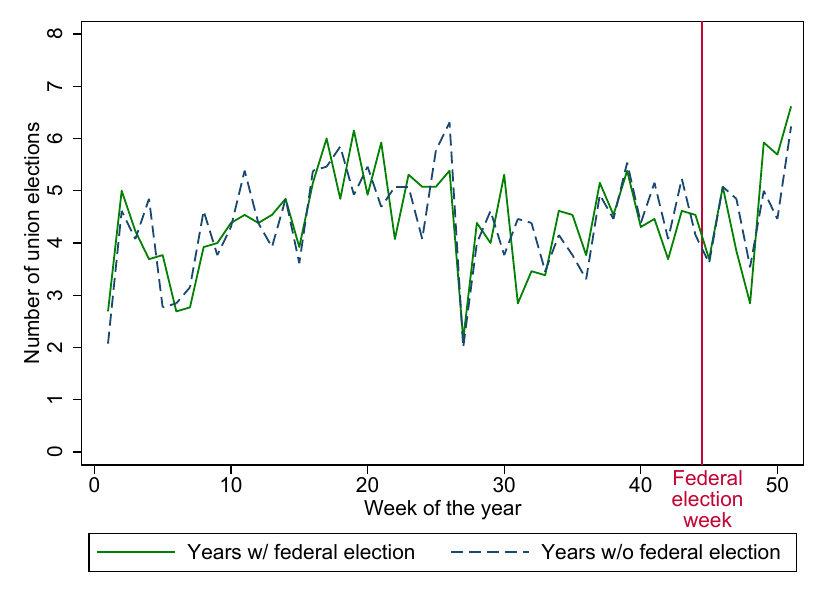} }} \\ %
		\subfloat[\centering Share of Won Union Elections per Week of the Year]{{\includegraphics[width=.7\linewidth]{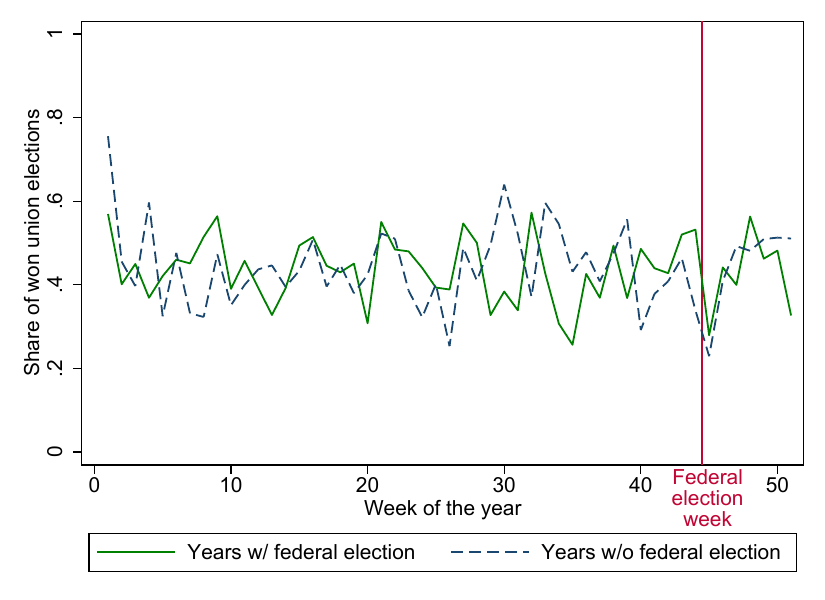} }} \\ %
	\end{center}
	\vspace{-1em}
	\begin{minipage}{\textwidth} {\scriptsize \smallskip 
 \textbf{Notes:} The graphs show the mean number of elections (Panel (a)) and mean share of won union elections (Panel (b)) per week of the year across all years in our period of analysis, i.e., between 1985 and 2010. The means are based on our matched estimation sample. We distinguish between years with and without federal elections. The red line highlights the week of federal elections, which is calendar week 44 or 45.\par}
    \end{minipage}
\end{figure}

\clearpage

\begin{figure}%
	\caption{\label{fig:rdd_tests_total}\centering Vote Share Tests for Effect of Unionization on Total Contribution Amounts}%
	\subfloat[\centering Vote Share Heterogeneity]{{\includegraphics[width=.475\linewidth]{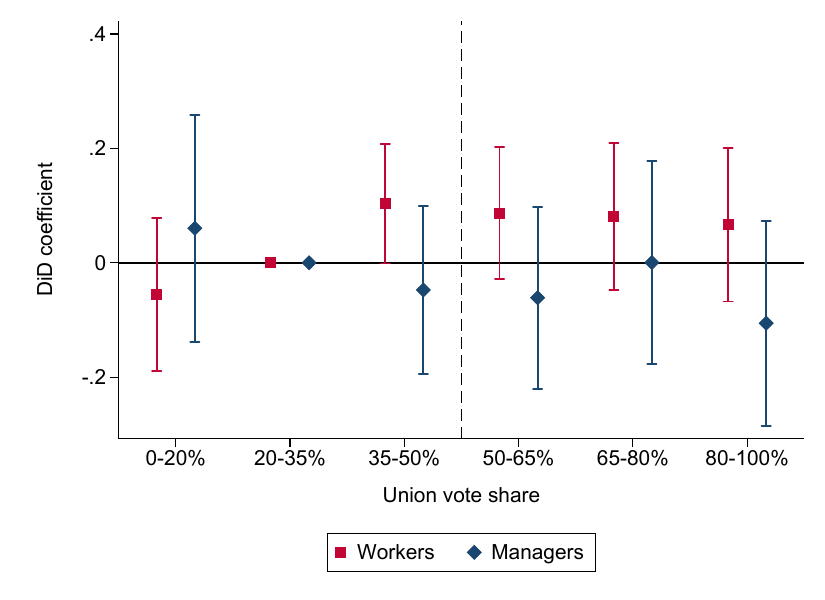} }}%
	\qquad
	\subfloat[\centering Vote Share Bandwidth Sample Restrictions]{{\includegraphics[width=.475\linewidth]{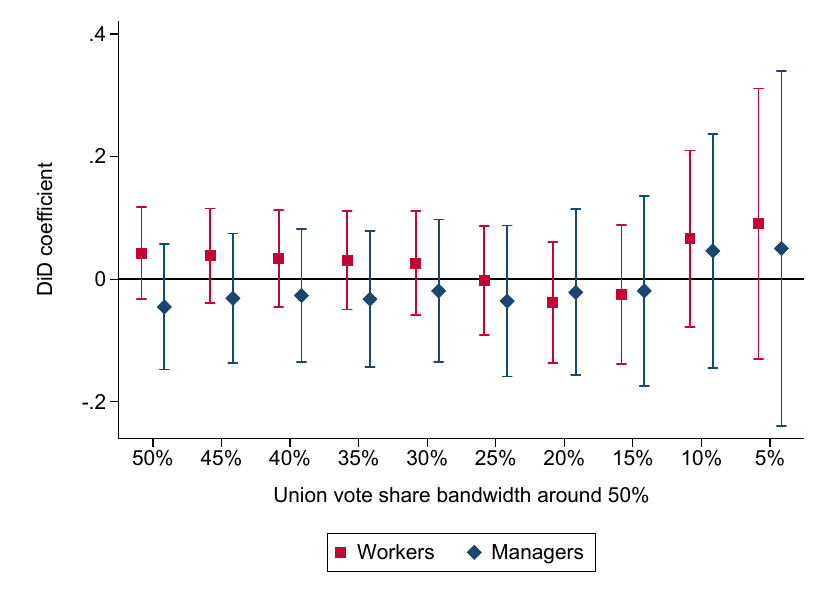} }}%
	\vspace{0.1em}
	\begin{minipage}{\textwidth} {\scriptsize \smallskip 
 \textbf{Notes:} The graphs show RDD-type placebo and robustness tests for the effect of unionization on the IHS-transformed total amount contributed. Panel (a) reports the $\delta_g$ coefficients estimated in model (\ref{model3}). The vote share distribution is partitioned into six bins, indicated on the x-axis. The omitted reference group is 20-35\%. Panel (b) reports DiD coefficients estimated in model (\ref{model1}). Each dot refers to a single DiD coefficient that is estimated among elections with a union vote share in a given bandwidth around the 50\% cutoff. Estimates from smaller bandwidths compare changes between increasingly close elections. Results are always shown separately for contributions from non-managerial workers (``workers") and from managers and supervisors (``managers"). 95\% confidence intervals are depicted for standard errors clustered at the establishment level.\par}
    \end{minipage}
	\vspace{1em}
\end{figure}

\begin{figure}%
	\caption{\label{fig:rdd_tests_prepost}\centering Vote Share Heterogeneity in Pre- versus Post-Effects}%
	\subfloat[\centering Workers]{\includegraphics[width=.475\linewidth]{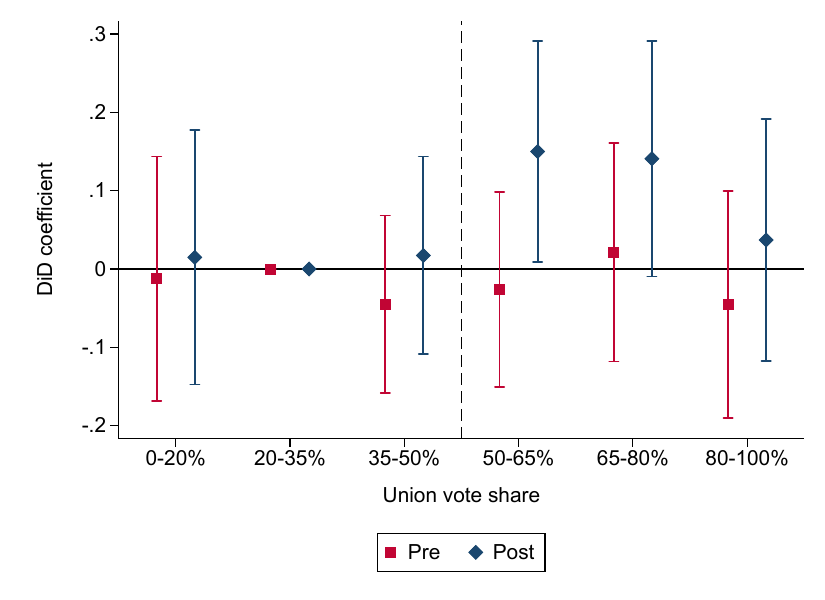} }%
	\qquad
	\subfloat[\centering Managers]{\includegraphics[width=.475\linewidth]{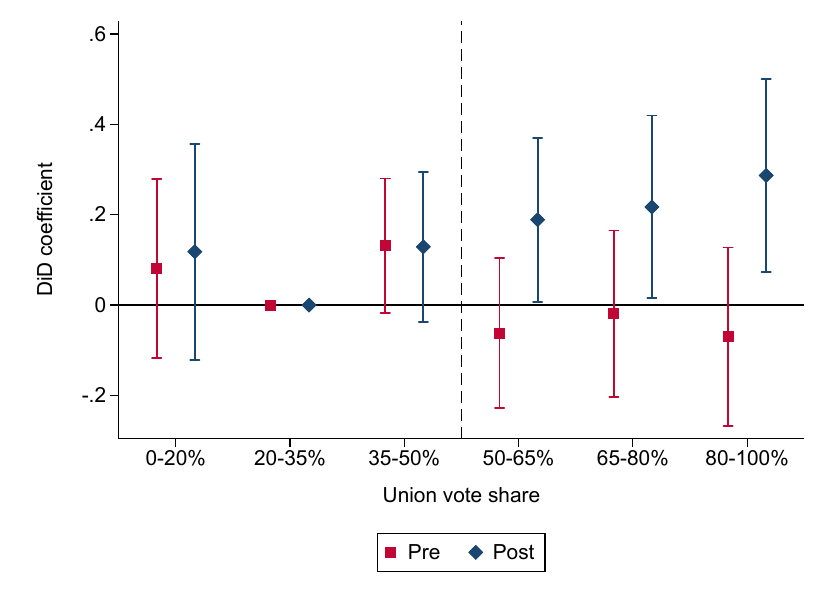} }%
	\vspace{0.1em}
	\begin{minipage}{\textwidth} {\scriptsize \smallskip 
 \textbf{Notes:} The graphs report coefficients for interactions between union win, six vote share categories, and two dummies for pre- versus post-union election periods. The regressions modify model (\ref{model3}) by including an additional interaction with a pre-period dummy (three and two cycles before the union election). The reference event time is the cycle before the union election and the reference vote share category is 20-35\%. The outcome variable is the difference between the IHS-transformed amounts contributed to Democratic and Republican candidates. Results are shown separately for contributions from non-managerial workers (``workers") and from managers and supervisors (``managers"). 95\% confidence intervals are depicted for standard errors clustered at the establishment level.\par}
    \end{minipage}
\end{figure}

\begin{figure}[htbp]
	\begin{center}
	\caption[gbadolite]{\label{fig:event_study_composition} Composition Effects - Event Study Results}
	\centering
	\vspace{-1em}
	\mbox{\includegraphics[width=\linewidth]{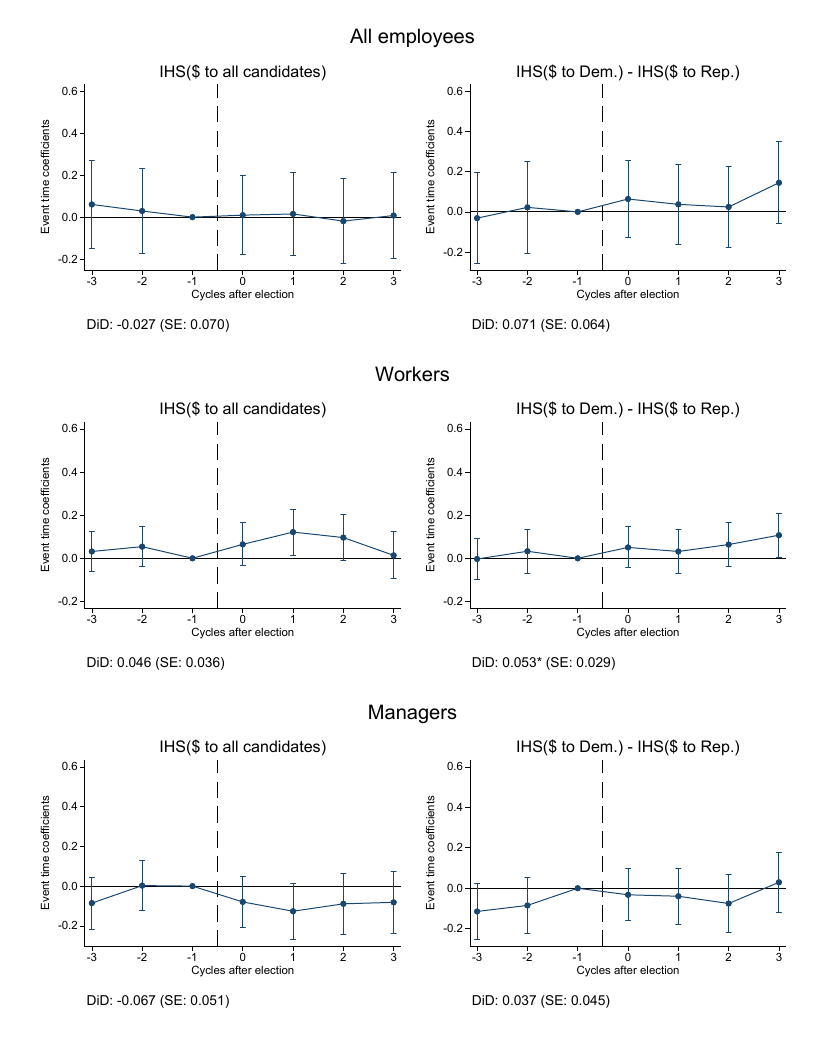}}
	\vspace{-1cm}
	\end{center}
	\begin{minipage}{\textwidth} {\scriptsize \smallskip 
 \textbf{Notes:} The figures report event-study coefficients $\delta_s$, estimated in model (\ref{model2}), for the composition effects of unionization. The establishment outcomes for the post-election periods are computed from the pre-election contributions of donors matched to an establishment in the respective post-election period. Outcomes for the pre-election periods are constructed as before from the actual contributions in those periods. Below each graph the DiD coefficient from model (\ref{model1}) is reported. In the graphs on the left side, the outcome is the IHS-transformed total amount contributed to all candidates. In the graphs on the right side, the outcome is the difference between the IHS-transformed amounts contributed to Democratic and Republican candidates. Results are reported for contributions from all employees (top part), from only non-managerial workers (middle part), and from only managers and supervisors (lower part). The sample includes all establishments with a pro-union vote share between 20\% and 80\% and covers three election cycles (six years) before and after the union election. $N=33,103$ establishment-cycle observations. 95\% confidence intervals are depicted for standard errors clustered at the establishment level.\par}
    \end{minipage}
\end{figure}

\begin{figure}[htbp]
	\begin{center}
	\caption[gbadolite]{\label{fig:event_study_individual} Individual-level Effects - Event Study Results}
	\centering
	\vspace{-1em}
	\mbox{\includegraphics[width=\linewidth]{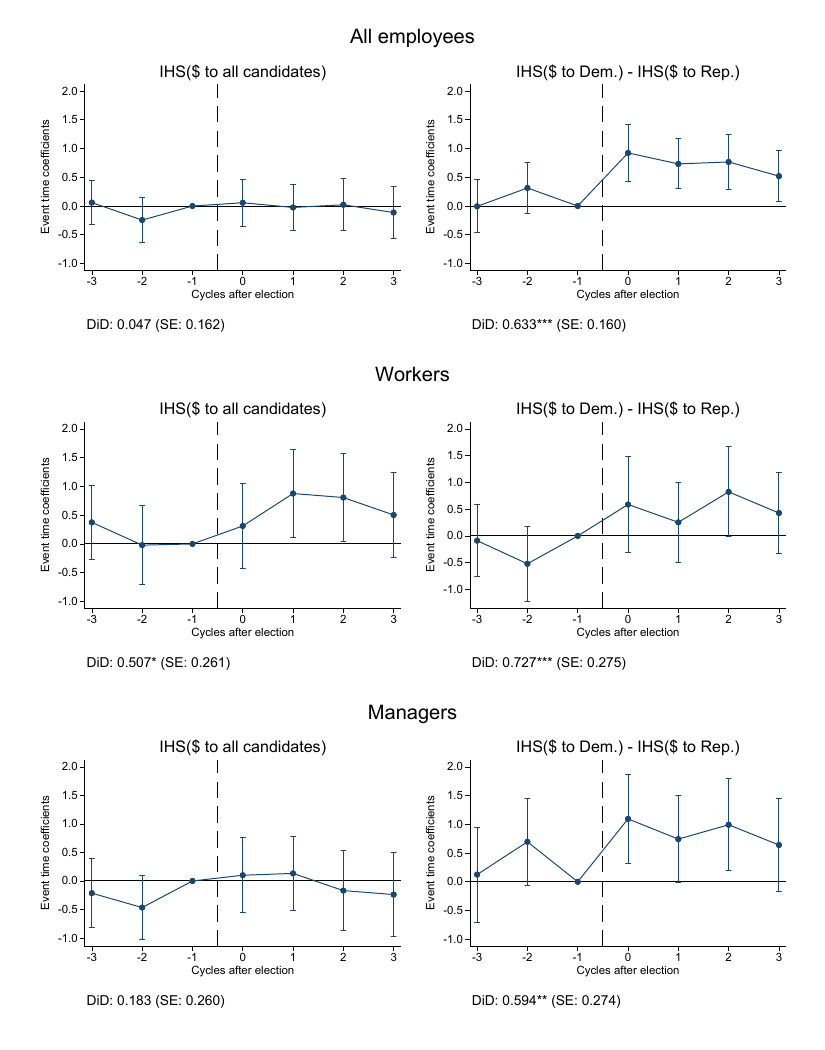}}
	\vspace{-1cm}
	\end{center}
	\begin{minipage}{\textwidth} {\scriptsize \smallskip 
 \textbf{Notes:} The figures report event-study coefficients $\delta_s$, estimated in model (\ref{event_study_individual}), for the individual-level effects of unionization. The sample includes incumbent employees who made at least one contribution to a PAC in the year before the union election that is matched to the establishment. The individual $\times$ cycle panel aggregates all the contributions of these donors three cycles before to three cycles after the union election, irrespective of whether or not they are matched to the initial establishment. In the regressions, each incumbent employee is weighted by the inverse of the total number of donating incumbent employees in the establishment. Below each graph the DiD coefficient from model (\ref{did_individual}) is reported. In the graphs on the left side, the outcome is the IHS-transformed total amount contributed to all candidates. In the graphs on the right side, the outcome is the difference between the IHS-transformed amounts contributed to Democratic and Republican candidates. Results are reported for contributions from all employees (top part), from only non-managerial workers (middle part), and from only managers and supervisors (lower part). The sample includes all establishments with a pro-union vote share between 20\% and 80\% and covers three election cycles (six years) before and after the union election. 95\% confidence intervals are depicted for standard errors clustered at the establishment level.\par}
    \end{minipage}
\end{figure}

\clearpage

\begin{table}[h]\centering
	\caption{Characteristics of Matched and Non-Matched Union Elections}
	\label{tab:nonmatched}
	\begin{tabular}{l*{2}{c}}
		\toprule[1.5pt]
		& Matched & Not matched\\
		\midrule
		Number of elections		& 6,063 	& 22,760  \\
		[1em]
		Union win (dummy)		& .4397	& .4405\\
		Union vote share		& .4950 & .4955 \\
		Number of votes  		& 119.37 & 81.92 \\
		Number of eligible voters & 139.27 & 94.01 \\
		Industry: mining		& .0397	& .0388 \\
		Industry: manufacturing & .3338	& .3731 \\
		Industry: transport		& .1785 & .1731 \\
		Industry: trade			& .1397 & .1251 \\
		Industry: finance		& .1008 & .0584 \\
		Industry: services		& .1834 & .2192 \\
		Years 1985-89			& .1618 & .2795 \\ 
		Years 1990-94			& .1908 & .2529 \\
		Years 1995-99			& .2319 & .2261 \\
		Years 2000-04			& .2547 & .1617 \\
		Years 2005-10			& .1608 & .0798 \\
		\hline\hline
		\multicolumn{3}{l}
		{
			\begin{minipage}{0.7\linewidth} \scriptsize \smallskip 
				\textbf{Notes:} The table reports mean characteristics of matched and non-matched union elections. Matched elections form our estimation sample and are defined as those for whom we were able to match at least one employee contribution in any of the seven election cycles around the union election (three before, cycle of union election, three after).  
			\end{minipage}
		}\\
	\end{tabular}
\end{table}

\clearpage

\begin{table}[h!]\centering
	\def\sym#1{\ifmmode^{#1}\else\(^{#1}\)\fi}
	\caption{Effect of Unionization on Any Contribution in Cycle}
	\label{tab:any_contribution}
		\begin{tabular}{l*{3}{c}}
			\toprule[1.5pt]
			&All&Workers&Managers\\	
			&(1) &(2)  &(3) \\	\vspace{-1em}
            & \hspace{8em} & \hspace{8em} & \hspace{8em} \\
            
			\midrule
			$\delta_{\textrm{DiD}}$    
                 &    -0.002         &     0.002         &    -0.003         \\
                &   (0.003)         &   (0.002)         &   (0.002)         \\
            [0.5em]
    		\hline\hline
			\multicolumn{4}{l}
			{
				\begin{minipage}{0.7\linewidth} \scriptsize \smallskip 
					\textbf{Notes:} The table presents DiD coefficients estimated in model (\ref{model1}). The sample includes all 28,823 establishments (159,026 establishment-cycle observations) without any restriction related to matched contributions. The outcome variable in column (1) is an indicator for observing any matched contribution from any employee in the establishment in a given cycle. We have used this indicator to restrict our baseline estimation sample to establishments with at least one matched contribution in any of the seven considered election cycles. Columns (2) and (3) also report effects on having any matched contribution from workers and managers. Standard errors clustered at the establishment level are in parentheses. \sym{*} \(p<0.10\), \sym{**} \(p<0.05\), \sym{***} \(p<0.01\)
				\end{minipage}
			}\\
		\end{tabular}
\end{table}

\vspace{5em}

\begin{table}[h!]\centering
	\caption{Descriptive Statistics for Contributions by Workers and Managers}
	\label{tab:sumstat_worker_manager}
	\begin{tabular}{lccc}
		\toprule[1.5pt]
		& All employees & Workers & Managers \\
        \midrule
		Amount (in million 2010 USD) & 105.82 & 13.31 & 56.84 \\
		Number of contributions & 357,436 & 71,782 & 214,593 \\
        Average amount per contribution (in 2010 USD) & 296.04 & 185.53 & 264.89\\
		Number of donors & 46,719 & 13,335 & 19,830 \\
		Number of recipients & 9,942 & 3,977 & 6,108 \\
		[0.5em]
		\hline\hline
		\multicolumn{4}{l}{
		\begin{minipage}{0.95\linewidth} \smallskip \scriptsize
		\textbf{Notes:} The table reports descriptive statistics on all matched employee contributions that were donated three cycles before to three cycles after the union election. The difference in the amounts from all employees and the total from workers and managers is driven by contributions for which we were unable to classify the occupation.   
		\end{minipage}} \\
	\end{tabular}
\end{table}

\begin{table}[h]\centering
    \def\sym#1{\ifmmode^{#1}\else\(^{#1}\)\fi}
    \caption{DiD Results}
    \label{tab:did_main_results}
    \begin{tabular}{lcccc}
        \toprule[1.5pt]
        &\multicolumn{4}{c}{IHS(\$ donated to ...)} \\ \cmidrule(lr){2-5}
        &All candidates&Dem.&Rep.&Dem. $-$ Rep.\\
        & (1) & (2) & (3) & (4)\\
        \midrule
        \multicolumn{5}{l}{[A] All employees} \\
        $\delta_{\textrm{DiD}}$     
        &      0.033         &      0.092         &      -0.147\sym{**} &       0.239\sym{***}\\
        &    (0.079)         &    (0.063)         &    (0.065)         &    (0.079)         \\
        [0.5em]
        \midrule
        \multicolumn{5}{l}{[B] Workers} \\
        $\delta_{\textrm{DiD}}$    
        &      0.026         &      0.073\sym{**} &     -0.050         &       0.123\sym{***}\\
        &    (0.043)         &    (0.035)         &    (0.032)         &    (0.040)         \\	
        [0.5em]
        \midrule
        \multicolumn{5}{l}{[C] Managers} \\
        $\delta_{\textrm{DiD}}$     
        &     -0.019         &      0.074         &      -0.130\sym{***}&       0.204\sym{***}\\
        &    (0.060)         &    (0.047)         &    (0.049)         &    (0.056)         \\			
        [0.5em]
        \hline\hline
			\multicolumn{5}{l}
			{
				\begin{minipage}{0.6\linewidth} \scriptsize \smallskip 
					\textbf{Notes:} The table reports DiD coefficients, estimated in model (\ref{model1}), for the effect of unionization on employees' campaign contributions. The outcome in column (1) is the IHS-transformed total amount contributed to all candidates. Columns (2) and (3) separately report IHS-transformed contributions to Democratic and Republican candidates, respectively, while column (4) reports the difference in IHS-transformed contributions to Democratic versus Republican candidates. Panel A includes contributions from all employees, Panel B focuses on non-managerial workers, and Panel C on managers and supervisors. The sample includes establishments with a pro-union vote share between 20 and 80\%. $N=33,103$ establishment-cycle observations. Standard errors clustered at the establishment level are in parentheses. \sym{*} \(p<0.10\), \sym{**} \(p<0.05\), \sym{***} \(p<0.01\)
				\end{minipage}
			}\\
    \end{tabular}
\end{table}

\begin{table}[h]\centering
	\def\sym#1{\ifmmode^{#1}\else\(^{#1}\)\fi}
	\caption{Extensive Margin Effects}
        \tabcolsep=0.4cm
	\label{tab:extensive}
		\begin{tabular}{l*{6}{c}}
			\toprule[1.5pt]
			&\multicolumn{3}{c}{1(\$ to all candidates $> 0$)}&\multicolumn{3}{c}{1(\$ to Dem. $> 0$) $-$ 1(\$ to Rep. $> 0$)} \\ \cmidrule(lr){2-4} \cmidrule(lr){5-7}
			&All&Workers&Managers&All&Workers&Managers \\	
			&(1)&(2)&(3)&(4)&(5)&(6) \\	
			\midrule
			$\delta_{\textrm{DiD}}$    
                &       0.009         &       0.003         &      -0.000         &       0.028\sym{***}&       0.017\sym{***}&       0.026\sym{***}\\
                &     (0.010)         &     (0.006)         &     (0.008)         &     (0.010)         &     (0.006)         &     (0.007)         \\
			[0.5em]
			\hline\hline
			\multicolumn{7}{l}
			{\footnotesize 
				\begin{minipage}{0.9\linewidth} \scriptsize \smallskip 
					\textbf{Notes:} The table reports DiD coefficients, estimated in model ((\ref{model1}), for the effect of unionization on the extensive margin of campaign contributions. In columns (1) - (3), the outcome variable is an indicator for observing any matched contribution from any employee in the establishment in a given cycle. In columns (4) - (6), the outcome variable is the difference between an indicator for any matched contribution to a Democratic candidate and an indicator for any matched contribution to a Republican candidate. The sample includes establishments with a pro-union vote share between 20 and 80\%. $N=33,103$ establishment-cycle observations. Standard errors clustered at the establishment level are in parentheses. \sym{*} \(p<0.10\), \sym{**} \(p<0.05\), \sym{***} \(p<0.01\)
				\end{minipage}
			}\\
		\end{tabular}
\end{table}

\begin{table}[h]\centering
	\def\sym#1{\ifmmode^{#1}\else\(^{#1}\)\fi}
	\caption{Valuing Extensive versus Intensive Margin}
	\label{tab:extensive_intensive}
		\begin{tabular}{l*{6}{c}}
			\toprule[1.5pt]
			&\multicolumn{3}{c}{m(\$ to all candidates)}&\multicolumn{3}{c}{m(\$ to Dem.) $-$ m(\$ to Rep.)} \\ \cmidrule(lr){2-4} \cmidrule(lr){5-7}
			&All&Workers&Managers&All&Workers&Managers \\	
			&(1)&(2)&(3)&(4)&(5)&(6) \\	
			\midrule
			\multicolumn{7}{l}{[A] Extensive-margin value $x = 0$} \\
			$\delta_{\textrm{DiD}}$    
			&       0.027         &       0.024         &      -0.019         &       0.220\sym{***}&       0.111\sym{***}&       0.186\sym{***}\\
                &     (0.073)         &     (0.039)         &     (0.054)         &     (0.072)         &     (0.036)         &     (0.051)         \\
			[0.5em]
			\midrule 
			\multicolumn{7}{l}{[B] Extensive-margin value $x = 0.1$} \\
			$\delta_{\textrm{DiD}}$    
			&       0.028         &       0.024         &      -0.019         &       0.222\sym{***}&       0.113\sym{***}&       0.189\sym{***}\\
                &     (0.074)         &     (0.040)         &     (0.055)         &     (0.073)         &     (0.036)         &     (0.052)         \\
			[0.5em]
			\midrule 
			\multicolumn{7}{l}{[C] Extensive-margin value $x = 0.7$} \\
			$\delta_{\textrm{DiD}}$    
			&       0.033         &       0.026         &      -0.019         &       0.239\sym{***}&       0.123\sym{***}&       0.204\sym{***}\\
                &     (0.079)         &     (0.043)         &     (0.060)         &     (0.079)         &     (0.040)         &     (0.056)         \\
			[0.5em]
			\midrule 
			\multicolumn{7}{l}{[D] Extensive-margin value $x = 1$} \\
			$\delta_{\textrm{DiD}}$    
			&       0.036         &       0.027         &      -0.019         &       0.248\sym{***}&       0.128\sym{***}&       0.211\sym{***}\\
                &     (0.082)         &     (0.045)         &     (0.062)         &     (0.082)         &     (0.041)         &     (0.058)         \\
			[0.5em]
			\midrule 
			\multicolumn{7}{l}{[E] Extensive-margin value $x = 3$} \\
			$\delta_{\textrm{DiD}}$    
			&       0.053         &       0.033         &      -0.020         &       0.304\sym{***}&       0.162\sym{***}&       0.263\sym{***}\\
                &     (0.102)         &     (0.057)         &     (0.077)         &     (0.103)         &     (0.052)         &     (0.073)         \\
			[0.5em]
			\hline\hline
			\multicolumn{7}{l}
			{\footnotesize 
				\begin{minipage}{0.73\linewidth} \scriptsize \smallskip 
					\textbf{Notes:} The table presents results under different valuations of extensive and intensive margin effects. Reported are the DiD coefficients estimated in model (\ref{model1}) for the effect of unionization on the total amount contributed (columns (1) - (3)) and on the difference between the amounts contributed to Democratic and Republican candidates (columns (4) - (6)). Amounts are transformed by the function that sets
                    \mbox{$m(\$) = \ln(\$) \, \text{if} \, \$ > 0$} and \mbox{$m(0) = -x$}. Panels A to E present results for varying the extensive-margin value $x$. The sample includes establishments with a pro-union vote share between 20 and 80\%. $N=33,103$ establishment-cycle observations. Standard errors clustered at the establishment level are in parentheses. \sym{*} \(p<0.10\), \sym{**} \(p<0.05\), \sym{***} \(p<0.01\)
				\end{minipage}
			}\\
		\end{tabular}
\end{table}

\begin{table}[h]\centering
	\def\sym#1{\ifmmode^{#1}\else\(^{#1}\)\fi}
	\caption{Robustness to Outcome Transformations, DiD Estimators, and Sample Restrictions}
	\label{tab:robust}
		\begin{tabular}{l*{6}{c}}
			\toprule[1.5pt]
			&\multicolumn{3}{c}{\$ to all candidates}&\multicolumn{3}{c}{\$ to Dem. $-$ \$ to Rep.} \\ \cmidrule(lr){2-4} \cmidrule(lr){5-7}
			&All&Workers&Managers&All&Workers&Managers \\	
			&(1)&(2)&(3)&(4)&(5)&(6) \\	
			\midrule
			\multicolumn{7}{l}{[A] Baseline} \\
			$\delta_{\textrm{DiD}}$    
			&      0.033         &      0.026         &     -0.019         &       0.239\sym{***}&       0.123\sym{***}&       0.204\sym{***}\\
			&    (0.079)         &    (0.043)         &    (0.059)         &    (0.079)         &    (0.040)         &    (0.056)         \\
			[0.5em]
			\midrule 
			\multicolumn{7}{l}{[B] Log(amount+1)} \\
			$\delta_{\textrm{DiD}}$    
            &       0.027         &       0.024         &      -0.019         &       0.220\sym{***}&       0.111\sym{***}&       0.186\sym{***}\\
            &     (0.073)         &     (0.039)         &     (0.054)         &     (0.072)         &     (0.036)         &     (0.051)         \\
			[0.5em]
			\midrule 
			\multicolumn{7}{l}{[C] Amount$^{1/4}$} \\
			$\delta_{\textrm{DiD}}$    
            &    -0.002         &      0.020         &     -0.032         &       0.219\sym{***}&      0.095\sym{***}&       0.178\sym{***}\\
            &    (0.068)         &    (0.035)         &    (0.051)         &    (0.064)         &    (0.030)         &    (0.046)         \\
			[0.5em]
			\midrule 
			\multicolumn{7}{l}{[D] Untransformed amounts} \\
			$\delta_{\textrm{DiD}}$    
			&      -27.618         &       2.414         &      -22.951         &       116.721\sym{***}&       15.581\sym{**} &       65.375\sym{***}\\
			&     (60.179)         &     (10.338)         &     (33.018)         &     (36.878)         &     (6.223)         &     (20.129)         \\
			[0.5em]
			\midrule  
			\multicolumn{7}{l}{[E] Borusyak, Jaravel, and Spiess (2021)} \\
			$\delta_{\textrm{DiD}}$    
			&      0.090         &      0.042         &     0.009         &       0.236\sym{***}&       0.130\sym{***}&       0.183\sym{***}\\
			&    (0.075)         &    (0.042)         &    (0.058)         &    (0.074)         &    (0.039)         &    (0.054)         \\
			[0.5em]
			\midrule 
			\multicolumn{7}{l}{[F] Callaway and Sant'Anna (2021)} \\
			$\delta_{\textrm{DiD}}$    
            &       0.015         &       0.042         &      -0.038         &       0.243\sym{***}&       0.137\sym{***}&       0.135\sym{**} \\
            &     (0.083)         &     (0.044)         &     (0.061)         &     (0.087)         &     (0.045)         &     (0.062)         \\
			[0.5em]
			\midrule 
            \multicolumn{7}{l}{[G] Sample restriction: at least one matched contribution in pre-period} \\
			$\delta_{\textrm{DiD}}$    
            &    -0.008         &      0.050         &      -0.132         &       0.311\sym{**} &       0.223\sym{***}&       0.292\sym{***}\\
            &     (0.128)         &    (0.077)         &     (0.103)         &     (0.134)         &    (0.070)         &     (0.100)  \\
			[0.5em]              
			\midrule 
			\multicolumn{7}{l}{[H] Sample restriction: period between first and last matched contribution} \\
			$\delta_{\textrm{DiD}}$    
            &      0.030         &       0.117         &      -0.126         &       0.469\sym{**} &       0.299\sym{***}&       0.432\sym{***}\\
            &     (0.156)         &     (0.115)         &     (0.142)         &     (0.206)         &     (0.115)         &     (0.161)         \\
			[0.5em]
			\hline\hline
			\multicolumn{7}{l}
			{
				\begin{minipage}{0.8\linewidth} \scriptsize \smallskip 
					\textbf{Notes:} The table presents robustness checks for our DiD estimates of the effect of unionization on the total amount contributed (columns (1) - (3)) and on the difference between the amounts contributed to Democratic and Republican candidates (columns (4) - (6)). Panel A shows the baseline results from the stacked DiD model (\ref{model1}) with IHS-transformed amounts. In Panels B and C, outcomes are transformed as log(amount+1) and amount$^{1/4}$, respectively, while in Panel D we use untransformed amounts after winsorizing at the top and bottom percentile. Panel E presents results from the imputation approach introduced by \cite{Borusyak2021}, and Panel F implements the DiD estimator of \cite{Callaway2021}, where we use both never-treated establishments (i.e., lost elections) and not-yet-treated establishments (i.e., won elections in later cycles) as comparison units. In Panels A to F, $N=33,103$ establishment-cycle observations. In Panel G, we constrain the sample to establishments for which we have matched at least one contribution in the three cycles before the union election ($N=15,792$). Panel H restricts the observation window for each establishment to the cycles between the first matched contribution before the union election and the last matched contribution after the union election ($N=17,911$). Standard errors clustered at the establishment level are in parentheses. \sym{*} \(p<0.10\), \sym{**} \(p<0.05\), \sym{***} \(p<0.01\)
				\end{minipage}
			}\\
		\end{tabular}
\end{table}

\begin{table}[h]\centering
	\def\sym#1{\ifmmode^{#1}\else\(^{#1}\)\fi}
	\caption{Robustness to Alternative Worker-Manager Classifications}
	\label{tab:robust_classification}
	\begin{tabular}{l*{4}{c}}
		\toprule[1.5pt]
		&\multicolumn{2}{c}{IHS(\$ to all candidates)}&\multicolumn{2}{c}{IHS(\$ to Dem.) $-$ IHS(\$ to Rep.)} \\ \cmidrule(lr){2-3} \cmidrule(lr){4-5}
		&Workers&Managers&Workers&Managers \\	
		&(1)&(2)&(3)&(4) \\	\vspace{-1em}
        & \hspace{7em} & \hspace{7em} & \hspace{7em} & \hspace{7em} \\
		\midrule
		\multicolumn{5}{l}{[A] 80\textsuperscript{th} percentile of supervisor tasks (baseline)} \\
		$\delta_{\textrm{DiD}}$    
		&      0.026         &     -0.019         &       0.123\sym{***}&       0.204\sym{***}\\
		&    (0.043)         &    (0.059)         &    (0.040)         &    (0.056)         \\
		[0.5em]
		\midrule 
		\multicolumn{5}{l}{[B] 90\textsuperscript{th} percentile of supervisor tasks} \\
		$\delta_{\textrm{DiD}}$    
		&      0.043         &     -0.041         &       0.140\sym{***}&       0.201\sym{***}\\
		&    (0.046)         &    (0.059)         &    (0.042)         &    (0.055)         \\
		[0.5em]	
		\midrule 
		\multicolumn{5}{l}{[C] Supervisor tasks ``very important" (4 out of 5 in ranking)} \\
		$\delta_{\textrm{DiD}}$    
		&      0.027         &     -0.022         &       0.131\sym{***}&       0.203\sym{***}\\
		&    (0.043)         &    (0.060)         &    (0.039)         &    (0.056)         \\
		[0.5em]
		\midrule
		\multicolumn{5}{l}{[D] Non-managerial supervisors as workers } \\
		$\delta_{\textrm{DiD}}$    
		&      0.040         &     -0.051         &       0.163\sym{***}&       0.183\sym{***}\\
		&    (0.048)         &    (0.057)         &    (0.045)         &    (0.053)         \\
		[0.5em]
		\midrule
		\multicolumn{5}{l}{[E] 80\textsuperscript{th} percentile of census-tract median income} \\
		$\delta_{\textrm{DiD}}$    
		&       0.059         &      -0.055         &       0.156\sym{**} &       0.216\sym{***}\\
            &     (0.066)         &     (0.065)         &     (0.064)         &     (0.062)         \\
		[0.5em]
		\midrule
		\multicolumn{5}{l}{[F] 90\textsuperscript{th} percentile of census-tract median income} \\
		$\delta_{\textrm{DiD}}$    
		&       0.031         &      -0.018         &       0.243\sym{***}&       0.100\sym{*}  \\
            &     (0.071)         &     (0.057)         &     (0.070)         &     (0.054)         \\
		[0.5em]
		\hline\hline
		\multicolumn{5}{l}
		{
			\begin{minipage}{0.83\linewidth} \scriptsize \smallskip 
				\textbf{Notes:} The table presents robustness checks for alternative worker-manager classifications. Reported are the DiD coefficients estimated in model (\ref{model1}) for the effect of unionization on the IHS-transformed total amount contributed (columns (1) and (2)) and on the difference between the IHS-transformed amounts contributed to Democratic and Republican candidates (columns (3) and (4)). $N=33,103$ establishment-cycle observations. Panel A shows the baseline results in which ``managers" are defined as donors in ``Management occupations" (SOC group 11) or in occupations above the 80\textsuperscript{th} percentile of supervisor tasks and independent judgment. ``Workers" are all remaining donors with a classified occupation. In Panel B, we increase the cutoff for supervisor tasks and independent judgment to the 90\textsuperscript{th} percentile. Panel C, instead, uses an absolute cutoff for the importance of supervisor tasks and independent judgment (both need to be ``very important", i.e., have a score of 4 or above in the 5-score ranking). In Panel D, we only consider ``Management occupations" (SOC group 11) as ``managers" and treat all other classified occupations as ``workers" (including those with high importance in supervisor tasks and independent judgment). In Panels E and F, we define ``managers'' (``workers'') as individuals residing in a census tract with a median income above (below) the 80\textsuperscript{th} or 90\textsuperscript{th} percentile of the state-specific distribution of census-tract median incomes. See Appendix \ref{sec:app_occclassify} for more details on the classifications. Standard errors clustered at the establishment level are in parentheses. \sym{*} \(p<0.10\), \sym{**} \(p<0.05\), \sym{***} \(p<0.01\)
			\end{minipage}
		}\\
	\end{tabular}
\end{table}

\begin{table}[h]\centering
	\def\sym#1{\ifmmode^{#1}\else\(^{#1}\)\fi}
	\caption{Federal versus Local Candidates}
	\label{tab:federal_local}
		\begin{tabular}{l*{6}{c}}
			\toprule[1.5pt]
			&\multicolumn{3}{c}{IHS(\$ to all candidates)}&\multicolumn{3}{c}{IHS(\$ to Dem.) $-$ IHS(\$ to Rep.)} \\ \cmidrule(lr){2-4} \cmidrule(lr){5-7}
			&All&Workers&Managers&All&Workers&Managers \\	
			&(1)&(2)&(3)&(4)&(5)&(6) \\	
			\midrule
			\multicolumn{7}{l}{[A] All candidates (baseline)} \\
			$\delta_{\textrm{DiD}}$    
			&      0.033         &      0.026         &     -0.019         &       0.239\sym{***}&       0.123\sym{***}&       0.204\sym{***}\\
			&    (0.079)         &    (0.043)         &    (0.059)         &    (0.079)         &    (0.040)         &    (0.056)         \\
			[0.5em]
			\midrule 
			\multicolumn{7}{l}{[B] Federal candidates} \\
			$\delta_{\textrm{DiD}}$    
			&      0.048         &      0.026         &     -0.018         &       0.207\sym{***}&      0.098\sym{***}&       0.182\sym{***}\\
			&    (0.075)         &    (0.039)         &    (0.053)         &    (0.076)         &    (0.036)         &    (0.052)         \\
			[0.5em]
			\midrule 
			\multicolumn{7}{l}{[C] Local candidates} \\
			$\delta_{\textrm{DiD}}$    
			&     -0.047         &      0.024         &     -0.034         &       0.158\sym{***}&      0.045\sym{*}  &       0.130\sym{***}\\
			&    (0.050)         &    (0.028)         &    (0.043)         &    (0.044)         &    (0.024)         &    (0.038)         \\
			[0.5em]
			\hline\hline
			\multicolumn{7}{l}
			{
				\begin{minipage}{0.76\linewidth} \scriptsize \smallskip 
					\textbf{Notes:} The table presents estimates for the effect of unionization on contributions to federal and local candidates. Reported are the DiD coefficients estimated in model (\ref{model1}) for the effect of unionization on the IHS-transformed total amount contributed (columns (1) - (3)) and on the difference between the IHS-transformed amounts contributed to Democratic and Republican candidates (columns (4) - (6)). $N=33,103$ establishment-cycle observations. Standard errors clustered at the establishment level are in parentheses. \sym{*} \(p<0.10\), \sym{**} \(p<0.05\), \sym{***} \(p<0.01\)
				\end{minipage}
			}\\
		\end{tabular}
\end{table}

\clearpage

\begin{table}[h]\centering
	\def\sym#1{\ifmmode^{#1}\else\(^{#1}\)\fi}
	\caption{Alternative Sample Restrictions for Individual-level Analysis}
        \tabcolsep=0.4cm
	\label{tab:robust_individual}
		\begin{tabular}{l*{6}{c}}
			\toprule[1.5pt]
			&\multicolumn{3}{c}{IHS(\$ to all candidates)}&\multicolumn{3}{c}{IHS(\$ to Dem.) $-$ IHS(\$ to Rep.)} \\ \cmidrule(lr){2-4} \cmidrule(lr){5-7}
			&All&Workers&Managers&All&Workers&Managers \\	
			&(1)&(2)&(3)&(4)&(5)&(6) \\	
			\midrule
			\multicolumn{7}{l}{[A] At least one matched PAC donation one year before union election (baseline)} \\  
			$\delta_{\textrm{DiD}}$    
			  &       0.047         &       0.507\sym{*}  &       0.183         &       0.633\sym{***}&       0.727\sym{***}&       0.594\sym{**} \\
                &     (0.162)         &     (0.261)         &     (0.260)         &     (0.160)         &     (0.275)         &     (0.274)         \\
			[0.5em]
                N           &      22,799         &       5,243         &      13,104         &      22,799         &       5,243         &      13,104         \\
			\midrule 
			\multicolumn{7}{l}{[B] At least one matched PAC donation two years before union election} \\
			$\delta_{\textrm{DiD}}$    
			  &      -0.041         &       0.461\sym{*}  &       0.035         &       0.437\sym{***}&       0.652\sym{***}&       0.320         \\
                &     (0.137)         &     (0.235)         &     (0.238)         &     (0.133)         &     (0.231)         &     (0.245)         \\
			[0.5em]
                N          &      30,849         &       7,007         &      17,038         &      30,849         &       7,007         &      17,038         \\
			\midrule 
			\multicolumn{7}{l}{[C] At least one matched PAC or candidate donation one year before union election} \\
			$\delta_{\textrm{DiD}}$    
			&       0.196         &       0.128         &       0.183         &       0.317\sym{**} &       0.319         &       0.494\sym{**} \\
                &     (0.135)         &     (0.248)         &     (0.200)         &     (0.135)         &     (0.258)         &     (0.211)         \\
			[0.5em]
                N          &      36,463         &       8,687         &      19,250         &      36,463         &       8,687         &      19,250         \\
			\midrule 
			\multicolumn{7}{l}{[D] At least one matched PAC or candidate donation two years before union election} \\
			$\delta_{\textrm{DiD}}$    
			  &       0.143         &       0.176         &       0.022         &       0.231\sym{**} &       0.473\sym{**} &       0.234         \\
                &     (0.118)         &     (0.216)         &     (0.178)         &     (0.113)         &     (0.220)         &     (0.178)         \\
			[0.5em]
                N         &      52,339         &      12,453         &      25,935         &      52,339         &      12,453         &      25,935         \\
			\hline\hline
			\multicolumn{7}{l}
			{\footnotesize 
				\begin{minipage}{0.9\linewidth} \scriptsize \smallskip 
					\textbf{Notes:} The table presents alternative sample restrictions for the individual-level effects of unionization on the IHS-transformed total amount contributed (columns (1) - (3)) and on the difference between the IHS-transformed amounts contributed to Democratic and Republican candidates (columns (4) - (6)). Panel A shows the baseline results, for which we consider donors who have at least one PAC contribution in the year before the union election that is matched to the establishment. In Panel B, we increase the sample to donors with at least one matched PAC contribution in the two years before the union election. In Panel C [D], the sample includes donors who have at least one matched donation to a PAC or to a candidate in the year [two years] before the union election. In all panels, we estimate an individual-level version of model (\ref{model1}) with individual and cohort $\times$ event time fixed effects and where each individual is weighted by the inverse of the total number of donating individuals in the establishment. All samples include establishments with a pro-union vote share between 20 and 80\%. Standard errors clustered at the establishment level are in parentheses. \sym{*} \(p<0.10\), \sym{**} \(p<0.05\), \sym{***} \(p<0.01\)
				\end{minipage}
			}\\
		\end{tabular}
\end{table}

\begin{table}[htbp]\centering
        \vspace*{-0.5cm}
	\caption{Contributions of Union Organizations}
	\label{tab:union_ideo}
	\small{
	\begin{tabular}{lcc}
		\toprule[1.5pt]
		Union organization&\# of&\% of contr. \\
		&elections&to Dem.\\
		\midrule
Teamsters Union&1605&91.0 \\
United Steelworkers&481&98.0 \\
United Food \& Commercial Workers Union&434&97.7 \\
Service Employees International Union&407&93.6 \\
International Brotherhood of Electrical Workers&320&94.4 \\
United Auto Workers&249&98.0 \\
Machinists/Aerospace Workers Union&217&98.5 \\
Operating Engineers Union&208&86.5 \\
Communications Workers of America&170&95.8 \\
UNITE HERE&136&94.0\\
Laborers Union&119&93.3 \\
Carpenters \& Joiners Union&110&89.6\\
American Federation of State/Cnty/Munic Employees&91&79.9 \\
Office and Professional Employees International Union&51&99.3 \\
Plumbers/Pipefitters Union&51&91.8 \\
Amalgamated Transit Union&50&92.8 \\
National Union of Hospital and Health Care Employees&47&96.7 \\
Security, Police and Fire Professionals of America&44&100.0 \\
International Longshore/Warehouse Union&43&94.2 \\
Bakery, Confectionery, Tobacco \& Grain Union&40&99.6 \\
International Alliance Theatrical Stage Employees&40&95.0\\
American Nurses Association&38&83.7 \\
Sheet Metal, Air, Rail \& Transportation Union&35&92.9\\
United Mine Workers&33&92.2\\
Utility Workers Union of America&33&96.8 \\
Transport Workers Union&27&94.1 \\
Bridge, Structural, Ornamental and Reinforcing Iron Workers&26&94.0 \\
Boilermakers Union&25&94.6 \\
Painters \& Allied Trades Union&25&89.1 \\
United Electrical, Radio and Machine Workers of America&22&100.0\\
American Federation of Teachers&19&96.3\\
Glass, Molders, Pottery, Plastics and Allied Workers&18&99.2 \\
International Union of Journeymen and Allied Trades&18&96.8 \\
Operative Plasterers and Cement Masons&17&91.2 \\
Seafarers International Union&16&71.2 \\
National Nurses United&15&98.3 \\
Roofers Union&14&92.7\\
International Guards Union of America&13&82.9 \\
American Federation of Government Employees&12&95.9 \\
SAG-AFTRA&9&100.0\\
American Postal Workers Union&9&96.5 \\
Marine Engineers Beneficial Assn&7&85.2 \\
International Association of Firefighters&6&84.2 \\
American Federation of Musicians&6&91.8\\
Bricklayers Union&5&95.7\\
Insulators Union&4&94.3 \\
Intl Fedn of Prof \& Technical Engineers&2&87.4 \\
International Longshoremens Assn&1&91.5 \\
National Education Assn&1&86.3 \\
		\midrule
		Total & 5,369 & 93.5  \\
		\hline\hline
		\multicolumn{3}{l}
		{
			\begin{minipage}{0.95\linewidth} \scriptsize \smallskip 
				\textbf{Notes:} The table reports the party composition of campaign contributions donated by union organizations in our sample of union elections. We consider all contributions from PACs associated with a union, including local union branches. For 685 out of the 6,063 elections in our estimation sample, we are not able to match any PAC contribution. Totals in the last row give the weighted average over all union organizations, where the weights are the number of elections in our sample.    
			\end{minipage}
		}\\
	\end{tabular}
	}
\end{table}

\begin{table}[h]\centering
	\caption{Contributions from Firm PACs}
	\label{tab:firm_pac}
		\begin{tabular}{l*{4}{c}}
			\toprule[1.5pt]
			&\multicolumn{2}{c}{Main sample} &\multicolumn{2}{c}{Ever-contributed sample} \\ \cmidrule(lr){2-3} \cmidrule(lr){4-5}
			&IHS(\$ to all &IHS(\$ to Dem.) &IHS(\$ to all &IHS(\$ to Dem.) \\
                & candidates) & $-$ IHS(\$ to Rep.) & candidates) & $-$ IHS(\$ to Rep.) \\
			& (1) & (2) & (3) & (4) \\
			\midrule
			$\delta_{\textrm{DiD}}$     
            &     -0.0935         &    -0.00756         &      0.0207         &     -0.0300         \\
            &    (0.0626)         &    (0.0407)         &     (0.192)         &     (0.138)         \\
			[0.5em]
			N & 33,103 			  & 33,103		 &    9,947             &       9,947 \\
			\hline\hline
			\multicolumn{5}{l}
			{
				\begin{minipage}{0.77\linewidth} \scriptsize \smallskip 
					\textbf{Notes:} The table reports DiD coefficients, estimated in model (\ref{model1}), for the effect of unionization on contributions from firm PACs. In columns (1) and (2), we consider all establishments in our main estimation sample and assign zero contribution amounts to all unmatched observations. In columns (3) and (4), we restrict the sample to establishments for which we observe at least one firm PAC contribution over our observation period. All samples include establishments with a pro-union vote share between 20 and 80\%. Standard errors clustered at the establishment level are in parentheses. \sym{*} \(p<0.10\), \sym{**} \(p<0.05\), \sym{***} \(p<0.01\)
				\end{minipage}
			}\\
		\end{tabular}
\end{table}

\clearpage
\renewcommand{\thesection}{B}
\renewcommand{\thetable}{B.\arabic{table}}
\setcounter{table}{0}
\renewcommand{\thefigure}{B.\arabic{figure}}
\setcounter{figure}{0}

\section{Data Appendix}

\subsection{Union Election Data}  \label{sec:app_dataunion}

\noindent \textbf{Data sources.}\quad We start by accessing data on NLRB union representation elections between 1961 and 2009 from the replication package of \cite{Knepper2020}. The data were originally compiled by \cite{Farber2016}. Then, we add data on elections between 2010 and 2018 from NLRB election reports available \href{https://www.nlrb.gov/reports/agency-performance/election-reports}{here}. Together, our data cover the universe of union elections between 1961 and 2018 and includes information on vote counts, voting outcome, petition filing and election date, establishment name, address, and industry, as well as the name of the union organization. 

\vspace{1em} \noindent \textbf{Sample restrictions.}\quad Before matching campaign contributions, we impose the following restrictions on the sample of union elections:
\begin{compactitem}
	\item We only consider elections where a union seeks to be certified and drop elections that stem from petitions of either employers or employees seeking to remove an existing union.
	\item We delete duplicate entries (multiple records of the same election). 
	\item For multiple entries that reflect elections where more than one union was on the ballot or where different worker groups formed different bargaining units, we follow \cite{Frandsen2020} and retain only the entry with the largest union vote share. 
	\item We further drop a few elections where the voting outcome (won or lost) is not consistent with the vote counts.
	\item Following the RDD literature on union elections, we restrict the sample to union elections where at least 20 votes were cast. 
	\item We only keep the first union election in each establishment. For this, we identify an establishment as a unique address or a unique combination of the standardized firm name and commuting zone. For a firm that has multiple establishments within the same commuting zone, we thus only consider the first election among these establishments. 
	\item Finally, we only use elections held between 1985 and 2010 to be able to observe employee contributions for three election cycles before and after each union election. 
\end{compactitem}
After these restrictions, we are left with 28,823 union elections. 

\clearpage
\subsection{Matching of Union Elections and Campaign Contributions}  \label{sec:app_matching}

We link the campaign contributions from employees to union elections in their employing establishment by combining a spatial match with a fuzzy match of firm names.

\vspace{1em} \noindent \textbf{Geocode commuting zones.}\quad In preparation for the spatial match, we first geocode all union election establishments based on their city and state (using the Open Street Map and Google Maps APIs) and assign 1990 commuting zones. For the employees' campaign contributions, we rely on donor addresses geocoded by \cite{Bonica2019} up to 2016.\footnote{\cite{Bonica2019} contains campaign contributions until 2018 but geocodes are only provided until 2016.} We use these geocodes to match to them 1990 commuting zones.

\vspace{1em} \noindent \textbf{Firm name cleaning.}\quad Firm names in both the union election and the contribution data are cleaned and harmonized using the \texttt{stnd\_compname} Stata command developed by \cite{Wasi2015}. The algorithm removes non-standard characters and whitespaces, doing-as-business and FKA names, as well as business entity types (e.g., CORP, INC, LLC). Moreover, it abbreviates common strings in firm names (e.g., Manufacturing $\rightarrow$ MFG, Professional $\rightarrow$ PROF). 

\begin{figure}[htbp]
	\begin{center}	
		\caption[gbadolite]{\label{fig:match_example} Example of Spatial Matching Procedure}
		\centering
		\mbox{\includegraphics[width=0.8\linewidth]{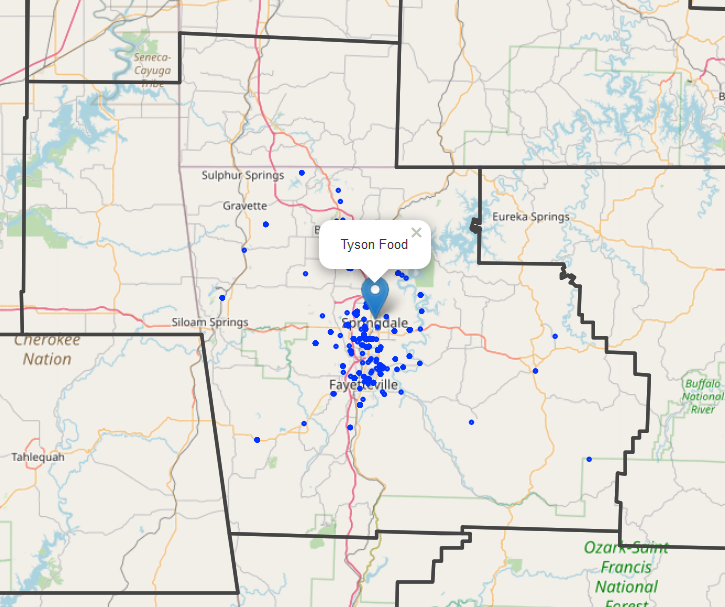}}
	\end{center}
	\vspace{-0.5em} 
	\begin{minipage}{\textwidth} {\footnotesize \textbf{Notes:} The map shows the location of the establishment ``Tyson Foods" in Springdale (Arkansas), which held a union election on 22/06/2006. Blue dots represent the location of all campaign contributions matched to the establishment. Black lines are 1990 commuting zone borders.\par}
    \end{minipage}  
\end{figure}

\vspace{1em} \noindent \textbf{Linkage algorithm.}\quad For each commuting zone, we create lists of all cleaned firm names from the union election and the contribution data. Then, we use the \texttt{reclink2} Stata command from \cite{Wasi2015} to compare the string similarity of firm names.\footnote{\texttt{reclink2} builds on \texttt{reclink} written by \cite{Blasnik2010}.} For each possible pair of firm names within the commuting zone, the command computes modified bigram scores. We keep potential matches with a score of at least .98 and manually review all of them. We identify roughly 70\% of them as correct matches.\footnote{The share of matches identified as correct is strongly increasing in the bigram score. For scores between .995 and 1, we keep 90\% of the potential matches, while for scores between .98 and .985 this share is only 34\%. We also tried keeping potential matches with a lower score (.95), but a manual review of a subsample of those revealed that a very low share of them represented correct matches.} In our review, we generally took a conservative approach and were more tolerant of possibly rejecting a true match than retaining an incorrect match. This means that we measure a lower bound for the sum of contributions from all employees of an establishment. To demonstrate the spatial dimension of the matching procedure, Figure \ref{fig:match_example} shows an example for the location of a union election establishment and all campaign contributions matched to it.

\vspace{1em} \noindent \textbf{GPT vs. manual linkage comparison.}\quad The rapid advancements in large language models (LLMs) have made alternative linkage procedures feasible that were unavailable when we initiated this project. Specifically, the low cost of going through a large number of potential matches possibly changes the cost-benefit calculation that is considered when setting the reclink2 threshold above which we review potential matches. To test the benefits of lowering the threshold and employing an LLM-based approach, we draw a random sample of 100 union elections, compute modified bigram scores again between all potential combinations of firm names in the union election and contribution data within the same commuting zone, and keep all potential matches with a score of at least 0.5. This yields 10,954 firm name pairs. We send these pairs to OpenAI's API and instruct it to identify correct matches using the GPT-4o model. To establish the ground truth, we also manually review all 10,954 potential matches. Finally, we merge our original classification with the test sample for comparison. The LLM-based approach fails to improve the accuracy of our linkage procedure. The original procedure yields 99.2\% agreement with the ground truth, while the LLM-based procedure yields 99.1\%.\footnote{To assess the quality of the linkage procedure, we also calculate Cohen's Kappa \citep{cohen1960coefficient}, which takes on the value 0 when the agreement amount is expected by chance and 1 when there is perfect agreement. Both procedures achieve a Kappa score of 0.9, indicating an ``almost perfect'' level of agreement with the ground truth, as defined by \citet{landis1977one, landis1977measurement}.} Although the LLM classifies the vast majority of cases correctly, it misclassifies some matches as correct, which outweighs its benefit in detecting rare matches with a bigram score below 0.98. 

\vspace{1em} \noindent \textbf{Establishment-level aggregation.}\quad As a last step, we use all matched contributions and sum them up at the establishment-election cycle level. Our period of analysis covers three cycles before to three cycles after each union election, i.e., we observe each establishment over a period of seven cycles (14 years). While we generally keep establishment-cycle observations without any matched contribution and code them as zero, we retain only establishments for which we observe at least one matched contribution over the 14-year period. Out of the initial 28,823 union election establishments, we thereby keep 6,063 matched establishments which form our final estimation sample. Table \ref{tab:nonmatched} compares the characteristics of matched and non-matched establishments. 

\bigskip
\vspace{-1em}

\clearpage
\subsection{Occupation Classification}  \label{sec:app_occclassify}

\noindent \textbf{NRLA definitions.}\quad We rely on the definition of the National Labor Relations Act (NLRA) to differentiate between employees eligible for unionization and employees banned from unionizing. The NLRA passed by Congress in 1935 sets rules for the unionization of private sector employees. It establishes who can and who cannot join a union. Section 7 describes the right of employees to join a union: 
\begin{quote}
	``Employees shall have the right to self-organization, to form, join, or assist labor organizations, to bargain collectively through representatives of their own choosing [...] and shall also have the right to refrain from any or all of such activities [...]." (29 U.S.C. § 157) 
\end{quote}

The NRLA explicitly restricts the right to unionize to employees. It does not extend it to individuals with management and supervisory responsibilities, as they are part of the company's management: The term `employee' ``shall include any employee [...] but shall not include any individual [...] employed as a supervisor" (29 U.S.C. § 152(3)). The distinction between supervisors and employees, however, is not clear-cut, and the NLRA goes on to define supervisors as follows: 
\begin{quote}
	``The term `supervisor' means any individual having authority, in the interest of the employer, to hire, transfer, suspend, lay off, recall, promote, discharge, assign, reward, or discipline other employees, or responsibly to direct them, or to adjust their grievances, or effectively to recommend such action, if in connection with the foregoing the exercise of such authority is not of a merely routine or clerical nature, but requires the use of independent judgment." (29 U.S.C. § 152(11))
\end{quote}
To differentiate between the labor force eligible for unionization and the company's management, we follow two steps: First, we harmonize occupations, and second, we calculate the supervisory element of each occupation based on the NLRA definition.

\vspace{1em} \noindent \textbf{Occupation harmonization.}\quad The free-text occupations reported in DIME are not standardized. Thus, we map them to the 6-digit Standard Occupation Classification. For this, we combine an ensemble classifier called SOCcer \citep{Russ2016}, fuzzy string matching to an extensive crosswalk of laymen's occupation titles from O*NET, as well as manual reviews from \cite{Dreher2020} and manual reviews of the most common occupation titles. In particular, we implement the following steps to identify good matches between a free-text occupation and a SOC code. First, we keep a match determined by SOCcer if the score of the first best match is higher than 0.3 and the difference to the second best match is larger than 0.1. Secondly, we search for exact matches of any substring of the free-text occupations and a list of laymen's occupation titles, abbreviations and reported titles by experts obtained from O*NET. Thirdly, we fuzzy match the lists from O*NET with the free-text occupations and keep matches with a score above 0.99. Fourthly, we add matches from \cite{Dreher2020}, which are based on a manual review. Finally, we manually review the free-text occupations that appear more than 50 times in our database of candidate contributions. With that procedure, we are able to assign a SOC code to 72\% of all candidate contributions in our matched sample. 

Since the share of non-classified occupations is not negligible, we seek to understand whether non-classification can impact our results on the effects of unionization. For this, we use the contribution-level dataset and estimate our baseline model (\ref{model1}) with an indicator for missing occupation classification as the dependent variable. The model yields an insignificant DiD coefficient of .0058 (p-value = 0.76). Thus, the likelihood of occupation non-classification does not appear to be related to unionization.

\vspace{1em} \noindent \textbf{Manager/supervisor versus worker classification.}\quad We follow the NLRA and classify an individual as a supervisor if \textit{independent judgment} and a \textit{supervisor task} are important for her occupation. In order to identify occupations with these characteristics, we merge the Occupational Information Network database (O*NET, version 26.3) containing task- and skill-content of 6-digit SOC occupations to our DIME occupations. The information in O*NET is supported by the U.S. Department of Labor and based on surveys of workers working in the respective occupation. Only the importance of specific skills and abilities for an occupation is determined by occupational analysts. We select six variables that closely resemble at least one work activity of a supervisor as defined in the NLRA to identify occupations with \textit{supervisor tasks}. The variables are listed in Table \ref{tab:supervisor_tasks} and measure the importance of the activity in each occupation. We classify an occupation as containing \textit{supervisor tasks} if the importance of at least one listed task is equal or above the 80\textsuperscript{th} percentile of all 6-digit SOC occupations.\footnote{In our robustness checks, we also use the 90\textsuperscript{th} percentile as cutoff and an absolute scale classifying any occupation as supervisor where a supervisor task is at least ``very important'' (a score of 4 or above in the 5-score ranking).} We then go on to evaluate whether the occupation requires \textit{independent judgment}, the second condition that we identify in the NLRA definition of a supervisor. We evaluate whether an occupation requires \textit{independent judgment} based on the following four variables: Independence (Work Styles), Leadership (Work Styles), Structured versus Unstructured Work (Work Context), and Freedom to Make Decisions (Work Context). Again, we classify an occupation as requiring \textit{independent judgment} if the importance of at least one of the listed variables is equal or above the 80\textsuperscript{th} percentile.\footnote{Again, in our robustness checks we also use the 90\textsuperscript{th} percentile as the cutoff and an absolute scale classifying any occupation as supervisor where independence is at least ``very important'' (a score of 4 or above in the 5-score ranking).} Finally, we classify individuals as managers or supervisors if their occupation is classified as ``Management Occupation'' in SOC (SOC group 11) or contains a \textit{supervisor task} \underline{and} \textit{independent judgment} as defined above.\footnote{We were not able to assign a 6-digit SOC code for some of the individuals in our data in cases where the free-text occupation was vague. Instead, we assigned 4-, 3- or 2-digit SOC codes. We classify a 2-digit SOC code occupation as supervisor if all 6-digit SOC code occupations have been classified as supervisors. We proceed accordingly for 3- and 4-digit SOC code occupations. We are thereby conservative and allow for some attenuation bias if supervisors are consequently incorrectly coded as workers.} Examples of occupations in the top 95\textsuperscript{th} percentile of both the \textit{independent judgment} and \textit{supervisor task} score are \textit{Chief Executives}, \textit{Human Resource Managers} and \textit{First-Line Supervisors of Retail Sales Workers}. Non-managerial workers are then identified as all remaining donors to whom we were able to assign a SOC code. With these definitions, we obtain the following occupational composition in our sample of candidate contributions: 42\% of contributions originate from managers and supervisors, 30\% from non-managerial workers, and for 28\% we are unable to obtain a classification.

\bigskip
\begin{table}[h]\centering
	\def\sym#1{\ifmmode^{#1}\else\(^{#1}\)\fi}
	\caption{Supervisor Tasks in NLRA and O*NET Occupations}
	\label{tab:supervisor_tasks}
		\resizebox{\textwidth}{!}{
	\begin{tabular}{ll}
		\toprule[1.5pt]
		\multicolumn{1}{l}{Tasks of a \textit{supervisor} defined in NLRA} & \multicolumn{1}{l}{Corresponding O*NET work activity / skill / context} \\
		\midrule
		Hire / transfer / suspend / lay off / discharge & Staffing organizational units \\
		[1em]
		Recall / assign									& Management of personnel resources \\
														& Coordinating the work and activities of others \\
		[1em]
		Promote / reward / discipline					& Guiding, directing, and motivating subordinates \\
														& Resolving conflicts and negotiating with others \\
		[1em]
		Direct employees / adjust their grievances 		& Management of personnel resources \\
														& Guiding, directing, and motivating subordinates \\
														& Coordinating the Work and Activities of Others \\
														& Coordinate or Lead Others \\
		\hline\hline												
	\end{tabular}
		}
\end{table}

\vspace{1em} \noindent \textbf{Alternative residence-based classification.}\quad Given the share of occupations that cannot be classified, as a robustness check we draw on an alternative classification that uses information on donors' exact location of residence. \cite{Bonica2019} geocodes donor addresses in the DIME data, which we link to median household income at the census tract level using data from the 1980, 1990, and 2000 Censuses, as well as the 5-year American Community Survey samples from 2010 onward.. With that, we identify managers (workers) as donors who live in a census tract that has a median income above (below) the 80\textsuperscript{th} or 90\textsuperscript{th} percentile of the state-specific distribution of census-tract median incomes.

\clearpage
\subsection{Measures of Labor Relations at the Establishment}  \label{sec:app_otherdata}

\noindent \textbf{Unfair labor practice charges.}\quad
We collect data on unfair labor practice (ULP) charges reported to the NLRB from the CHIPS archive for 1984-2000 (compiled by Forest Gregg and available \href{https://labordata.bunkum.us/chips-4185bba}{here}), from the CATS archive for 1999-2011 (available \href{https://catalog.archives.gov/id/6341122}{here}), and from the NLRB website for 2007-2020 (available \href{https://www.nlrb.gov/search/case?f[0]=case_type:C}{here}). The charges include complaints against employers for restraining employees in their rights to organize in a union or collectively bargain (NLRA § 8(a)(1)), dominating or controlling a union (§ 8(a)(2)), discharging or otherwise discriminating workers involved in organizing (§§ 8(a)(3) and 8(a)(4)), and failing to bargain in good faith with the union (§ 8(a)(5)). To focus on ULP committed by employers in the process of the organizing drive, we limit the data to charges related to violations of §§ 8(a)(1), 8(a)(3), and 8(a)(4).

The data contains the name, state, city, and address of the employer. Firm name and address are cleaned and harmonized using the Stata commands \texttt{stnd\_compname} and \texttt{stnd\_address}, respectively \citep{Wasi2015}. Then, we match a ULP charge to a union election in our estimation sample if one of the following conditions is fulfilled.
\begin{itemize}
    \item Exact match of state, firm name, and city 
    \item Exact match of state, firm name, and address
    \item Exact match of state, city, and address
    \item Exact match of state and match score above 0.9 from fuzzy merge on firm name, city, and address using \texttt{reclink2} (if address information is missing in both datasets or not missing in both datasets)
    \item Exact match of state and match score above 0.8 from fuzzy merge on firm name, city, and address using \texttt{reclink2} (if address information is missing in one of both datasets)
\end{itemize}

Since charges must be filed within six months of an alleged violation, we identify all ULP charges that were filed between six months before to six months after a union election. With that, we find that 44\% of all union elections in our sample involve a ULP charge, similar to the shares reported by \cite{Bronfenbrenner2009} and \cite{McNicholas2019}.

\vspace{1em} \noindent \textbf{Collective bargaining contracts.}\quad We combine bargaining contract data compiled by \cite{Holmes2006} for 1985-2003 (available \href{https://users.econ.umn.edu/~holmes/data/geo_spill/}{here}) and by \cite{Gregg2024} for 2004-2020 (available \href{https://labordata.bunkum.us/f7-3b56a21}{here}). The data contains notices about both initial contracts, i.e. first-time negotiations after an election, and about the renegotiation or reopening of existing contracts. It is reported to the Federal Mediation and Conciliation Service (FMCS) for it to be able to prepare mediation services for the negotiation of contracts. 

In the data, we observe the name, state, city, and street of the reporting establishment. We clean this information and match contract notices to establishments in our union election estimation sample using the same algorithm as above for the matching of ULP charges. We find a contract expiration notice in the five years following the union election for 48\% of the won union elections in our sample, which is comparable to the numbers presented in \cite{DiNardo2004}, \cite{Ferguson2008}, and \cite{Frandsen2020}.

\clearpage
\renewcommand{\thesection}{C}
\renewcommand{\thetable}{C.\arabic{table}}
\setcounter{table}{0}
\renewcommand{\thefigure}{C.\arabic{figure}}
\setcounter{figure}{0}

\renewcommand{\baselinestretch}{1.5}
\setlength{\baselineskip}{20pt}
\renewcommand{\arraystretch}{0.93}

\section{Instrumental Variable Strategy}
\label{sec:app_iv}

\bigskip

In this section, we describe our IV approach, which complements the DiD strategy with arguably exogenous variation in union support driven by spikes in work-related fatalities. Specifically, we use as instrument sector-level fatal work accidents in the 30 days before a union election.\footnote{The median time between petition and union election in our sample is 45 days. Less than 5\% of all elections are held within 30 days after the petition. Thus, for almost all elections, the election date is fixed during our time of instrument exposure.} Safety at work is a fundamental concern to all workers, especially when one's life is in danger. Work-related fatalities are still common in the United States. In 2023, the Occupational Safety and Health Administration (OSHA) reported 5,283 deaths at work, more than 14 per day on average. Workplace safety is a central theme in union campaigns, and unions can be successful in improving safety conditions \citep[e.g.,][]{AFLCIO2022, Hagedorn2016, Li2022}. It is therefore plausible that increased attention to serious workplace hazards -- such as fatal accidents -- leads to a surge in support for unionization.

\begin{figure}[h]%
\begin{center}
	\caption{\label{fig:iv_fat}\centering Seasonally Adjusted Fatal Work Accidents, 1984-2012}%
	\includegraphics[width=.9\linewidth]{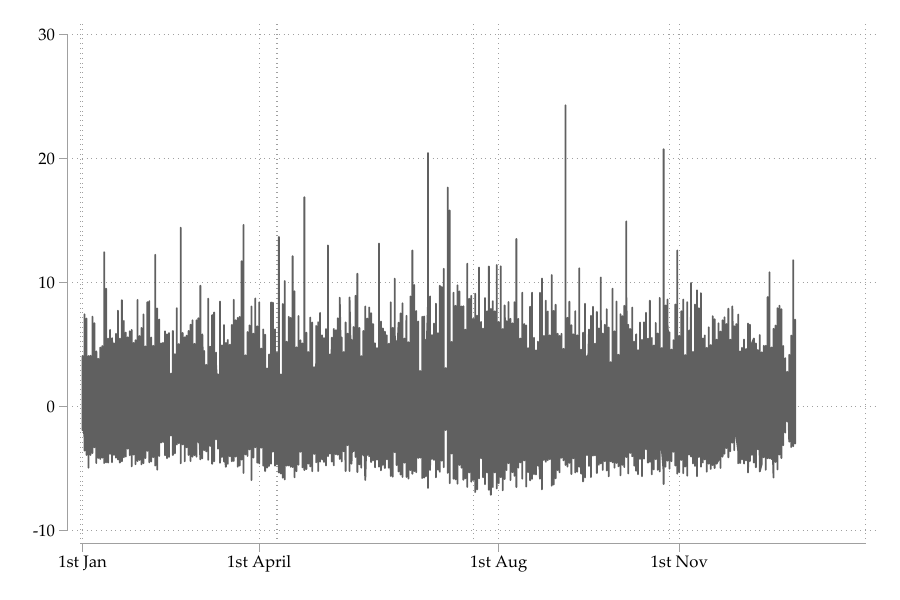} 
	\end{center}
        \vspace{-1em}
	\begin{minipage}{\textwidth} {\scriptsize \smallskip 
 \textbf{Notes:} The graph shows the number of fatalities caused by work accidents on a given day of a year (e.g., January 1st) for all years in our sample period after the mean number of fatalities on that given day over our sample period (e.g., mean number of fatalities on January 1st between 1984 and 2012) is subtracted.\par}
    \end{minipage}
\end{figure}

We implement the IV approach by estimating the following two-stage model:
\begin{equation}\label{1st_stage}
V_{j} = \alpha_1 + \alpha_{2} A_{st} + \alpha_{3} A_{st}^2 + \alpha_{4} A_{st} \times FR_{s} + \alpha_{5} FR_{s} + \alpha_{6} X_{j} + \gamma_t + \mu_m + \epsilon_{j}
\end{equation}
\begin{equation}\label{2nd_stage}
\Delta y_{j} = \beta_1 + \beta_2 \mathbbm{1}[\widehat{V_j} > .5] + \beta_{3} FR_{s} + \beta_{4} X_{j} + \gamma_t + \mu_m + \epsilon_{j},
\end{equation}
where $\Delta y_{j}$ denotes the change in campaign contribution patterns in the three cycles after the union election relative to the three cycles before (excluding the cycle of the union election). By using changes as the outcome variable, the specification builds on the DiD approach and accounts for time-invariant differences between establishments that may affect the level of campaign contributions. Our main instrument is $A_{st}$, which represents the number of fatal accidents in 2-digit sector $s$ in the 30 days prior to the election after accounting for seasonal variation.\footnote{Data on fatal work accidents is obtained from OSHA in the form of Fatality and Catastrophe Investigation Summaries (OSHA form 170). Figure \ref{fig:iv_fat} depicts the exploited time variation.} We allow for a non-linear effect by including $A_{st}^2$ and for a larger impact of fatalities in sectors where fatalities are common and where workers may be more concerned about workplace safety by the interaction term $A_{st} \times FR_{s}$ ($FR_{s}$ denotes the share of fatal work accidents occurring in a given sector out of all fatal work accidents in the sample). Importantly, instead of directly instrumenting union victory in a standard 2SLS approach, the first stage explains the continuous pro-union vote share $V_{j}$. In the second stage, we then use an indicator for predicted victory that is based on the predicted vote share in the first stage. This approach resembles the treatment assignment process and exploits the maximum available information. To account for the uncertainty from the first-stage regression, we compute standard errors by bootstrapping the two-stage model. The model includes several control variables: the main effect of $FR_s$, the number of fatalities and employees at the sector-year level (to account for sector-specific fatality trends), the number of eligible voters in the union election (as precision control), as well as year fixed effects $\gamma_t$ and month-of-the-year fixed effects $\mu_m$.

Results are reported in Table \ref{tab:iv}. The first-stage results show that sector-level fatal work accidents are a significant predictor of the union election outcome, with an F-statistic of 16.5. We find that the positive effect of spikes in work accidents on unionization is stronger in sectors where work accidents are more common, i.e., where workplace safety may be a greater concern for workers. The second-stage results confirm our main findings from the DiD model, highlighting a leftward shift in campaign contributions in response to unionization. The magnitude of the coefficients is comparable but slightly larger than in the DiD model. As compliers respond to information on fatal work accidents, we deem it plausible that they also react more strongly to information provided by unions and to changes to their work environment induced by unionization. The estimates are considerably less precise, however. While the effects on the party composition of contributions from managers are still significant at the 5\% level, the effects for workers are no longer significant.\footnote{We also verify the IV approach with a falsification exercise. We re-estimate model (\ref{2nd_stage}) using the change in campaign contribution patterns between $t-1$ and $t-2$ as the outcome. We do not find any evidence for pre-existing differential trends related to spikes in fatal work accidents.}

\clearpage

\clearpage

\begin{table}[h]\centering
	\def\sym#1{\ifmmode^{#1}\else\(^{#1}\)\fi}
	\caption{Instrumental Variable Results}
	\label{tab:iv}
		\begin{tabular}{l*{6}{c}}
			\toprule[1.5pt]
			&\multicolumn{3}{c}{IHS(\$ to all candidates)}&\multicolumn{3}{c}{IHS(\$ to Dem.) $-$ IHS(\$ to Rep.)} \\ \cmidrule(lr){2-4} \cmidrule(lr){5-7}
			&All&Workers&Managers&All&Workers&Managers \\	
			&(1)&(2)&(3)&(4)&(5)&(6) \\	
			\midrule
			\multicolumn{7}{l}{[A] OLS} \\
			$\mathbbm{1}[V_j > .5]$    
			&      -0.092         &      0.038         &     -0.072         &       0.227\sym{***}&       0.089\sym{**}&       0.232\sym{***}\\
			&    (0.082)         &    (0.044)         &    (0.062)         &    (0.079)         &    (0.041)         &    (0.056)         \\
			[0.5em]
			\midrule 
			\multicolumn{7}{l}{[B] 2nd stage} \\
			$\mathbbm{1}[\widehat{V_j} > .5]$    
			&      0.036         &      0.086         &     -0.042         &       0.334\sym{*}&       0.115&       0.260\sym{**}\\
			&    (0.174)         &    (0.097)         &    (0.134)         &    (0.176)         &    (0.086)         &    (0.125)         \\
			[0.5em]
			\midrule 
			\multicolumn{7}{l}{[C] 1st stage} \\
			$A_{st}$    
			&     0.002         &      0.002         &     0.002         &       0.002 &       0.002 &       0.002 \\
			&    (0.007)         &    (0.007)         &    (0.007)         &    (0.007)         &    (0.007)         &    (0.007)         \\
			[0.5em]
			$A_{st}^2$    
			&     -0.003\sym{**}         &      -0.003\sym{**}         &     -0.003\sym{**}         &       -0.003\sym{**} &       -0.003\sym{**} &       -0.003\sym{**} \\
			&    (0.001)         &    (0.001)         &    (0.001)         &    (0.001)         &    (0.001)         &    (0.001)         \\
			[0.5em]
			$A_{st} \times FR_{s}$    
			&     0.223\sym{***}         &      0.223\sym{***}         &     0.223\sym{***}         &       0.223\sym{***} &       0.223\sym{***} &       0.223\sym{***} \\
			&    (0.055)         &    (0.055)         &    (0.055)         &    (0.055)         &    (0.055)         &    (0.055)         \\
			[1em]
   			K-P F-stat & 16.50 			  & 16.50		 &         16.50           &       16.50  &       16.50 &       16.50    \\
                [0.5em]
			\midrule 
			\multicolumn{7}{l}{[D] 2nd stage falsification: pre-trend} \\
			$\mathbbm{1}[\widehat{V_j} > .5]$    
			&      -0.007         &      0.093         &     0.033         &       0.124&       -0.020&       0.046\\
			&    (0.207)         &    (0.094)         &    (0.116)         &    (0.230)         &    (0.100)         &    (0.129)         \\
			[0.5em]
			\hline\hline
			\multicolumn{7}{l}
			{
				\begin{minipage}{0.85\linewidth} \scriptsize \smallskip 
					\textbf{Notes:} The table reports IV results  for the effect of unionization on the IHS-transformed total amount contributed (columns (1) - (3)) and on the difference between the IHS-transformed amounts contributed to Democratic and Republican candidates (columns (4) - (6)). Panel A reports OLS coefficients, Panel B reports the second-stage coefficients from model (\ref{2nd_stage}), and Panel C reports the first-stage coefficients from model (\ref{1st_stage}). In Panels A and B, the outcome is the difference between the average outcome in the three cycles after and the average outcome in the three cycles before the union election (excluding the cycle of the union election). In Panel D, the outcome is the change between one and two cycles before the union election. $N=5,803$ establishments. Bootstrapped standard errors (with 500 replications), shown in parentheses, are clustered at the establishment level. \sym{*} \(p<0.10\), \sym{**} \(p<0.05\), \sym{***} \(p<0.01\)
				\end{minipage}
			}\\
		\end{tabular}
\end{table}

\clearpage
\renewcommand{\thesection}{D}
\renewcommand{\thetable}{D.\arabic{table}}
\setcounter{table}{0}
\renewcommand{\thefigure}{D.\arabic{figure}}
\setcounter{figure}{0}

\renewcommand{\baselinestretch}{1.5}
\setlength{\baselineskip}{20pt}
\renewcommand{\arraystretch}{0.93}

\section{Effects of Losing a Union Election}
\label{sec:app_losing}

\bigskip

We estimate the effects of losing a union election compared to holding no election by using establishments that hold and lose an election in the future as a control group. Consider the treatment cohort of elections that were held and lost in the cycle 1985/86. Given that we observe each establishment only up to three cycles before the union election, we can use elections held and lost in the next two cycles as control cohorts. The untreated pre-election observations of the 1987/88 control cohort refer to the cycles 1981/82, 1983/84, and 1985/86 (event times $k=\{-2,-1,0\}$ of the treated cohort), and those of the 1989/90 control cohort refer to the cycles 1983/84, 1985/86, and 1987/1988 (event times $k=\{-1,0,1\}$ of the treated cohort). Note that later cohorts are not observed before the treated cohort hold their election and can therefore not be used in a DiD comparison. Consequently, we only have untreated observations that we can compare to the treated cohort's observations in cycles 1981/82, 1983/84, 1985/86, and 1987/88 (event times $k=\{-2,-1,0,1\}$). This means we can only identify short-term effects.  

Given these considerations, we implement a stacked DiD model as follows. For each cohort of lost elections in cycle c, we create a cohort-specific dataset that is built from cycles in event times $k=\{-2,-1,0,1\}$ of the treated cohort $c_j=c$ and from the three pre-election cycles of lost elections in the control cohorts $c_j=\{c+1,c+2\}$. Then, the stacked DiD model is estimated as:
\begin{equation}\label{model_losing}
y_{jk} = \alpha_{jc} + \beta_{kc} + \delta_{\textrm{DiD}} \times \Big( \mathbbm{1}[k\geq 0] \times \mathbbm{1}[c_j = c] \Big) + \epsilon_{jk}
\end{equation}
where $k$ now denotes the number of cycles relative to the cycle when the treated cohort held its union election. Establishment fixed effects are now saturated with indicators for the cohort-specific dataset $c$ to account for the fact that establishments enter several datasets. The DiD coefficient $\delta_{\textrm{DiD}}$ is given by the interaction between a dummy for post-election cycles of the treated cohort ($k\geq 0$) and a dummy for the treated cohort ($c_i = c$). Results are reported in Panel A of Table \ref{tab:losing}. 

In Panels B and C of Table \ref{tab:losing}, we also show results for the alternative staggered DiD estimators by \cite{Borusyak2021} and \cite{Callaway2021}. In line with our stacking implementation, in settings with no never-treated units, both estimators use not-yet-treated observations as controls. The methods differ from the stacked DiD model in the number of pre-treatment periods used and the aggregation of unit- or cohort-specific effects. In our results, however, the estimates are very similar to those of the stacked DiD model.  

\clearpage

\begin{table}[h]\centering
	\caption{Effects of Losing a Union Election}
	\label{tab:losing}
	\begin{tabular}{l*{6}{c}}
		\toprule[1.5pt]
		&\multicolumn{3}{c}{IHS(\$ to all candidates)}&\multicolumn{3}{c}{IHS(\$ to Dem.) $-$ IHS(\$ to Rep).} \\ \cmidrule(lr){2-4} \cmidrule(lr){5-7}
		&All&Workers&Managers&All&Workers&Managers \\	
		&(1)&(2)&(3)&(4)&(5)&(6) \\	
		\midrule
		\multicolumn{7}{l}{[A] Stacking} \\
		[0.5em]
		$\delta_{\textrm{DiD}}$    
		&     -0.049         &     -0.026         &      0.070         &      0.057         &     -0.013         &      0.037         \\
		&    (0.088)         &    (0.040)         &    (0.053)         &    (0.097)         &    (0.043)         &    (0.057)         \\
		[0.5em]
		N           &       31,501         &       31,501         &       31,501         &       31,501         &       31,501         &       31,501         \\			
		\midrule
		\multicolumn{7}{l}{[B] Borusyak, Jaravel, and Spiess (2021)} \\
		[0.5em]
		$\delta_{\textrm{DiD}}$    
		&     -0.048         &     -0.029         &      0.075         &      0.080         &    -0.007         &      0.048         \\
		&    (0.090)         &    (0.045)         &    (0.059)         &     (0.100)         &    (0.049)         &    (0.064)         \\
		[0.5em]
		N           &       16,658         &       16,658         &       16,658         &       16,658         &       16,658         &       16,658         \\
		\midrule
		\multicolumn{7}{l}{[C] Callaway and Sant'Anna (2021)} \\
		[0.5em]
		$\delta_{\textrm{DiD}}$    
		&     -0.043         &     -0.038         &      0.061         &      0.076         &    -0.007         &      0.053         \\
		&    (0.095)         &    (0.047)         &    (0.064)         &     (0.105)         &    (0.052)         &    (0.070)         \\		
		[0.5em]
		N           &       16,658         &       16,658         &       16,658         &       16,658         &       16,658         &       16,658         \\
					[0.5em]		
		\hline\hline
		\multicolumn{7}{l}{
			\begin{minipage}{0.78\linewidth} \scriptsize \smallskip 
				\textbf{Notes:} The table presents DiD estimates for the effect of losing a union election versus holding no election. We compare establishments with a lost union election in a given cycle (treated cohort) with establishments with a lost union election in one of the next two cycles (control cohorts) in a DiD design. Thereby, we estimate short-term effects of losing an election (i.e., for event times $k=\{0,1\}$). Panel A shows results from a stacked DiD model, and Panels B and C implement the staggered DiD estimators of \cite{Borusyak2021} and \cite{Callaway2021}. See Appendix \ref{sec:app_losing} for details of the implementation. Standard errors clustered at the establishment level are in parentheses. \sym{*} \(p<0.10\), \sym{**} \(p<0.05\), \sym{***} \(p<0.01\)
			\end{minipage}
		}\\
	\end{tabular}
\end{table}

\end{appendix}

\end{document}